\documentclass[%
 preprint,
showpacs,
 amsmath,amssymb, showkeys
 aps,
]{revtex4-1}

\newcommand{\change}{}
\usepackage{epsfig}
\usepackage{epstopdf}

\usepackage{epstopdf}
\usepackage{graphicx}
\usepackage{amssymb}
\usepackage{mathrsfs}
\usepackage{amsthm}
\usepackage{mhchem}
\usepackage{subcaption}
\usepackage{circuitikz}
\usepackage{float}
\usepackage{amsmath}
\DeclareMathOperator\erfc{erfc}
\DeclareMathOperator\erf{erf}
\DeclareMathOperator\erfi{erfi}
\DeclareMathOperator\sgn{sgn}
\DeclareMathOperator\arctanh{arctanh}
\DeclareMathOperator\arcsinh{arcsinh}
\usepackage{dcolumn}% Align table columns on decimal point
\usepackage{bm}% bold math
\usepackage[normalem]{ulem}

\begin{document}

\preprint{APS/123-QED}

\title{Theory of linear sweep voltammetry with diffuse charge: unsupported electrolytes, thin films, and leaky membranes}%

\author{David Yan}
 \email{dyan@ece.utoronto.ca}
\affiliation{Department of Electrical and Computer Engineering, University of Toronto}
\author{Martin Z. Bazant}%

\affiliation{%
 Department of Chemical Engineering and Department of Mathematics, Massachusetts Institute of Technology
}%

\author{P. M. Biesheuvel}
\affiliation{
Wetsus, European Centre of Excellence for Sustainable Water Technology, The Netherlands
}%
\author{Mary C. Pugh}
\affiliation{%
Department of Mathematics, University of Toronto
}%
\author{Francis P. Dawson}
\affiliation{%
Department of Electrical and Computer Engineering, University of Toronto
}%

\date{\today}% It is always \today, today,

%\begin{frontmatter}

\begin{abstract}
Linear sweep and cyclic voltammetry techniques are important tools for electrochemists and have a variety of applications in engineering. Voltammetry has classically been treated with the Randles-Sevcik equation, which assumes an electroneutral supported electrolyte. %No general theory of linear-sweep voltammetry is available, however, for unsupported electrolytes and for other situations where diffuse charge effects play a role. 
In this paper, we provide a comprehensive mathematical theory of voltammetry in electrochemical cells with unsupported electrolytes and for other situations where diffuse charge effects play a role\change{, and present analytical and simulated solutions of the} time-dependent Poisson-Nernst-Planck (PNP) equations with generalized Frumkin-Butler-Volmer (FBV) boundary conditions for a 1:1 electrolyte and a simple reaction. \change{Using these solutions, we construct} theoretical and simulated current-voltage curves for liquid and solid thin films, membranes with fixed background charge, and cells with blocking electrodes. The full range of dimensionless parameters is considered, including the dimensionless Debye screening length (scaled to the electrode separation), Damkohler number (ratio of characteristic diffusion and reaction times) and dimensionless sweep rate (scaled to the thermal voltage per diffusion time). The analysis focuses on the coupling of Faradaic reactions and diffuse charge dynamics, \change{although} capacitive charging of the electrical double layers (EDL) is also studied, for early time transients at reactive electrodes and for non-reactive blocking electrodes. Our work highlights cases where diffuse charge effects are important in the context of voltammetry, and illustrate\change{s} which regimes can be approximated using simple analytical expressions and which require more careful consideration.
\end{abstract}
%\keywords{Linear Sweep Voltammetry, Diffuse charge, Double layers, Frumkin-Butler-Volmer, Poisson-Nernst-Planck}
%\begin{keyword}
%Linear Sweep Voltammetry \sep Diffuse charge \sep Double layers  \sep Frumkin-Butler-Volmer \sep Poisson-Nernst-Planck
%\end{keyword}
%\end{frontmatter}

%\linenumbers
\maketitle

%\section*{List of Symbols}
%Note: we take the general convention that, when both a dimensional and dimensionless version of a quantity appear, the upper case and/or script symbol denotes the dimensional quantity, whereas a lower case symbol denotes the nondimensional quantity. In certain cases, a tilde is used instead to denote the dimensionless quantity.
%\begin{multicols}{2}
%\\%
%\setlength{\parindent}{0cm}{
%$X$, $x$:  Length \\
%$\tau$, $t$: Time \\
%$C$, $c$: Concentration \\
%$\Phi$, $\phi$: Potential \\
%$z$: Charge number \\
%$D$: Diffusivity \\
%$R$: Gas constant \\
%$T$: Temperature \\
%$\varepsilon_s$: Dielectric permittivity \\
%$F$: Faraday's constant \\
%$\rho_s$, $\tilde{\rho_s}$: Background charge density \\
%$K_c$, $k_c$: Forward reaction rate \\
%$K_a$, $j_r$: Backward reaction rate \\
%$\alpha$: Transfer coefficient \\
%$\lambda_s$: Stern layer width \\
%$J$, $j$: Electrical current density \\
%$S$, $\tilde S$: Voltage sweep rate \\
%$V$, $v$: Cell potential \\
%$L$: Interelectrode width \\
%$\epsilon$: Nondimensional Debye length \\
%$\delta$: Ratio of Stern width to Debye length \\
%$\tilde {R}$: Rescaled resistance \\
%$\tilde {\mathcal C}$: Nondimensional capacitance \\
%$\tilde {C}$: Rescaled capacitance \\
%$\eta$: Overpotential}
%\end{multicols}

\section{Introduction}\label{introduction}
Polarography/linear sweep voltammetry (LSV)/cyclic voltammetry (CV) is the most common method of electro-analytical chemistry~\cite{Bard2001,Compton2011,Compton2014}, pioneered by Heyrovsky and honored by a Nobel Prize in Chemistry in 1959. The classical Randles-Sevcik theory of polarograms is based on the assumption of diffusion limitation of the active species in a neutral liquid electrolyte, driven by fast reactions at the working electrode \cite{Randles1947, Randles1948, Sevcik1948}. Extensions for slow Butler-Volmer kinetics (or other reaction models) are also available \cite{Bard2001}.  The half-cell voltage is measured at a well-separated reference electrode in the bulk liquid electrolyte, which is assumed to be electro-neutral, based on the use of a supporting electrolyte \cite{Newman2012}. 

Most classical voltammetry experiments and models featured a supported electrolyte: an electrolyte with an inert salt added to remove the effect of electromigration. In many electrochemical systems of current interest, however, the electrolyte is unsupported, doped, or strongly confined by electrodes with nanoscale dimensions. Examples include super-capacitors, \cite{Simon2008}, capacitive deionization \cite{Biesheuvel2010PRE}, pseudo-capacitive deionization and energy storage~\cite{Biesheuvel2011,Biesheuvel2012}, electrochemical thin films~\cite{Bazant2005,Chu2005,Biesheuvel2009,Soestbergen2012}, solid electrolytes used in Li-ion/Li-metal \cite{Luntz2015,Tarascon2011}, electrochemical breakdown of integrated circuits \cite{Muralidhar2014}, fuel cells \cite{BiesheuvelFranco2009, Franco2007}, nanofludic systems \cite{BazantSquires2004, Squires2005, Schoch2008}, electrodialysis \cite{Rubinstein1979, Rubinstein1988, Nikonenko2010}, and charged porous ``leaky membranes'' \cite{Dydek2011, Dydek2013} for shock electrodialysis~\cite{Deng2013,Schlumpberger2015} and shock electrodeposition~\cite{Han2014,Han2016}. In all of these situations, diffuse ionic charge must play an important role in voltammetry, which remains to be fully understood. 

Despite an extensive theoretical and experimental literature on this subject (reviewed below), to the authors' knowledge, there has been no comprehensive mathematical modeling of the effects of diffuse charge on polarograms, while fully taking into account time-dependent electromigration and Frumkin effects of diffuse charge on Faradaic reaction rates. The goal of this work is thus to construct and solve a general model for voltammetry for charged electrolytes. We consider the simplest Poisson-Nernst-Planck (PNP) equations for ion transport for dilute liquid and solid electrolytes and leaky membranes, coupled with ``generalized'' Frumkin-Butler-Volmer (FBV) kinetics for Faradaic reactions at the electrodes \cite{Bazant2005,Itskovich1977, Kornyshev1981, Bonnefont2001, He2006, Streeter2008}, as reviewed by Biesheuvel, Soestbergen and Bazant \cite{Biesheuvel2009} and extended in subsequent work~\cite{Biesheuvel2011, Biesheuvel2012, Soestbergen2012}. Unlike most prior analyses, however, we make no assumptions about the electrical double layer (EDL) thickness, Stern layer thickness, sweep rate, or reaction rates in numerical solutions of the full PNP-FBV model. We also define a complete set of dimensionless parameters and take various physically relevant limits to obtain analytical results whenever possible, which are compared with numerical solutions in the appropriate limits. 

The paper is structured as follows. We begin with a review of theoretical and experimental studies of diffuse-charge effects in voltammetry in Section \ref{history}. The PNP-FBV model equations are formulated and cast in dimensionless form in Section \ref{model}, followed by voltammetry simulations on a single electrode in a supported and unsupported electrolyte in Section \ref{bulk_liquid}. Section \ref{thinfilms} shows simulations on liquid and solid electrochemical thin films, considering both the thin and thick double layer limits. In Section \ref{leaky}, we simulate ramped voltages applied to porous ``leaky" membranes with fixed background charge. Finally, ramped voltages are applied to systems with one or two blocking electrodes in Section \ref{blocking}, and simulation results are compared with theoretical capacitance curves. We conclude with a summary and outlook for future work.

\section{Historical Review}
\label{history}

The modeling of diffuse charge dynamics in electrochemical systems has a long history, reviewed by Bazant, Thornton and Ajdari~\cite{Bazant2004}. Frumkin~\cite{Frumkin1933} and Levich~\cite{Levich1949} are credited with identifying the need to consider effects of diffuse charge on Butler-Volmer reaction kinetics at electrodes in unsupported electrolytes by pointing out that the potential drop which drives reactions is across the Stern layer rather than in relation to a reference electrode far away, although complete mathematical models for dynamical situations were not formulated until much later. In electrochemical impedance spectroscopy (EIS)~\cite{Barsoukov2005}, it is well known that the double-layer capacitance must be considered in parallel with the Faradaic reaction resistance in order to describe the interfacial impedance of electrodes. Since the pioneering work of Jaff{\' e}~\cite{Jaffe1933,Jaffe1952} for semiconductors and Chang and Jaff{\' e} for electrolytes~\cite{Chang1952}, Macdonald~\cite{Macdonald1973,Macdonald1978} and others have formulated microscopic PNP-based models using the Chang-Jaff{\' e} boundary conditions~\cite{Macdonald2011}, which postulate mass-action kinetics, proportional to the active ion concentration at the surface. This approximation includes the first ``Frumkin correction'', i.e. the jump in active species concentration across the diffuse layer, but not the second, related to the jump in electric field that drives electron transfer at the surface~\cite{Bazant2005,Biesheuvel2009}.  More importantly, all models used to interpret EIS measurements, whether based on microscopic transport equations or macroscopic equivalent circuit models, only hold for the small-signal response to an infinitesimal sinusoidal applied voltage or current.

The {\it nonlinear} coupling of diffuse charge dynamics with Faradaic reaction kinetics has received far less attention and, until recently, has been treated mostly by empirical macroscopic models, as reviewed by Biesheuvel, Soestbergen and Bazant~\cite{Biesheuvel2009}. Examples of {\it ad hoc} approximations include fixing the electrode charge or zeta potential, independently from the applied current or voltage, and assuming a constant double layer capacitance in parallel with a variable Faradaic reaction resistance given by the Butler-Volmer equation. Such empirical models are theoretically inconsistent, because the electrode charge and double-layer capacitance are not constant independent variables to be fitted to experimental data. Instead, the microscopic model must include a proper electrostatic boundary condition, relating surface charge to the jump in the normal component of the Maxwell dielectric displacement~\cite{Jackson1962}, and the Butler-Volmer equation, or another reaction-rate model, must be applied at the same position (the ``reaction plane"). This inevitably leads to nonlinear dependence of the electrode surface charge on the local current density~\cite{Biesheuvel2011, Bazant2005}, which includes both Faradaic current from electron-transfer reactions and Maxwell displacement current from capacitive charging~\cite{Bonnefont2001,Bazant2004,Soestbergen2010}.

Interest in voltammetry in low conductivity solvents and dilute electrolytes without little to no supporting electrolyte has slowly grown  since the 1970s. Since the work of Buck~\cite{Buck1973}, a variety of PNP-based microscopic models have been proposed~\cite{Norton1990,Murphy1992,Smith1993,Oldham2001}, which provide somewhat different boundary conditions than the FBV model described below. Since the 1980s, various experiments have focused on the role of supporting electrolyte~\cite{Compton2011}. For example,  Bond and co-workers performed linear sweep \cite{Bond1984A} and cyclic \cite{Bond1985, Bond2001} voltammetry using the ferrocene oxidation reaction on a microelectrode while varying supporting electrolyte concentrations.  
One of the primary theoretical concerns at that time was modeling the extra Ohmic drop in the solution from the  electromigration effects which entered into the physics due to low supporting electrolyte concentration. Bond \cite{Bond1984B} and Oldham \cite{Oldham1988} both solved the PNP equations for steady-state voltammetry in dilute solutions with either Nerstian or Butler-Volmer reaction kinetics to obtain expressions for the Ohmic drop. While they listed the full PNP equations, they analytically solved the system only with the electroneutrality assumption or for small deviations from electroneutrality, which is a feature of many later models as well \cite{Myland1999, Oldham1999, Oldham2001}. The bulk electroneutrality approximation is generally very accurate in macroscopic electrochemical systems \cite{Newman2012,Oldham2016}, although care must be taken to incorporate diffuse charge effects properly in the boundary conditions, as explained below. Bento, Thouin and Amatore \cite{Bento1998A, Amatore1999} continued this type of experimental work on voltammetry and studied the effect of diffuse layer dynamics and migrational effects in electrolytes with low support. They also performed voltammetry experiments on microelectrodes in solutions while varying the ratio of supporting electrolyte to reactant, and noted shifts in the resulting voltammograms and a change in the solution resistance \cite{Bento1998B}.

More recently, Compton and co-workers have done extensive work involving theory \cite{Dickinson2011, Dickinson2011rev, Batchelor-Mcauley2015}, simulations \cite{Streeter2008, Dickinson2011, Dickinson2010, Belding2012} (and see Chapter 7 of \cite{Compton2014} or Chapter 10 of \cite{Compton2011}) and experiments \cite{Limon-Petersen2008, Limon-Petersen2009, Belding2010, Dickinson2009} on the effect of varying the concentration of the supporting electrolyte on voltammetry.  Besides considering a different (hemispherical) electrode geometry motivated by ultramicroelectrodes~\cite{aoki1993},  there are two significant differences between this body of work and ours, from a modeling perspective. 

The first difference is that Compton and co-authors consider three or more
species in their models in order to account for the supporting
electrolyte and the effect of varying its concentration. The additional species in their work are sometimes neutral
\cite{Limon-Petersen2008, Limon-Petersen2009} and sometimes charged
\cite{Dickinson2011, Belding2012}, depending on the reaction being
modeled. Modeling of uncharged and supporting ionic species has also been
a feature of other models in the literature \cite{Bond1984B, Oldham1988,
Myland1999}. In this work, we only consider the two extremes: either an
unsupported binary electrolyte  or (briefly, for comparison) a 
fully supported electrolyte.

The second difference lies in their treatment of specific adsorption of ions and a related approximation used to simplify the full model. Although some studies involve numerical solutions of the full PNP equations with suitable electrostatic and reaction boundary conditions~\cite{Streeter2008}, many others employ the ``zero-field approximation'' for the thin double layers \cite{Streeter2008, Dickinson2010}, which is motivated by strong specific adsorption of ions. In this picture, the Stern layer~\cite{Stern1924} (outside the continuum region of ion transport) is postulated to have two parts: an outer layer of adsorbed ions that fully screens the surface charge, and an inner uncharged dielectric layer that represents solvent molecules on the electrode surface. The assumption of complete screening motivates imposing a vanishing normal electric field as the boundary condition for the neutral bulk electrolyte at the electrode outside the double layers. This assertion also has the effect of eliminating the electromigration term from the flux entering the double layers, despite the inclusion of this term in the bulk mass flux. Physically, this model assumes that the electrode always remains close to the potential of zero charge, even during the passage of transient large currents. 

Here, we base our analysis on the PNP equations with ``generalized FBV" boundary conditions~\cite{Biesheuvel2009, Soestbergen2012}, to describe Faradaic reactions at a working electrode, whose potential is measured relative to the point of zero charge, in the absence of specific adsorption of ions. The model postulates a charge-free Stern layer of constant capacitance, whose voltage drop drives electron transfer according to the Butler-Volmer equation, where the exchange current is determined by the local concentration of active ions at the reaction plane. This approach was first introduced by Frumkin and later applied by
Itskovich, Kornyshev and Vorontyntsev~\cite{Itskovich1977,Kornyshev1981}
in the context of solid electrolytes, with one mobile ionic species and
fixed background charge. For general liquid or solid electrolytes, the FBV
model based on the Stern boundary condition was perhaps first formulated
by Bazant and co-workers~\cite{Bazant2005, Biesheuvel2009, Bonnefont2001}
and independently developed by He et al.~\cite{He2006} and Streeter and
Compton~\cite{Streeter2008}. The FBV-Stern model was extended to porous
electrodes and multicomponent electrolytes by Biesheuvel, Fu and
Bazant~\cite{Biesheuvel2011, Biesheuvel2012}.

Most of the literature on modeling diffuse-charge effects in electrochemical cells has focused on either linear AC response (impedance) or nonlinear steady-state response (differential resistance), with a few notable exceptions.  The PNP-FBV model was apparently first applied to chronoampereometry (response to a voltage step) by Streeter and Compton~\cite{Streeter2008} and to  chonopotentiometry (response to a current step) by Soestbergen, Biesheuvel and Bazant~\cite{Soestbergen2010}, each in the limit of thin double layers for unsupported binary liquid electrolytes.  Here, we extend this work to voltammetry and consider a much wider range of conditions, including thick double layers, fixed background charge, and slow reactions.  

It is important to emphasize the generality of the PNP-FBV framework, which is not limited to thin double layers and stagnant neutral bulk electrolytes. Chu and Bazant~\cite{Chu2005} analyzed the model for a binary electrolyte under extreme conditions of over-limiting current, where  the double layer loses its quasi-equilibrium structure and expands into an extended bulk space charge layer, and performed simulations of steady transport across a thin film at currents over 25 times the classical diffusion-limited current (which might be achieved in solid, ultra-thin films).  Transient space charge has also been analyzed in this way for large applied voltages, for both blocking electrodes~\cite{Bazant2004,Olesen2010} and FBV reactions~\cite{Soestbergen2010}. A validation of the PNP-FBV theory was recently achieved by Soestbergen~\cite{Soestbergen2012a}, who fitted the model to experimental data of Lemay and co-workers for planar nano-cavities~\cite{Zeven2009a,Zeven2009b}.  The FBV model has also been applied to nonlinear ``induced-charge" electrokinetic phenomena~\cite{Bazant2009}, in the asymptotic limit of thin double layers. Olesen, Bruus and Ajdari~\cite{Olesen2006,Olesen_thesis} used the quasi-equilibrium FBV double-layer model in a theory of AC electro-osmotic flow over micro-electrode arrays.  Moran and Posner~\cite{Moran2011} applied the same approach to reaction-induced-charge-electrophoresis of reactive metal colloidal particles in electric fields~\cite{Moran2010}.

Many of these recent studies have exploited effective boundary conditions for thin double layers, which were systematically derived by asymptotic analysis and compared to full solutions of the PNP equations, rather than {\it ad hoc} approximations.  Matched asymptotic expansions were first applied to steady-state PNP transport in the 1960s without considering FBV reaction kinetics~\cite{Grafov1962, Newman1965, Smyrl1966, Macgillivray1968} and used extensively in theories of ion transport~\cite{Rubinstein1979} and electro-osmotic fluid instabilities~\cite{Rubinstein2001,Zaltzman2007} in electrodialysis, involving ion-exchange membranes rather than electrodes.  Baker and Verbrugge~\cite{Baker1997} analyzed a simplified problem with fast reactions, where the active species concentration vanishes at the electrode.  Bazant and co-workers first used matched asymptotic expansions to treat Faradaic reactions using the PNP-FBV framework, applied to steady conduction through electrochemical thin films~\cite{Bazant2005,Chu2005}.  Richardson and King~\cite{Richardson2007} extended this approach to derive effective boundary conditions for time-dependent problems, which provides rigorous justification for subsequent studies of transient electrochemical response using the thin double layer approximation, including ours.

Finally, for completeness, we note that Moya et al.~\cite{Moya2015} recently analyzed a model of double-layer effects on LSV for the case of a neutral electrolyte surrounding an ideal ion-exchange membrane with thin quasi-equilibrium double layers, and no electrodes with Faradaic reactions to sustain the current.

\section{Mathematical Model}\label{model}
\subsection{Modeling Domain}
\begin{figure}[htb!]
\centering
    \begin{subfigure}[htb!]{0.49\textwidth}
        \centering
        \includegraphics[width=\linewidth]{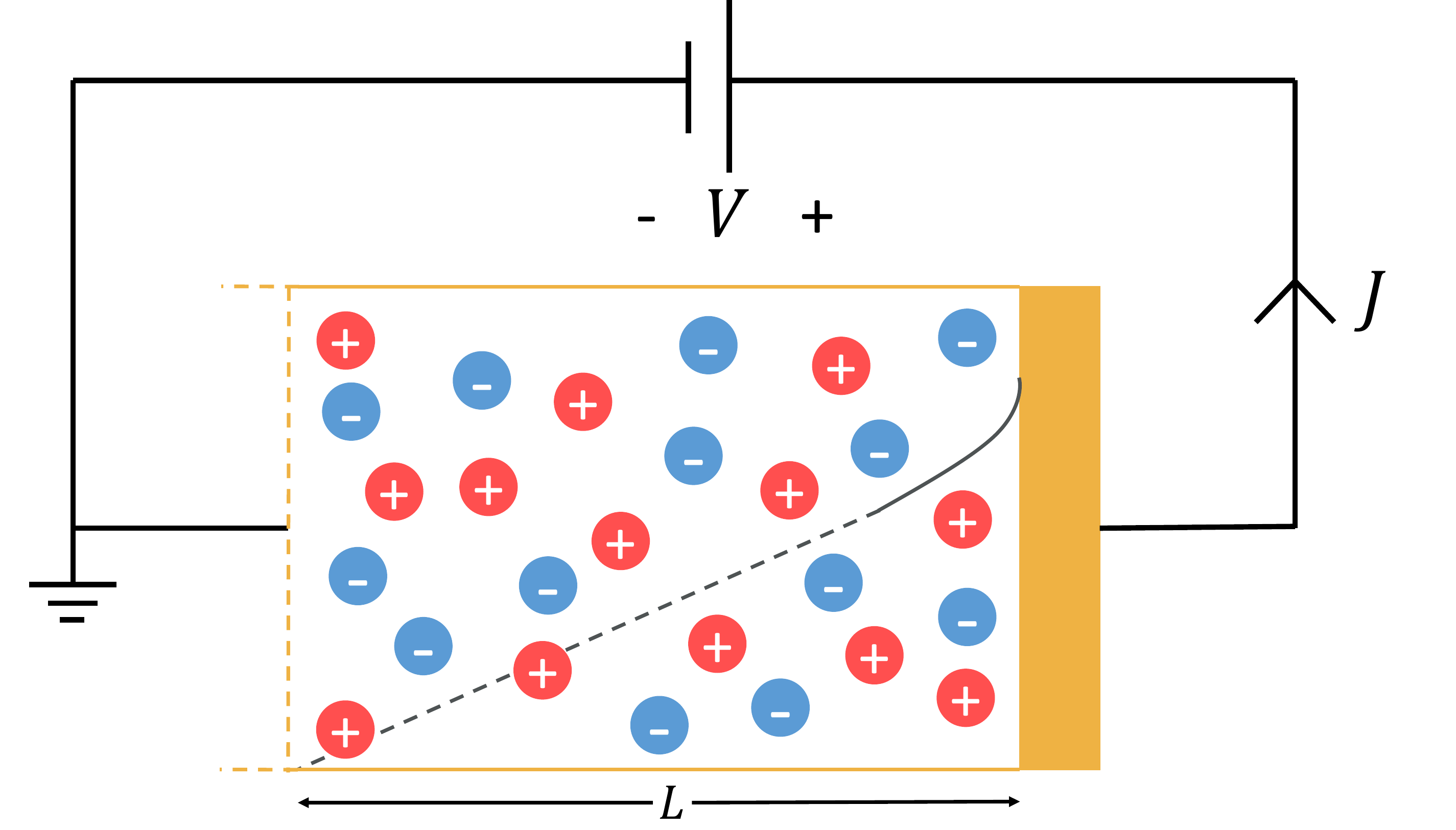}
        \caption{Single electrode}\label{batt_diagram_1electrode}
    \end{subfigure}
    \begin{subfigure}[htb!]{0.49\textwidth}
        \centering
        \includegraphics[width=\linewidth]{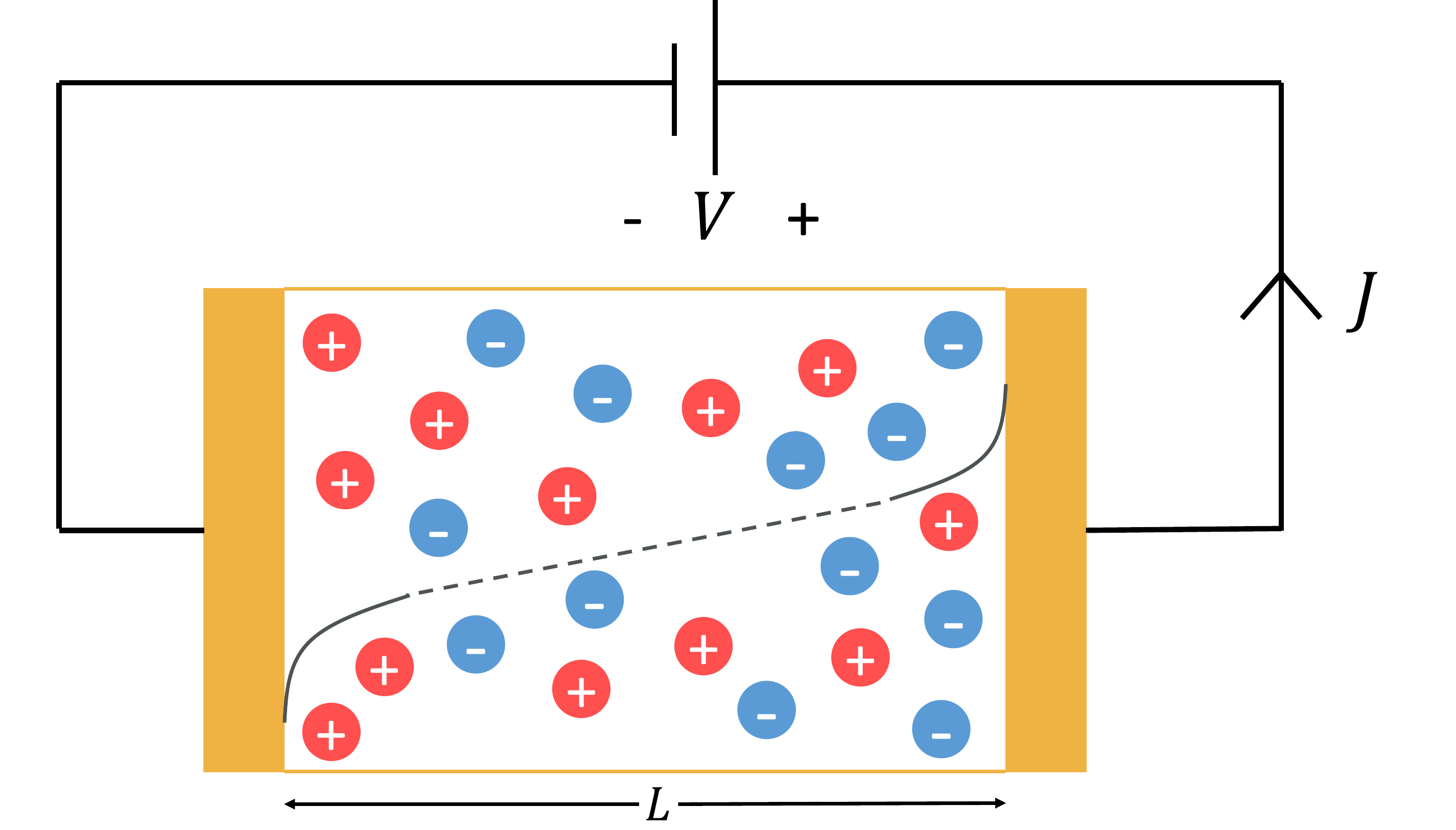}
        \caption{Two electrodes}\label{batt_diagram_2electrode}
    \end{subfigure}
\caption{Diagram showing electrochemical systems with applied voltage $V$ and resulting current $J$ for a system with (a) a single electrode and (b) two electrodes. In the single electrode case, the left hand boundary is modeled as an ideal reservoir with a fixed potential. Also shown is a sketch of the potential inside the cell: dashed line is the potential in the bulk, solid line is the potential in the diffuse region.}
\label{batt_diagram}
\end{figure}

In this work, we consider ramped voltages applied to electrochemical systems with one or two electrodes. Figure \ref{batt_diagram} shows a sketch of both systems, with the directions of the voltage and current density. The two electrode case models a battery or capacitor, whereas the one electrode case models an electrode under test at the right hand side of the domain with an ideal reservoir and potential set to zero at the left hand side.
\subsection{Poisson-Nernst-Planck Equations for Ion Transport}
Transport of ions in the bulk is described by a conservation equation for the concentration,
\begin{equation}
\frac{dC_i}{d\tau}=-\frac{\partial}{\partial X} J_{\text{NP},i}
\label{transport}
\end{equation}
and the definition of the flux,
\begin{equation}
J_{\text{NP},i}=- D_i \frac{\partial C_i}{\partial X}-\frac{z_i D_i F}{RT}C_i\frac{\partial \Phi}{\partial X} 
\label{flux_NP}
\end{equation}
where $C_i (X,\tau)$ denotes the concentration of the $i^\text{th}$ species and $z_i$ is its associated charge number, where we take the convention that the sign of the charge is attached to $z$ (so that a negative ion has negative $z$). $D_i$ is the diffusivity of the $i^\text{th}$ species, and $F$, $R$ and $T$ are Faraday's constant, the ideal gas constant and temperature, respectively. Substituting equation \eqref{flux_NP} into equation \eqref{transport} gives the time-dependent Nernst-Planck equation for ions in the bulk ($X\in(0,L)$), where we replace the index $i$ with either $+$ or $-$ for the positive and negative ion.
\begin{equation}
\label{nernst-planck}
\frac{\partial C_\pm}{\partial \tau} = \frac{\partial \;}{\partial X}\left[D_\pm \frac{\partial C_\pm}{\partial X} + \frac{z_\pm D_\pm F}{RT}C_\pm\frac{\partial \Phi}{\partial X}\right]
\end{equation}
Note that this
classical first approximation \cite{Kornyshev1981} neglects excluded
volume (``crowding") effects, which are important in crystalline ionic
solids \cite{Grimley1947} and ion insertion electrodes
\cite{Bazant2013ACR} especially at high voltages \cite{Bazant2009}. For
solid electrolytes in a lattice, Bikerman's model \cite{Bikerman1942} can be used, and for liquid electrolytes there are hard sphere
models such as the Carnahan-Starling model \cite{Carnahan1969}. Without
steric effects, the model might be better for polymeric solid electrolytes
with less severe volume constraints, or for doped solid crystals with low
carrier concentrations at sufficiently low voltages.

The potential $\Phi(X,\tau)$ is determined by Poisson's equation,
\begin{equation}
\label{poisson}
-\varepsilon_s\frac{\partial^2 \Phi}{\partial X^2}=F\left(z_+C_+ +z_- C_-\right)
\end{equation}
where $\varepsilon_s$ is the bulk permittivity of the electrolyte. We will be considering 1:1 binary electrolytes in this work, and equations \eqref{nernst-planck}-\eqref{poisson} describe a system with two mobile charges. For a solid electrolyte, negative ions take a fixed value of concentration, and so we set $C_-(X,\tau)=C_0$ and only solve one transport equation. For a `leaky membrane' \cite{Dydek2011}, we add a constant background charge $\rho_s$ to equation \eqref{poisson},
\begin{equation}
\label{poisson_lmm}
-\varepsilon_s\frac{\partial^2 \Phi}{\partial X^2}=F\left(z_+C_++ z_-C_- + \rho_s\right)
\end{equation}
Finally, for a supported electrolyte, we set $\Phi(X,\tau)=0$ and only solve the two Nernst-Planck equations for each charged species.

\subsection{Frumkin-Butler-Volmer Boundary Conditions for Faradaic Reactions}
The boundary condition for equation \eqref{nernst-planck} is given by
\begin{equation}
\label{NP_BC}
\pm\left(D_+\frac{\partial C_+}{\partial X}+\frac{z_+ DF}{RT}\frac{\partial \Phi}{\partial X} \right)\bigg |_{X=0,L} = J_F
\end{equation}
for the flux of the cations, where the $\pm$ in equation \eqref{NP_BC} varies depending on whether the boundary condition is applied at the left or right boundary. We assume that the anion ($C_-$) flux is zero at both boundaries. For the Faradaic current $J_F$, we use the generalized Frumkin-Butler-Volmer (FBV) equation,
\begin{equation}
\label{BV}
J_F = 
K_c\,
C_+e^{\left(\frac{-\alpha_cnF \Delta \Phi}{RT}\right)}
- K_a \, C_M \, e^{\left(\frac{\alpha_anF\Delta \Phi}{RT}\right)}
\end{equation}
In particular, equation \eqref{BV} models a reaction of the form
\begin{equation}
\ce{C^{z_+}_+ + ne^- <=> M}
\label{reaction}
\end{equation}
%{\setlength{\parindent}{0cm}
where $n=z_+$ is the number of electrons transfered to reduce the cation to the metallic state $M$. An example of a reaction of the form \eqref{reaction} is the Cu-CuSO$_4$ electrodeposition process, where $z_+=n=2$. $K_a$ and $K_c$ are the reaction rate parameters and $\alpha_{a,c}$ is the so-called transfer coefficient, with $\alpha_a+\alpha_c=1$. In the case of a blocking, or ideally polarizable electrode, or for a species which does not take part in the electrode reaction, $K_a=K_c=0$. Following \cite{Soestbergen2012} and other models using FBV kinetics, $\Delta \Phi$ is explicitly defined as being across the Stern layer (and always as the potential at the electrode minus the potential at the Stern plane), and the direction of $J_F$ is defined to be from the solution into the electrode. 

The boundary condition on $\Phi$ at an electrode is 
\begin{equation}
\label{phi_bc}
\mp\lambda_s \frac{\partial \Phi}{\partial X} = \Delta \Phi
\end{equation}
where again the sign $\mp$ is for the left hand and right side electrode and $\lambda_s$ is the effective width of the Stern layer, allowing for a different (typically lower) dielectric constant ~\cite{Bazant2005, Bonnefont2001}. In this model, the point of zero charge (pzc) occurs when $\Delta\Phi=0$ across the Stern layer, so the potential of the working electrode is defined relative to the pzc.  For a more general model with two electrodes having different pzc's, an additional potential shift must be added to equation \eqref{phi_bc} for at least one electrode~\cite{He2006,Streeter2008}. 

The system of equations is closed with a current conservation equation (see Equation 9 in \cite{Soestbergen2010}),
\begin{equation} \label{curr_cons}
\frac{d \Phi_X(L,\tau)}{d\tau}
= - \frac{1}{\varepsilon_s} \left\{
J(\tau) -F J_F \right\}
\end{equation}
where $J(\tau)$ can either be an externally set electrical current or (when the voltage is prescribed as is the case in this work) a post-processed electrical current, also defined to be from the solution into the electrode. In equation \eqref{curr_cons}, $\Phi_X$ denotes the $X$ derivative of $\Phi$.

%\change{\sout{As mentioned in the historical review, equations \eqref{nernst-planck}--\eqref{poisson} with boundary conditions \eqref{BV}--\eqref{curr_cons} allows for the nonlinear and time-dependent coupling of diffuse charge effects with the Faradaic and displacement currents at an electrode \cite{Soestbergen2012}.}}
In this work, we will be using this model to consider the effect of a ramped voltage at the right hand side electrode. The applied voltage takes the form
\begin{equation}
V(\tau)=S\tau
\end{equation}
where $S$ is the voltage scan or sweep rate. For some simulations in this work, we wish to model only one electrode, with the other approximating an ideal reference electrode or an ideal reservoir. In this case, we use the electrode at $X=L$ as the electrode under test, and use the boundary conditions
\begin{equation}
\label{single_electrode_bc}
C_\pm(0,\tau)=C_0, \,\,\,\, \Phi(0)=0
\end{equation}
at $X=0$, where $C_0$ is a reference concentration.
\subsection{Dimensionless Equations}
It is convenient to rescale the PNP equations so that distance, time, concentration and potential are scaled to interelectrode width, the diffusion time scale, a reference concentration and the thermal voltage, respectively:
\begin{equation}
\label{rescale}
x \equiv \frac{X}{L}, \qquad
t \equiv \frac{D\tau}{L^2}, \qquad
c_{\pm}\equiv \frac{C_{\pm}}{C_{0} } ,
\qquad
 \phi\equiv \frac{F\Phi}{RT}
\end{equation}
Rescaling the Poisson-Nernst-Planck
equations (\ref{nernst-planck})--(\ref{poisson}) yields the dimensionless equations. For the remainder of the paper, we will be assuming a 1:1 electrolyte ($z_+=-z_-=1$), equal diffusivities ($D_+=D_-=D$), and restricting ourselves to the electrode reaction in equation \eqref{reaction} with $z_+=n=1$. Other ionic solutions and reactions can be modeled by making the appropriate changes to equations \eqref{nernst-planck}--\eqref{poisson} and \eqref{BV}. The Nernst-Planck and Poisson equations become
\begin{equation}
\label{np_nondim}
\frac{\partial c_\pm}{\partial t}=\frac{d}{dx}\left[\frac{\partial c_\pm}{\partial x} \pm c_\pm\frac{\partial \phi}{\partial x} \right]
\end{equation}
\begin{equation}
\label{poisson_nondim}
-\epsilon^2\frac{\partial^2 \phi}{\partial x^2}=\frac{1}{2}\left(c_+ - c_-\right)
\end{equation}
where $\epsilon=\lambda_D/L$ is the ratio between the Debye length $\lambda_D\equiv \sqrt{\frac{\varepsilon_sRT}{2F^2C_{0}}}$ and $L$.
To nondimensionalize the boundary conditions, we introduce the following rescaled parameters 
\begin{equation}
\label{bc_rescale}
k_c = \frac{K_c \, L}{4D}, \qquad
j_r = \frac{K_a\, L \, C_M}{4DC_{0}},
\quad \delta = \frac{\lambda_s}{\lambda_D}  
\end{equation}
so that equation \eqref{BV} becomes (assuming $\alpha_c=\alpha_a=1/2$)
\begin{equation}
\label{bv_nondim}
\pm\left(c_+\frac{\partial \phi}{\partial x} + \frac{\partial
  c_+}{\partial x}\right)\bigg |_{x=0,1} = 4k_c \,
c_+ \, e^{-\Delta \phi/2} - 4 \, j_r \, e^{\Delta \phi/2}
\end{equation}
Equation \eqref{phi_bc} rescales to
\begin{equation}
\label{phi_bc_nondim}
\mp \epsilon \, \delta \; \phi_x = \Delta \phi
\end{equation}
and finally, equation \eqref{curr_cons} becomes
\begin{equation}
\label{current_conservation_nondim}
-\frac{\epsilon^2}{2} \, \frac{d \; }{dt} \phi_x(1,t)
=j - \left[k_c \, c_+\left(1,t\right) \, e^{-\Delta \phi/2} - j_r \, e^{\Delta \phi/2}\right]  
\end{equation}
where $j$ is scaled to the limiting current density $J_\text{lim}=\frac{4FDC_0}{L}$.

The rescaled equations contain two fundamental time scales: the diffusion time scale, $\tau_D=L^2/D$ and the reaction time scale, $\tau_R=L/K$, for a characteristic reaction rate $K$. There is also the imposed voltage scan rate time scale, $\tau_S = \frac{RT}{F}/S$. The ratios of these three time scales result in two dimensionless groups: the nondimensional voltage scan rate, 
\begin{equation}
\tilde{S} = \frac{\tau_D}{\tau_S} = \frac{S L^2 F }{D RT}
\end{equation}
and the nondimensional reaction rate,
\begin{equation}
k=\frac{\tau_D}{4 \tau_R} = \frac{ K L}{ 4 D}
\end{equation}
also known as the ``Damkohler number". In Eq. (\ref{bc_rescale}), the latter takes two forms, $k=k_c$ and $k=j_r$, for the forward (oxidation) and backward (reduction) reactions with the characteristic rates, $K=K_c$ and $K=K_a C_M/C_0$, respectively.

The dimensionless model also contains thermodynamic information about the electrochemical reaction, independent of the overall reaction rate~\cite{Bard2001,Biesheuvel2011,Bazant2013ACR}. 
For any choice of a single reaction-rate scaling, there is a third dimensionless group, which can be expressed as the (logarithm of) the ratio of the dimensionless forward and backward reaction rates,
\begin{equation}
\ln \frac{k_c}{j_r} = \ln \frac{K_c C_0}{ K_a C_M } = \Delta\phi^\Theta  
\label{dphi0}
\end{equation}
This is the dimensionless equilibrium interfacial voltage, where the net Faradaic current (equation \eqref{bv_nondim}) vanishes in detailed balance between the oxidation and reduction reactions for the reactive cation at the bulk reference concentration $C_0$. 

\subsection{Important Limits}
\subsubsection{Supported Electrolyte}
As mentioned in Section \ref{history}, we choose to model liquid electrolytes in the limit of high supporting salt concentrations by setting $\phi(x,t)=0$ and removing Poisson's equation (equation \eqref{poisson_nondim}) from the model. This has two effects: first, electromigration is no longer a part of the transport equations (equation \eqref{nernst-planck}). Second, the voltage difference $\Delta \phi=v-\phi(1)$ which enters into the FBV equation (equation \eqref{bv_nondim}) is just the electrode potential since $\phi(1)=0$. This results in a time-dependent version of the classical Randles-Sevcik system \cite{Bard2001}, with Butler-Volmer rather than Nernstian boundary conditions.
\subsubsection{Thin Double Layers (Electroneutral Limit)}
The dimensionless parameters $\epsilon$ and $\delta$ control how the model handles the double layer and diffuse charge dynamics, respectively. The dimensionless Debye length $\epsilon$ governs the amount of charge separation that is allowed to occur, or equivalently the thickness of the diffuse layer relative to device thickness. The parameter $\delta$, which is effectively a ratio of Stern and diffuse layer capacitance, controls the competition between the Stern and diffuse layers in overall double layer behavior. The electroneutral, or thin double layer limit, corresponds to $\epsilon \rightarrow 0$. In this limit, equation \eqref{poisson_nondim} is simply
\begin{equation}
\label{equal_c}
c_+=c_-
\end{equation}

Substituting equation \eqref{equal_c} into the Nernst-Planck equations (equation \eqref{np_nondim}) results in the ambipolar diffusion equation for the concentration $c=c_+=c_-$ (written here for equal diffusivities),
\begin{equation}
\label{ambipolar}
\frac{\partial c}{\partial t}=\frac{\partial^2 c}{\partial x^2}
\end{equation}
and the following equation for the potential
\begin{equation}
\label{ambipolar_phi}
\frac{\partial }{\partial x} \left(c\frac{\partial \phi}{\partial x}\right)=0
\end{equation}

In the limit of thin double layers, there are two further limits on $\delta$. The first is the ``Gouy-Chapman" (GC) limit, $\delta \rightarrow 0$, which corresponds to a situation where all of the potential drop in the double layer is across the diffuse region, and the second is the ``Helmholtz" (H) limit, $\delta \rightarrow \infty$, where all of the potential drop is across the Stern layer. In the former case, the Boltzmann distribution for ions is invoked ($c_\pm=\exp\left(\mp\Delta \phi_D\right)$) and equation \eqref{bv_nondim} becomes
\begin{equation}
\label{gc}
\pm\left(c_+\frac{\partial \phi}{\partial x} +\frac{\partial
  c_+}{\partial x}\right)\bigg |_{x=0,1} = 4k_c \,
c_+ \, e^{-\Delta \phi_D} - 4 \, j_r
\end{equation}
where $\Delta \phi_D$ is the potential drop across the diffuse region. In the latter case, equation \eqref{bv_nondim} becomes
\begin{equation}
\label{helmholtz}
\pm\left(c_+\frac{\partial \phi}{\partial x} +\frac{\partial
  c_+}{\partial x}\right)\bigg |_{x=0,1} = 4k_c \,
c_+ \, e^{-\Delta \phi_S/2} - 4 \, j_re^{\Delta \phi_S/2}
\end{equation}
where $\Delta \phi_S$ is the potential drop across the Stern layer.

\subsubsection{Fast Reactions}
The most common approximation for the boundary conditions in voltammetry is the Nernstian limit of fast reactions, $k_c, j_r \rightarrow \infty$, for a fixed ratio, $j_r/k_c$, corresponding to a given equilibrium half-cell potential for the reaction,  Eq. (\ref{dphi0}). In this limit, the left-hand sides of equations \eqref{bv_nondim} and \eqref{gc}--\eqref{helmholtz} approach zero, and all three boundary conditions reduce to the Nernst equation,
\begin{equation}
\label{nernst}
\Delta \phi = \Delta\phi^\Theta + \ln c_+
\end{equation}
where the bulk solution is used as the reference concentration.  In order to place the Nernst equation in the standard form~\cite{Bard2001,Bazant2013ACR}, 
\begin{equation}
E =\frac{RT}{nF} \Delta \phi^\text{eq} = E^\Theta + \frac{RT}{nF} \ln \frac{C_+}{C_\text{ref}}
\end{equation}
we must define the equilibrium interfacial voltage relative to a standard reference electrode (e.g. the Standard Hydrogen Electrode in aqueous systems at room temperature and atmospheric pressure with $C_\text{ref}=1$ M) by shifting the reference potential,
\begin{equation}
E^\Theta = \frac{RT}{nF} \left( \Delta \phi^\Theta - \ln \frac{C_0}{C_\text{ref}}\right) = \frac{RT}{nF}\ln \frac{K_c C_\text{ref}}{K_a C_M }
\end{equation}
In this way, tables of standard half-cell potentials can be used to determine the ratio of reaction rate constants, and measurements of exchange current density can then determine the individual oxidation and reduction rate constants.

%\subsection{\change{\sout{Numerical Method}}}
%\change{\sout{In space, we use a three-point variable mesh size finite difference scheme. In time, we use a variable step size implicit-explicit backwards differencing formula (VSSBDF2) published by Wang and Ruuth }\cite{Wang2008, Ascher1995}.\sout{ Time steps are chosen adaptively by controlling the error; our time stepping and computational method will be presented in a follow up work.}} %All simulations were performed using MATLAB running on Debian Linux with 16 cores and 64 GB of RAM.

\section{Bulk Liquid Electrolytes}\label{bulk_liquid}
\subsection{Model Problem}
The model problem for this section is a single electrode in an aqueous electrolyte with a reference electrode at $x=0$ and two mobile ions. Voltammetry on a single electrode in solution is one of the oldest problems in electrochemistry; experiments are usually conducted using the ``three-electrode setup", where one electrode acts as a voltage reference while another (the counter electrode) provides current. Computationally, this setup can be mimicked with equation \eqref{single_electrode_bc} applied at $X=0$. This models a cell with only one electrode, with the other approximating an ideal reference electrode or an ideal reservoir.

In order to ensure that dynamics do not reach the reference electrode, we restrict ourselves to the regime $\tilde{S} \gg 1$ and $t \ll 1$. The reaction rate parameter $k$ can either be fast ($k\gg 1$) or slow ($k\ll 1$), which corresponds to a diffusion-limited regime and reaction-limited regime, respectively.

\subsection{Supported Electrolytes} \label{supported_electrolytes}
A supported electrolyte is where an inert salt is added (a salt which does not take part in electrode or bulk reactions) in order to screen the electric field so as to render electromigration effects negligible (i.e. $\phi=0$ in the bulk). Voltammetry with supported electrolytes has classically been treated as a semi-infinite diffusion problem. For an aqueous system at a planar electrode with a reversible electrode reaction (reaction much faster than diffusion) involving two species denoted O (oxidized) and R (reduced), the current resulting from a slow ramped voltage is given, using the present notation, by (see Chapters 5 and 6 of \cite{Bard2001})
\begin{equation}
\label{voltammetry_diffusion6}
J(\tau) = nFD_O\left(\frac{\partial C_O(X,\tau)}{\partial X}\right)_{X=0} = nFAC_O^*\left(\pi D_O \sigma\right)^{\frac{1}{2}}\chi(\sigma \tau),
\end{equation}
where $\sigma \equiv nFS/RT$ ($S$ is the scan rate) and $\chi$ is the Randles-Sevcik function. The same system of equations is also often considered for irreversible (slow reactions compared to diffusion) and `quasi-reversible' reactions, with the appropriate changes to the boundary conditions.

\begin{figure}[htb!]
\centering
\includegraphics[width=0.65\linewidth]{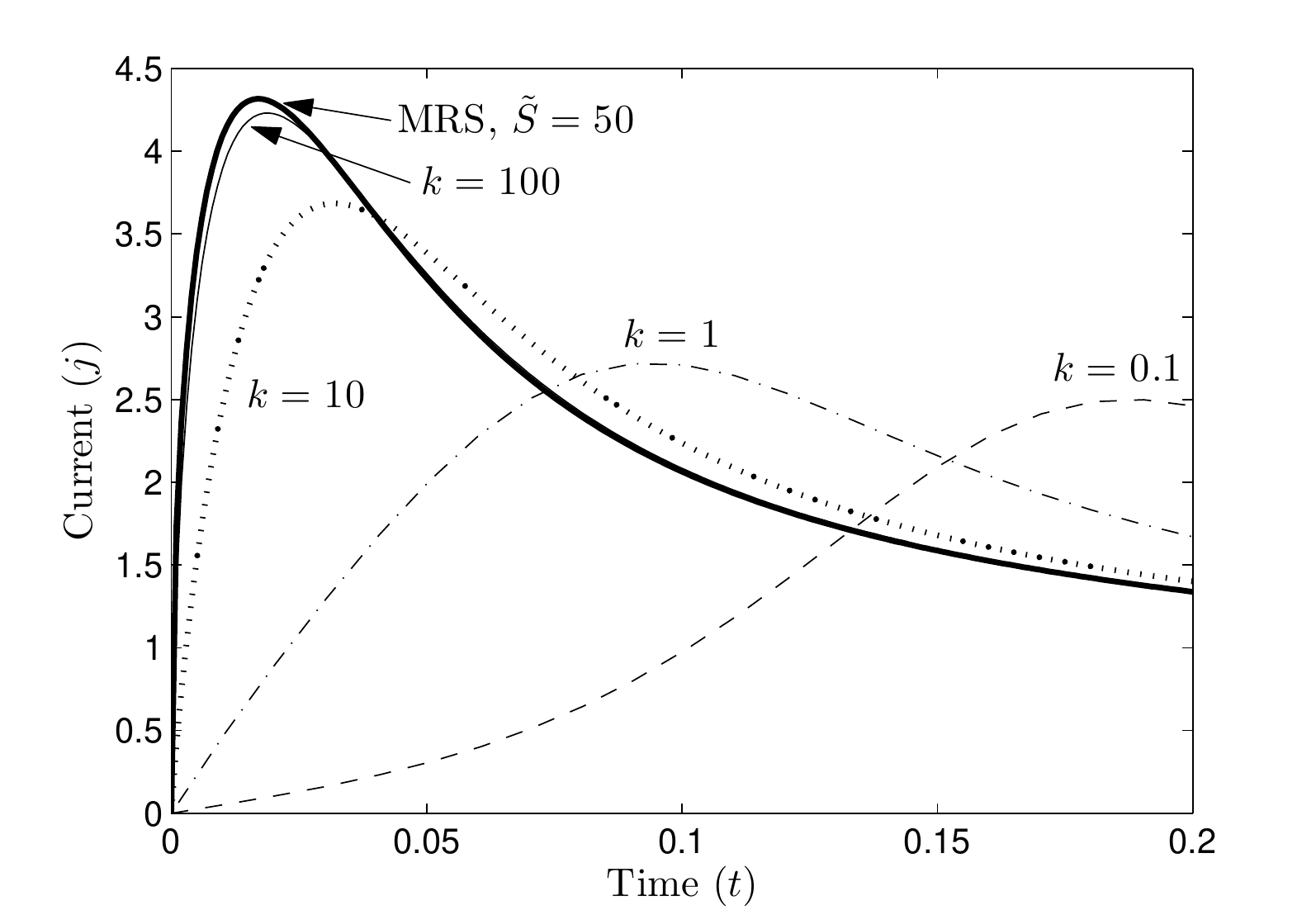}
\caption{Simulated current curves for a supported electrolyte with one electrode in response to a voltage ramp with $\tilde S=-50$, with various values of $k$. Also shown is the high reaction rate limit in equation \eqref{MRS_eqn}.}
\label{MRS}
\end{figure}

The reaction we will be considering is
\begin{equation}
\ce{C^+ + e^- <=> M}
\end{equation}
where M represents the electrode material. Since the reaction only involves one of the two ions as opposed to both as in the Randles-Sevcik case, the mathematics are simplified somewhat and we are able to obtain \change{a novel} analytical solution. The diffusion equation (equation \eqref{ambipolar}) needs to be solved on $x \in (0,\infty)$ with fast reactions (equation \eqref{nernst} with $\Delta \phi = v(t)$). In order to probe the forward, or reduction reaction, we apply a voltage is $v(t)=-\tilde{S}t$ and compute the resulting current via $j=\frac{\partial c}{\partial x}\big |_{x=0}$. These equations admit the solution
\begin{equation}
\label{MRS_eqn}
j(t)=\sqrt {\tilde{S}} e^{-\tilde{S}t} \erfi\left(\sqrt{\tilde{S} t}\right)
\end{equation}
where $\erfi (z) = \frac{2}{\sqrt \pi}\int_0^z  \exp x^2 \, dx$ is the imaginary error function. The derivation of equation \eqref{MRS_eqn} can be found in the appendix. We term equation \eqref{MRS_eqn} a ``modified" Randles-Sevcik equation, which applies to voltammetry on an electrode with fast reactions involving only one ionic species, and in a supported electrolyte. Figure \ref{MRS} shows simulated $j(t)$ curves with various values of $k$ compared to equation \eqref{MRS_eqn}, with $\tilde{S}=50$. The curves in Figure \ref{MRS} exhibit the distinguishing features of single-reaction voltammograms: current increases rapidly until most of the reactant at the electrode has been removed due to transport limitation. The peak represents the competition between the increasing rate of reaction and the decreasing amount of reactant at the electrode. After the peak, the lack of reactant wins out, and there is a decrease in the amount of current the electrode is able to sustain. Also worth noting is that at low reaction rate $k$, the start of the voltammogram is exponential rather than linear due to reaction limiting.

The simulated curves in Figure \ref{MRS} approach equation \eqref{MRS_eqn} in the limit of large $k$, which makes equation \eqref{MRS_eqn} a good approximation for the current response to a ramped voltage in a supported electrolyte with fast, single-species reaction. Note that due to definition differences, the simulated current must be multiplied by 4 because there is a difference of a factor of 4 between the current in equation \eqref{current_conservation_nondim} and the ion flux in equation \eqref{bv_nondim}.
\subsection{Unsupported Electrolytes with Thin Double Layers} \label{unsupported_thinedl}
\begin{figure}[htb!]
\centering
\includegraphics[width=0.65\linewidth]{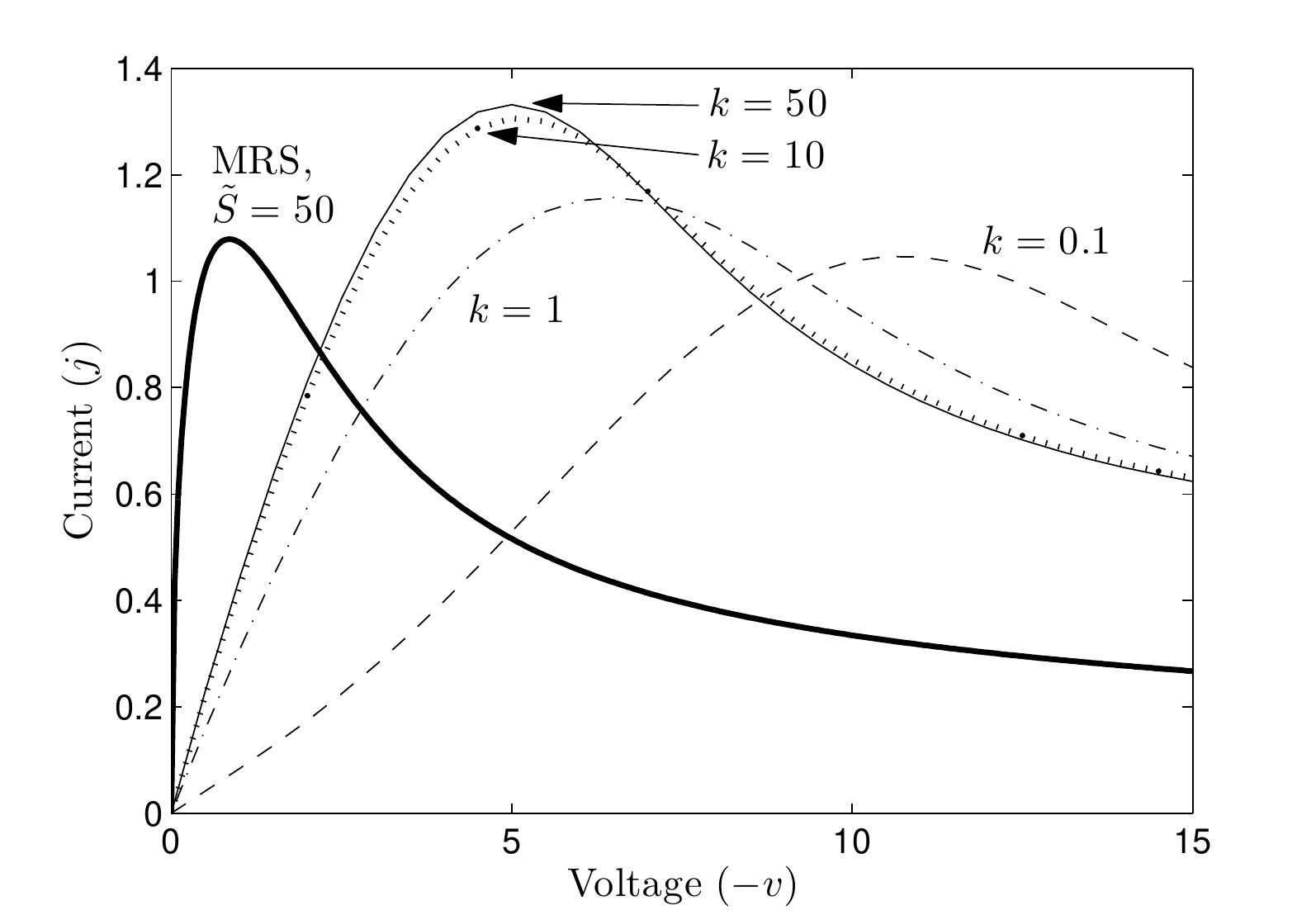}
\caption{Simulated $j$ vs. $v$ curves with various values of reaction rate $k$ for an unsupported electrolyte with one electrode, in response to a ramped voltage with scan rate $\tilde{S}=-50$ and with $\epsilon=0.001$ and $\delta=100$. The $k=50$ simulation gives results which are numerically equivalent to the large $k$ limit. Also shown is the theoretical result for a supported electrolyte from equation \eqref{MRS_eqn}.}
\label{vary_k_vs_MRS}
\end{figure}

For the single-electrode, thin-EDL, unsupported electrolyte problem, we use a value of $\epsilon=0.001$, and impose the restrictions $\tilde S \gg 1$ and $t \ll 1$ in order to remain in a diffusion-limited regime. Figure \ref{vary_k_vs_MRS} shows plots of voltammograms for $v(t)=-50t$ ($\tilde S=50$) with $\delta=100$, and the modified Randles-Sevcik plot is also shown for comparison. 

Note that while the situations leading to the currents in Figures \ref{MRS} and \ref{vary_k_vs_MRS} may seem superficially similar (both involve voltammetry on a single electrode with fast reactions), they do not produce the same results. 
The physical difference is that electromigration is included in the latter (i.e. equation \eqref{poisson_nondim} is solved along with the anion transport equation), which opposes diffusion, resulting in a slower response. This type of shift in the voltammogram for low support has been well documented in the experimental literature (see \cite{Bond1984A, Bento1998A, Amatore1999, Belding2012, Limon-Petersen2009}, among others).

\begin{figure}[htb!]
\centering
    \begin{subfigure}[htb!]{0.49\textwidth}
        \centering
        \includegraphics[width=\linewidth]{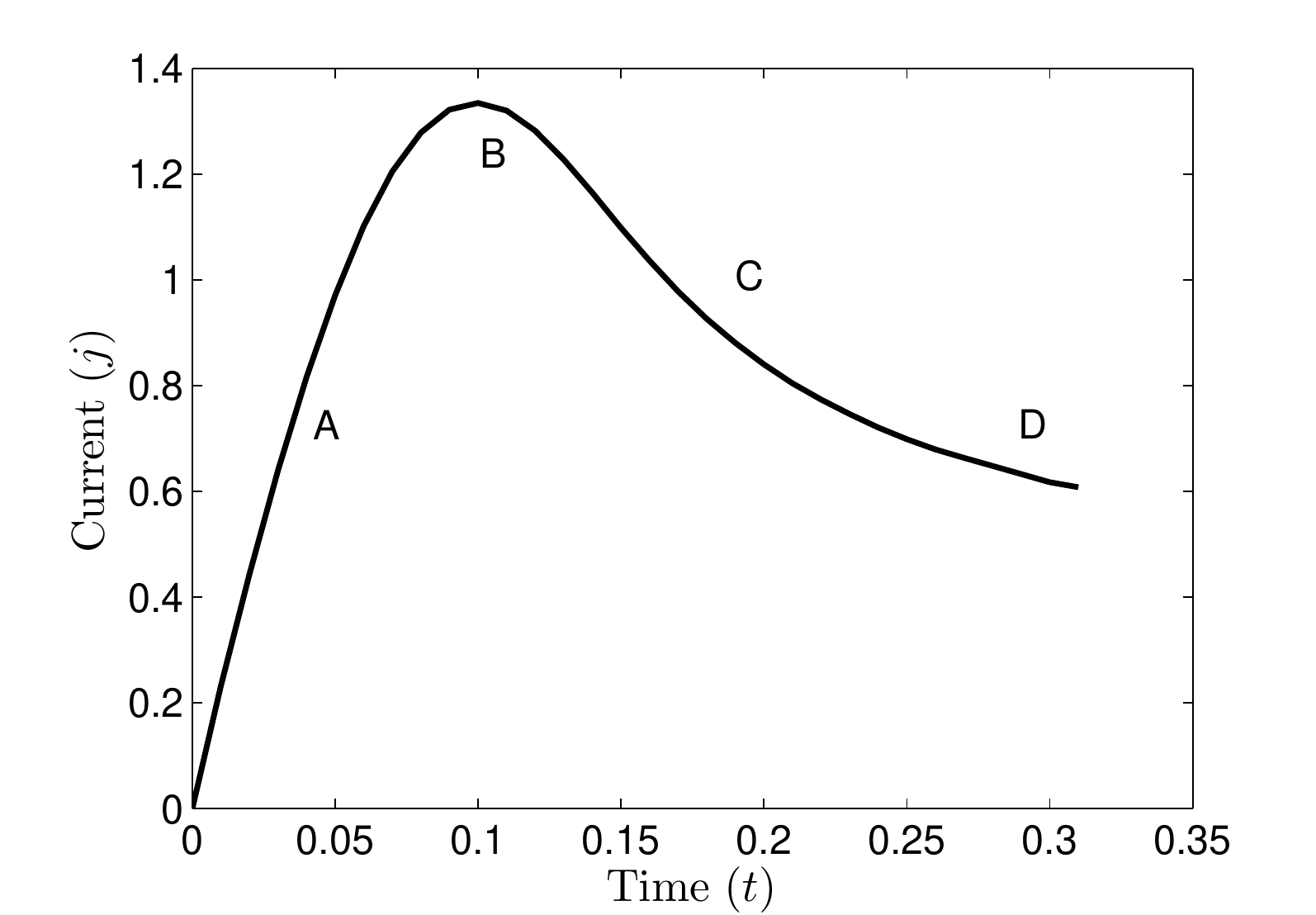}
        \caption{Current}\label{conc_volt_100_A}
    \end{subfigure}
    \begin{subfigure}[htb!]{0.49\textwidth}
        \centering
        \includegraphics[width=\linewidth]{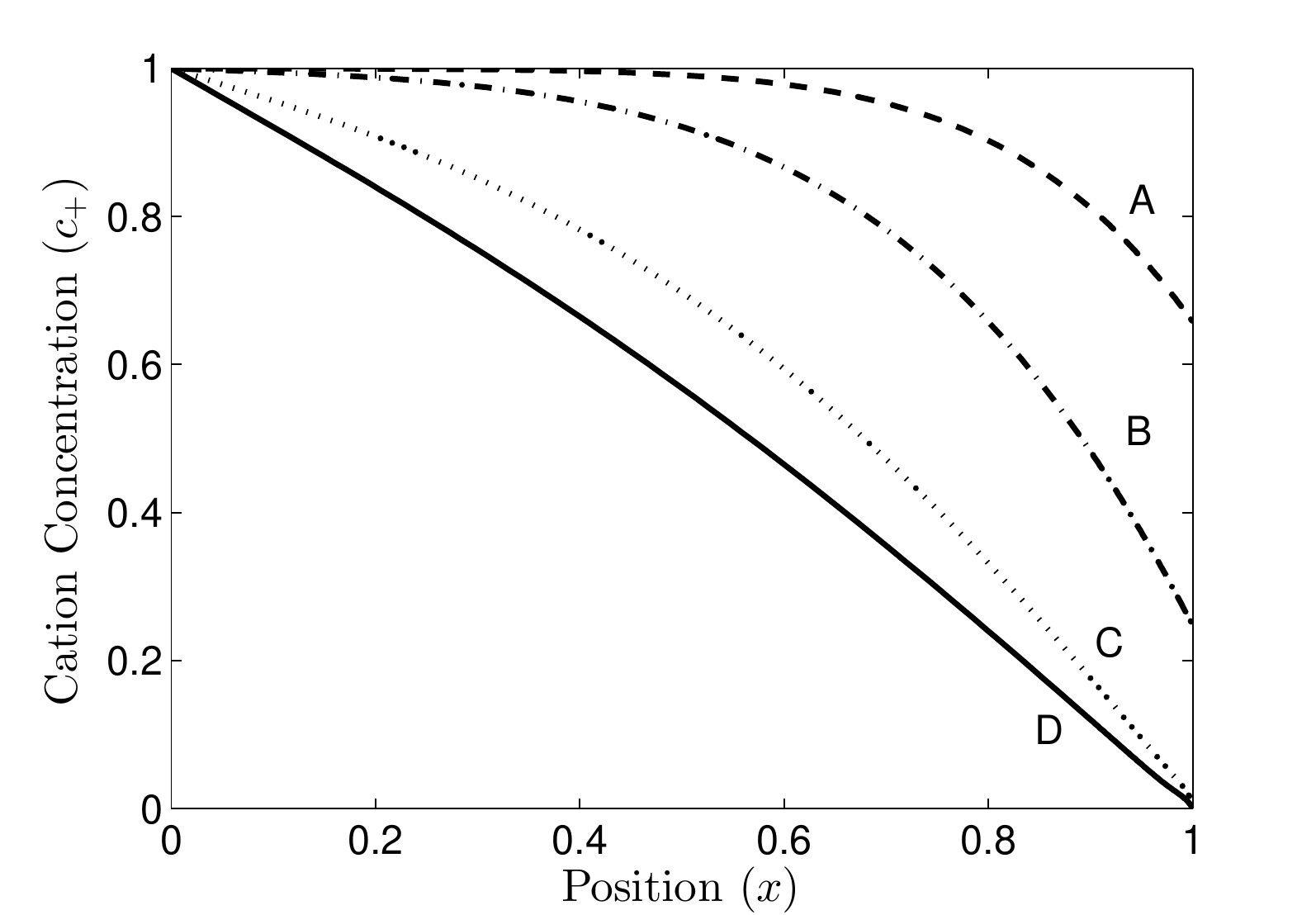}
        \caption{Concentrations}\label{conc_volt_100_B}
    \end{subfigure}
\caption{(a) Voltammogram and (b) concentrations with $k=50$ (diffusion limited) for an unsupported electrolyte with one electrode subjected to a ramped voltage with scan rate $\tilde S=-50$, with $\epsilon=0.001$ and $\delta=100$. Labels in the concentration plot correspond to snapshots of cation concentration $c_+$ at times labeled on the current vs. time plot.}
\label{conc_volt_100}
\end{figure}

Next, Figure \ref{conc_volt_100} show a voltammogram for the $k=50$ (diffusion limited) case, with accompanying concentration profiles ($c_+$ and $c_-$ are identical in the bulk but only $c_+$ is shown). %The current in Figure \ref{conc_volt_0point1} exhibits the initial exponential shape and slower response indicative of reaction limited dynamics compared to Figure \ref{conc_volt_100}.

%\begin{figure}[htb!]
%\centering
   % \begin{subfigure}[htb!]{0.49\textwidth}
   %     \centering
   %     \includegraphics[width=\linewidth]{conc_volt_0point1_A.pdf}
   %     \caption{Current}\label{conc_volt_0point1_A}
  %  \end{subfigure}
  %  \begin{subfigure}[htb!]{0.49\textwidth}
%\centering
   %     \includegraphics[width=\linewidth]{conc_volt_0point1_B.pdf}
   %     \caption{Concentrations}\label{conc_volt_0point1_B}
  %  \end{subfigure}
%\caption{(a) Voltammogram and (b) concentrations with $k=0.1$ (reaction limited) for an unsupported electrolyte with one electrode subjected to a ramped voltage with scan rate $\tilde S=50$, with $\epsilon=0.001$ and $\delta=100$. Labels in the concentration plot correspond to snapshots of cation concentration $c_+$ at times labeled on the current vs. time plot.}
%\label{conc_volt_0point1}
%\end{figure}

The parameter $\delta$ was chosen to be large for these simulations so that large diffuse layers do not form, making it easier to see a correspondence between the slope of the concentration at the electrode and the resulting current. The current and slope of $c_+$ both reach a maximum when the voltammogram peaks, followed by a gradual flattening of the concentration as transport limitation sets in. Furthermore, though we end our simulations in this section after the cation concentration reaches zero at the electrode, we will see in Section \ref{TSC_section} that the PNP-FBV equations admit solutions past this point with the formation of space charge regions.

\begin{figure}[htb!]
\centering
    \begin{subfigure}[htb!]{0.49\textwidth}
        \centering
        \includegraphics[width=\linewidth]{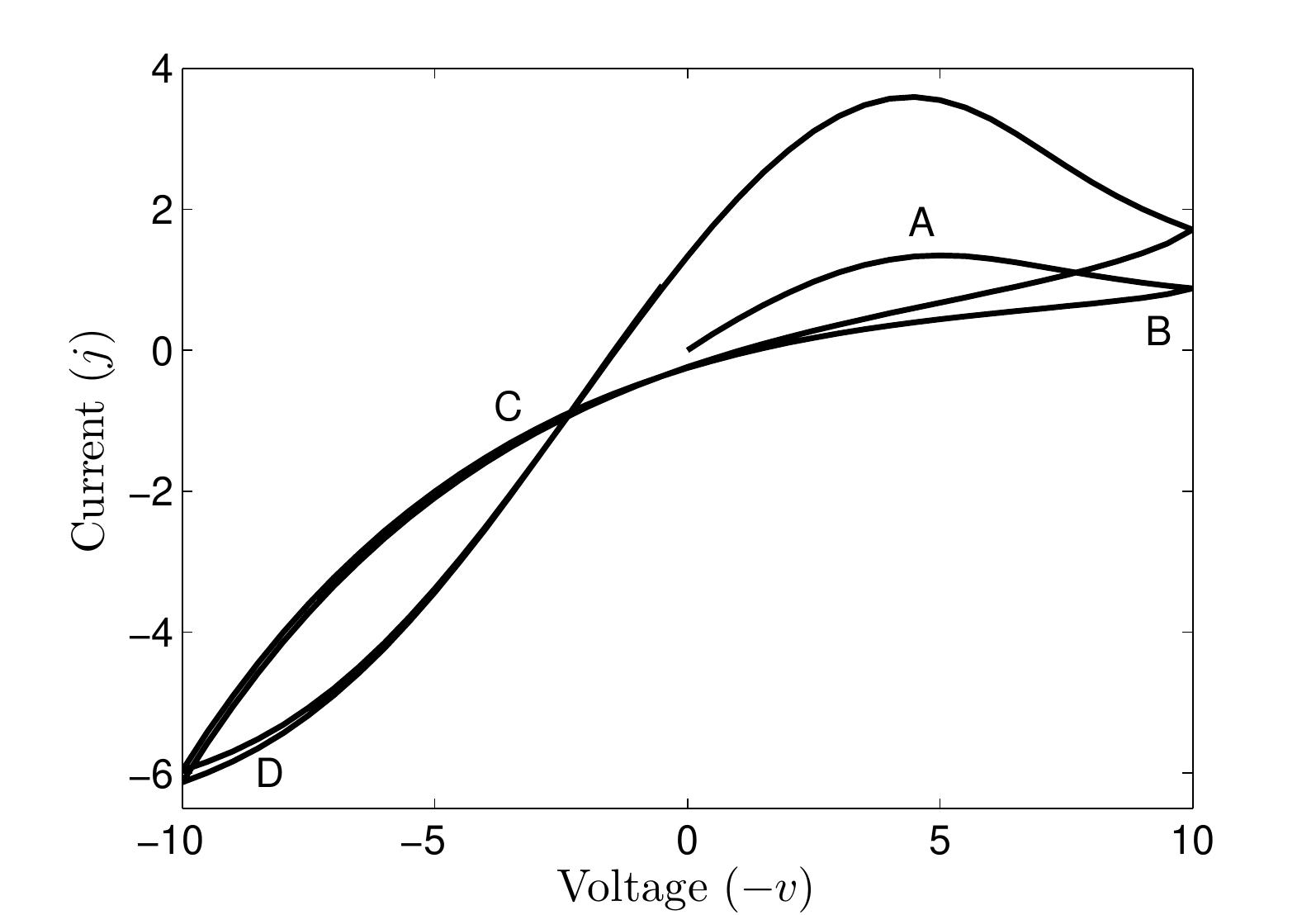}
        \caption{Current}\label{twocycle_A}
    \end{subfigure}
    \begin{subfigure}[htb!]{0.49\textwidth}
        \centering
        \includegraphics[width=\linewidth]{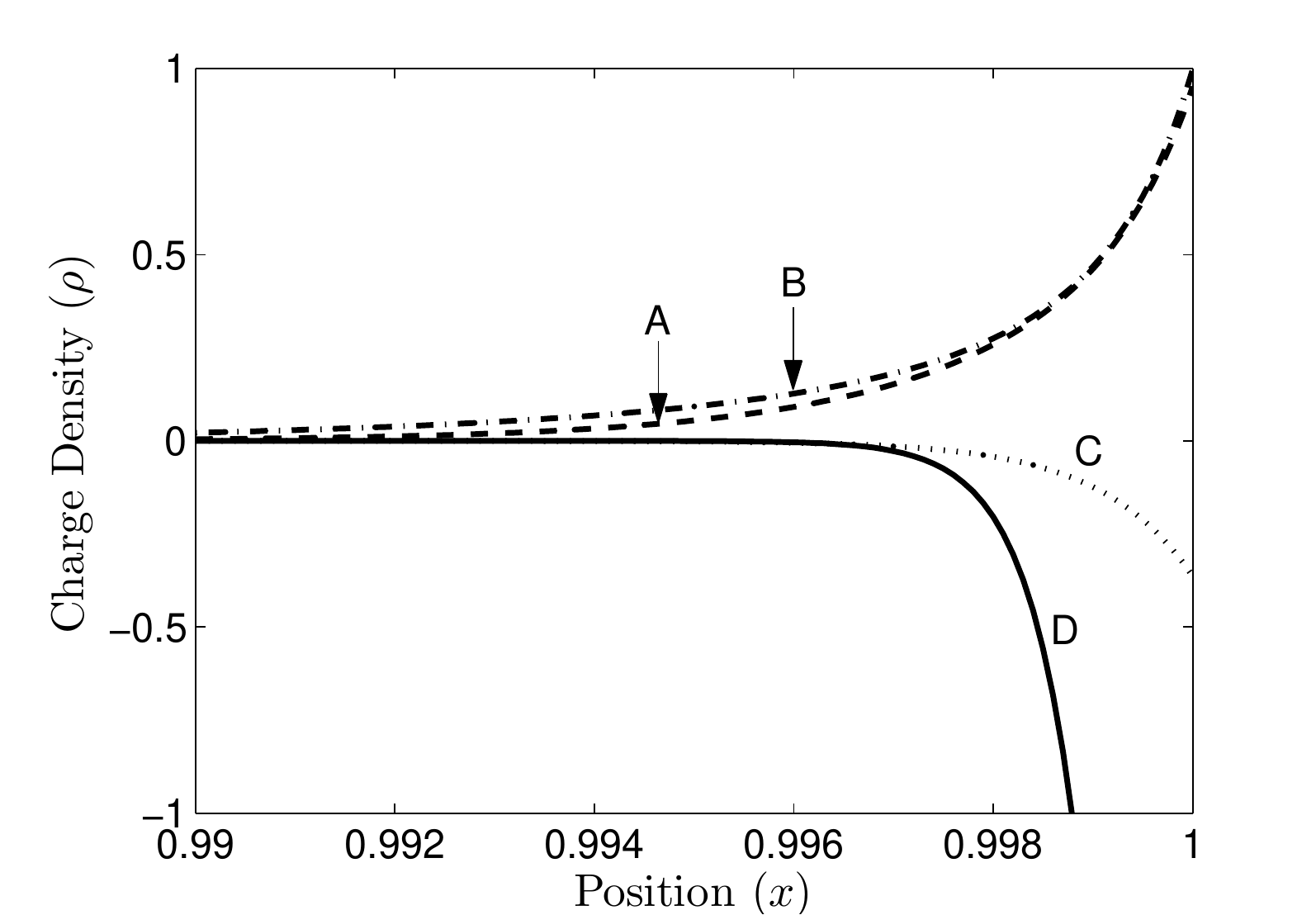}
        \caption{Charge densities}\label{twocycle_B}
    \end{subfigure}
\caption{(a) Voltammogram for an unsupported electrolyte with one electrode subjected to a triangular voltage with $|\tilde S|=50$, $\epsilon=0.001$, $k=50$ and $\delta=0.01$. (b) Net charge densities $\rho=c_+-c_-$. Labels on the charge density plot correspond to snapshots of $\rho$ in the double layer at times labeled on the voltammogram.}
\label{twocycle}
\end{figure}

We end this section with Figure \ref{twocycle}, which shows two voltammetry cycles on a system with $\epsilon=0.001$, $k=50$, $\delta=0.01$ and $|\tilde{S}|=50$. Due to the fact there is only one electrode and only one species takes part in the reaction, the voltammogram exhbits diode-like behavior: diffusion limiting in the direction of positive current and exponential growth in the direction of negative current. Also shown are the net charge densities ($\rho=c_+-c_-$) in the diffuse region ($x>0.99$) during the first cycle, which are allowed to form since $\delta$ is small, so that the double layer is dominated by the diffuse charge region. The amount of charge separation in the diffuse layer is very large at large voltages, and highlights the need to use the PNP-FBV equations to capture their dynamics.

Lastly, an interesting observation is that in Figure \ref{twocycle}, the current peak is much higher in the second (and subsequent) cycle(s) than in the first. We expect that this is due to charge dynamics near the electrode: concentration distributions do not return their initial distributions when the polarity of $v$ reverses. Due to the fast scan rate, there is still an excess of positive ions near the electrode when the voltage switches polarity at the start of the second cycle, allowing for a much longer time for the current to build before transport limitation occurs.

\section{Liquid and Solid Electrolyte Thin Films}\label{thinfilms}
\subsection{Model Problem}
In this section, we study voltammograms of liquid and solid electrolyte electrochemical thin films. General steady-state models for thin films have been previously presented in \cite{Bazant2005} and \cite{Chu2005} as well as in \cite{Biesheuvel2009}, with a time-dependent model considered in \cite{Soestbergen2010}. From a modeling perspective, the only difference between the two systems is that the counterion concentration is constant for solid electrolytes, i.e. $c_-(x,t)=1$. Furthermore, for some simulations in this section, we consider voltage ramps on systems with two dissimilar electrodes, i.e. values of $k_c$ and $j_r$ such that an equilibrium voltage develops across the cell.
\subsection{Simulation Results}
\subsubsection{Low Sweep Rates}\label{lowsweep}
At low sweep rates, the current-voltage relationship approaches the steady-state response, which for solid and liquid electrolytes was derived by Bazant et al.~\cite{Bazant2005} (for an electrolytic cell with two identical electrodes) and Biesheuvel et al. \cite{Biesheuvel2009} (for a galvanic cell with two different electrodes). For a liquid electrolyte, the cell voltage is given in the GC ($\delta\rightarrow 0$) limit by
\begin{equation}
\label{liquidgc}
v(j)=v_0-4\arctanh(j) + \ln \frac{1-j/j_{r,A}}{1+j/j_{r,C}}
\end{equation}
and in the H limit ($\delta \rightarrow \infty$) by
\begin{equation}
\label{liquidh}
v(j)=v_0-4\arctanh(j) - 2\arcsinh \frac{j}{\sqrt{\beta_A(1+j)}} - 2\arcsinh \frac{j}{\sqrt{\beta_C(1-j)}}
\end{equation}
where $v_0=\ln\frac{k_{c,C}j_{r,A}}{k_{c,A}j_{r,C}}$ is the equilibrium voltage, $\beta_A=4k_{c,A}j_{r,A}$ and $\beta_C=4k_{c,C}j_{r,C}$. The subscripts $A$ and $C$ denote parameter values at the anode and cathode, respectively. For a solid electrolyte in the thin EDL limit, the cell voltage is given in the GC limit by
\begin{equation}
\label{solidgc}
v(j)=v_0-4j + \ln \frac{1-j/j_{r,A}}{1+j/j_{r,C}}
\end{equation}
and in the H limit by
\begin{equation}
\label{solidh}
v(j)=v_0-4j - 2\arcsinh \frac{j}{\sqrt{\beta_A}} - 2\arcsinh \frac{j}{\sqrt{\beta_C}}
\end{equation}

\begin{figure}[htb!]
\centering
\begin{subfigure}[htb!]{0.49\textwidth}
\includegraphics[width=\linewidth]{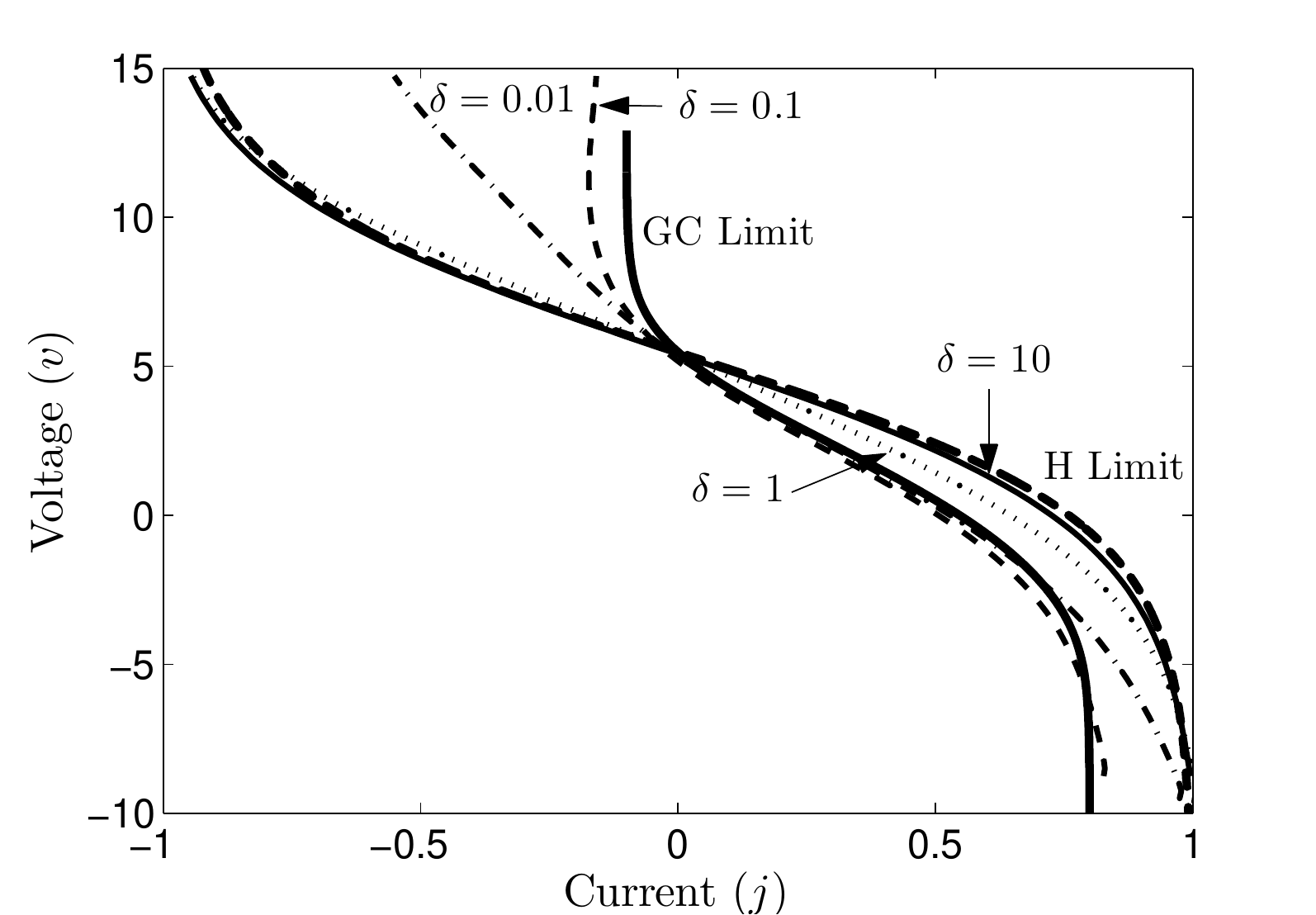}
\caption{Liquid electrolyte}
\label{liquid_vi}
\end{subfigure}
\begin{subfigure}[htb!]{0.49\textwidth}
\includegraphics[width=\linewidth]{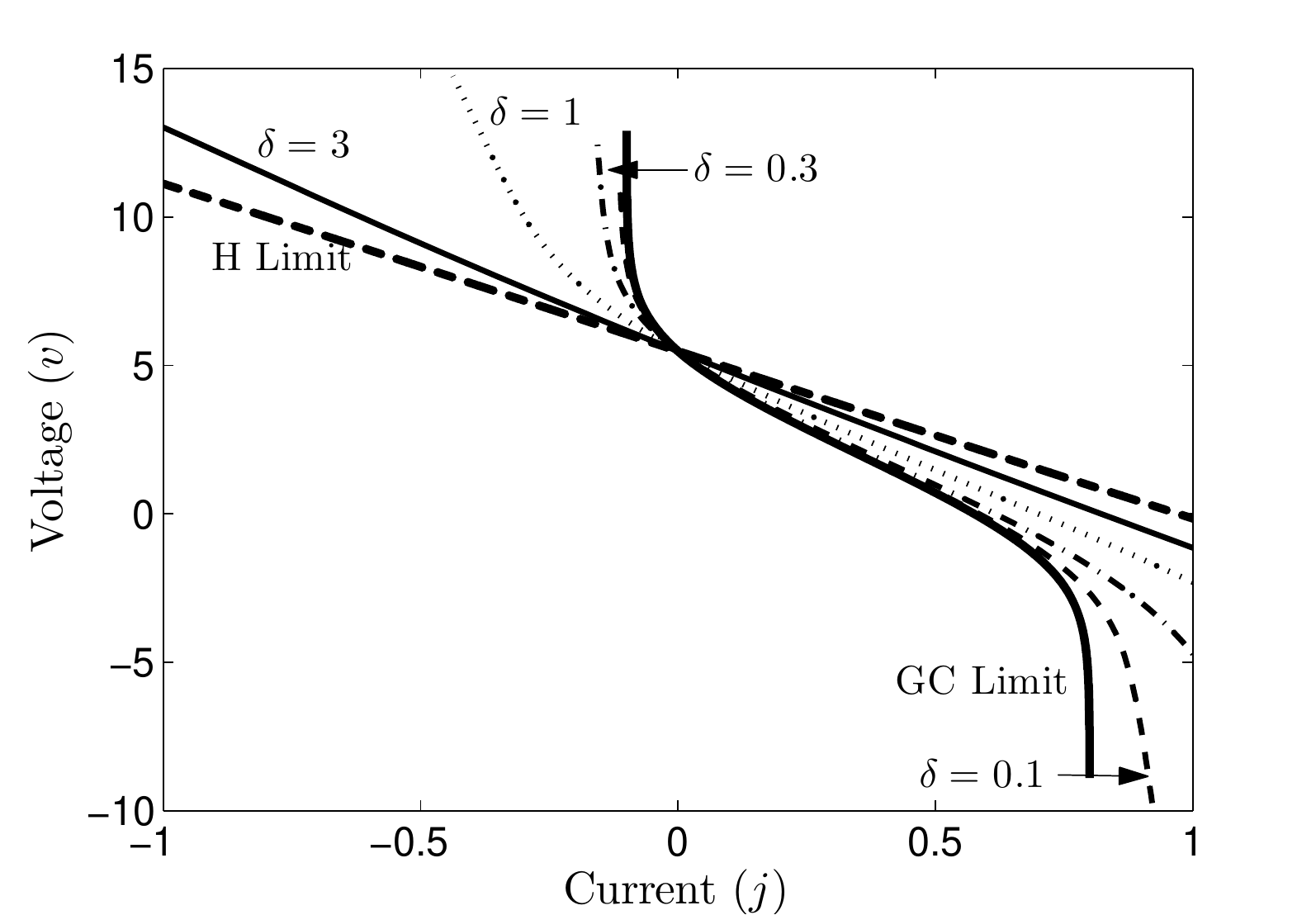}
\caption{Liquid electrolyte}
\label{solid_vi}
\end{subfigure}
\caption{$v$ vs $j$ curves for (a) thin EDL liquid electrolyte and (b) thin EDL solid electrolyte with two electrodes and with parameters $\epsilon=0.001$, $k_{c,C}=30$, $k_{c,A}=1$, $j_{r,C}=0.1$ and $j_{r,A}=0.8$ ($v_0 \approx 5.5$). Also shown are the steady-state curves in the GC and H limits from equations \eqref{liquidgc}--\eqref{liquidh} and \eqref{solidgc}--\eqref{solidh}. Simulated curves were created using a voltage scan rate of $\tilde S=2.5$.}
\label{vi}
\end{figure}

Figure \ref{vi} shows $v$ vs $j$ curves for liquid and solid electrolytes for $k_{c,C}=30$, $k_{c,A}=1$, $j_{r,C}=0.1$, $j_{r,A}=0.8$ (the same as Figure 3 in \cite{Biesheuvel2009}) for an open circuit voltage of $v_0 \approx 5.5$ and various values of $\delta$ along with the GC and H limits from equations \eqref{liquidgc}--\eqref{liquidh} and \eqref{solidgc}--\eqref{solidh}. The curves were created with a voltage ramp from -10 to +15 with a scan rate of $\tilde S=2.5$. For the liquid electrolyte case, the $\delta=1$ and $\delta=10$ curves were generated with $\epsilon=0.001$, while the $\delta=0.1$ and $\delta=0.01$ curves with $\epsilon=0.005$.

As expected, the voltage-current response for a liquid electrolyte is seen in Figure \ref{liquid_vi} to have reaction limits at $j=-j_{r,C}=-0.1$ and $j=j_{r,A}=0.8$ in the GC limit, and diffusion limits at $j=\pm 1$ in the H limit. The limiting cases from equations \eqref{liquidgc}--\eqref{liquidh} also do not strictly bound the simulated results in Figure \ref{liquid_vi} due to the non-monotonic dependence of the cell voltage $v$ on $\delta$ \cite{Biesheuvel2009}. Compared to the liquid electrolyte case, the two limits on $\delta$ for fixed countercharge are seen in Figure \ref{solid_vi} to have a reaction limits at $j=-j_{r,C}=-0.2$ and $j=j_{r,A}=0.8$ in the GC limit, but no diffusion limit in the H limit, which is consistent with the expected behavior for a solid electrolyte.
%\subsubsection{Thick EDL}
%Figure \ref{liquid_thickedl} shows $j$ vs $v$ for a thick EDL liquid electrolyte with two electrodes, with various values of $k_{c,C}$. $\epsilon=0.1$, $\delta=1$, scan rate $\tilde S=2.5$, with other reaction rate parameters the same as in Figure \ref{liquid_vi}. Since the voltage sweep rate is slow compared to both reaction and diffusion and $\epsilon$ is large enough that the double layers are able to overlap and interact with each other, the current vs. voltage curve straightens out compared to a thin double layer system, becoming effectively Ohmic in nature. 
%\begin{figure}[htb!]
%\centering
%\includegraphics[width=0.65\linewidth]{liquid_thickedl.pdf}
%\caption{$j$ vs $v$ for a thick EDL liquid electrolyte with two electrodes, with various values of $k_{c,C}$. $\epsilon=0.1$, $\delta=1$, scan rate $\tilde S=2.5$ and other reaction rate parameters the same as in Figure \ref{liquid_vi}.}
%\label{liquid_thickedl}
%\end{figure}

\subsubsection{Diffusion and Reaction Limitating} \label{RDL}
When sweep rates are fast, current-voltage curves will differ from the slow sweep results in Section \ref{lowsweep} due to physical limitation of the speed at which current can be produced at electrodes. This nonlinear interdependence of current and voltage when current flows into an electrode is described in electrochemistry by the blanket term \emph{polarization}, not to be confused with dielectric polarization. Generally speaking, when current flows across a cell, its cell potential, $v$, will change. The difference between the equilibrium value of $v$ and its value when current is applied is commonly referred to as the \emph{overpotential} or \emph{overvoltage} and labeled $\eta$. 

Very briefly, there are three competing sources of overpotential in an electrochemical cell:
\begin{enumerate}
\item Ohmic polarization is caused by the slowness of electromigration in the bulk. When a cell behaves primarily Ohmically, it is characterized by a linear j-v curve which can be written as $\eta_\text{ohm}=r_\text{cell}j$. When this behavior is modeled by a circuit, it is usually represented as a single resistor between the two electrodes.
\item Kinetic polarization is due to the slowness of electrode reactions ($k$ small). Using the overpotential version of the Butler-Volmer equation as a starting point, and assuming fast transport of species to and from the electrode ($C_\text{electrode}=C_\text{bulk}$), we can invert the equation to find $\eta_\text{kin}=\arcsinh\left(\frac{j}{j_0}\right)$, where $j_0$ is the nondimensional exchange current density. Thus, when kinetic polarization is the primary cause of overpotential, the j-v curve takes on an exponential characteristic. This type of polarization is represented by a charge-transfer resistance in circuit models.
\item Transport, or concentration polarization is due to slowness in the supply of reactants or removal of products from the electrode, resulting in a depletion of reactants at the electrode. Concentration polarization is characterized by a saturation of the current-voltage relationship, and is represented by the frequency-dependent Warburg element in circuit models.
\end{enumerate}

In terms of electrode polarization, the key difference between liquid and solid electrolytes is that the imposed constant counterion concentration associated with a solid electrolyte does not allow diffusion limiting to occur (except, perhaps, with very large forcings), since the reacting species is not allowed to be depleted at the electrodes. The trade-off is that current is only carried by one species in solid electrolytes, increasing the electrolyte resistance.

\begin{figure}[htb!]
\centering
\begin{subfigure}[htb!]{0.49\textwidth}
\includegraphics[width=\linewidth]{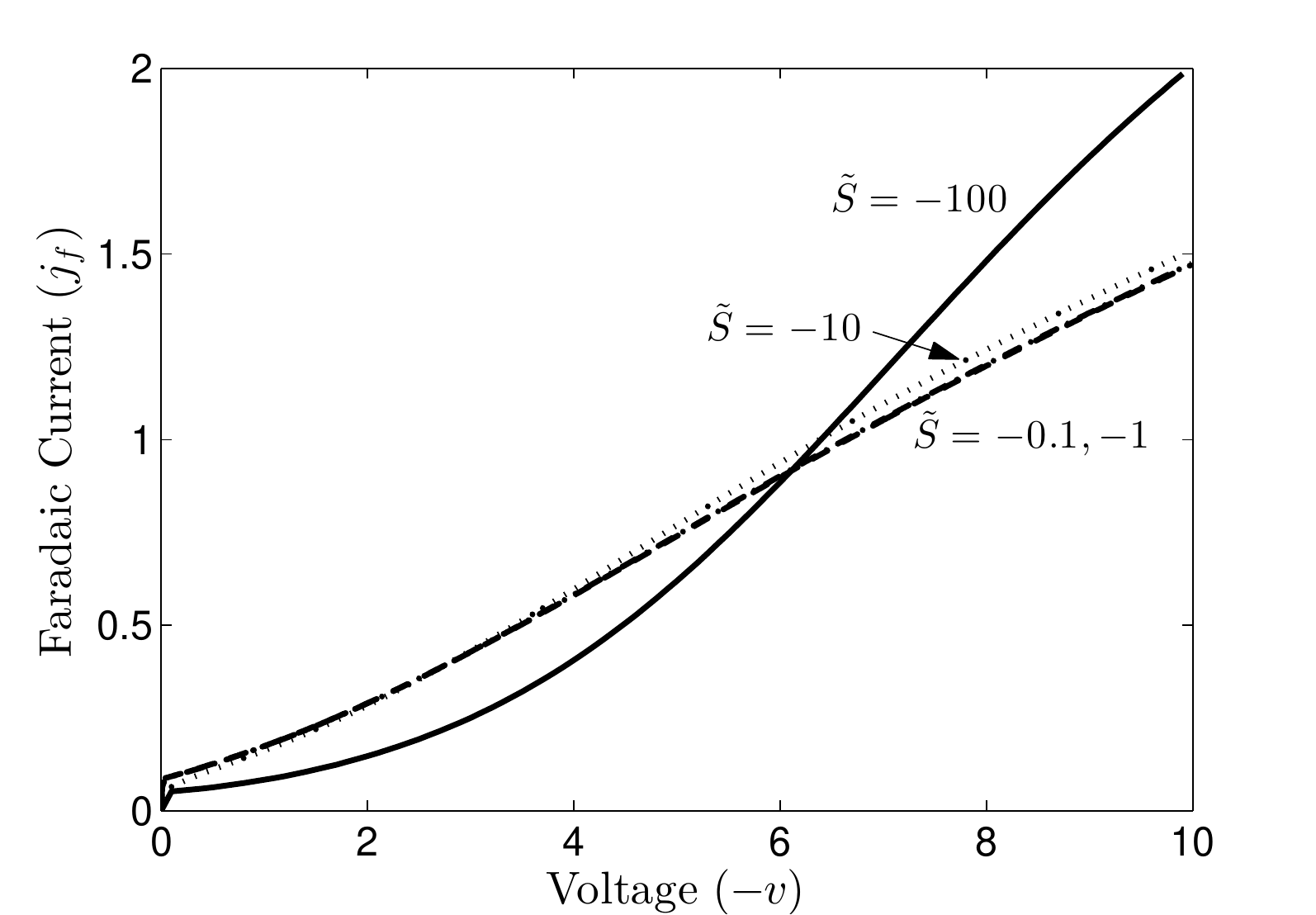}
\caption{Liquid electrolyte}\label{liquid_RL}
\end{subfigure}
\begin{subfigure}[htb!]{0.49\textwidth}
\includegraphics[width=\linewidth]{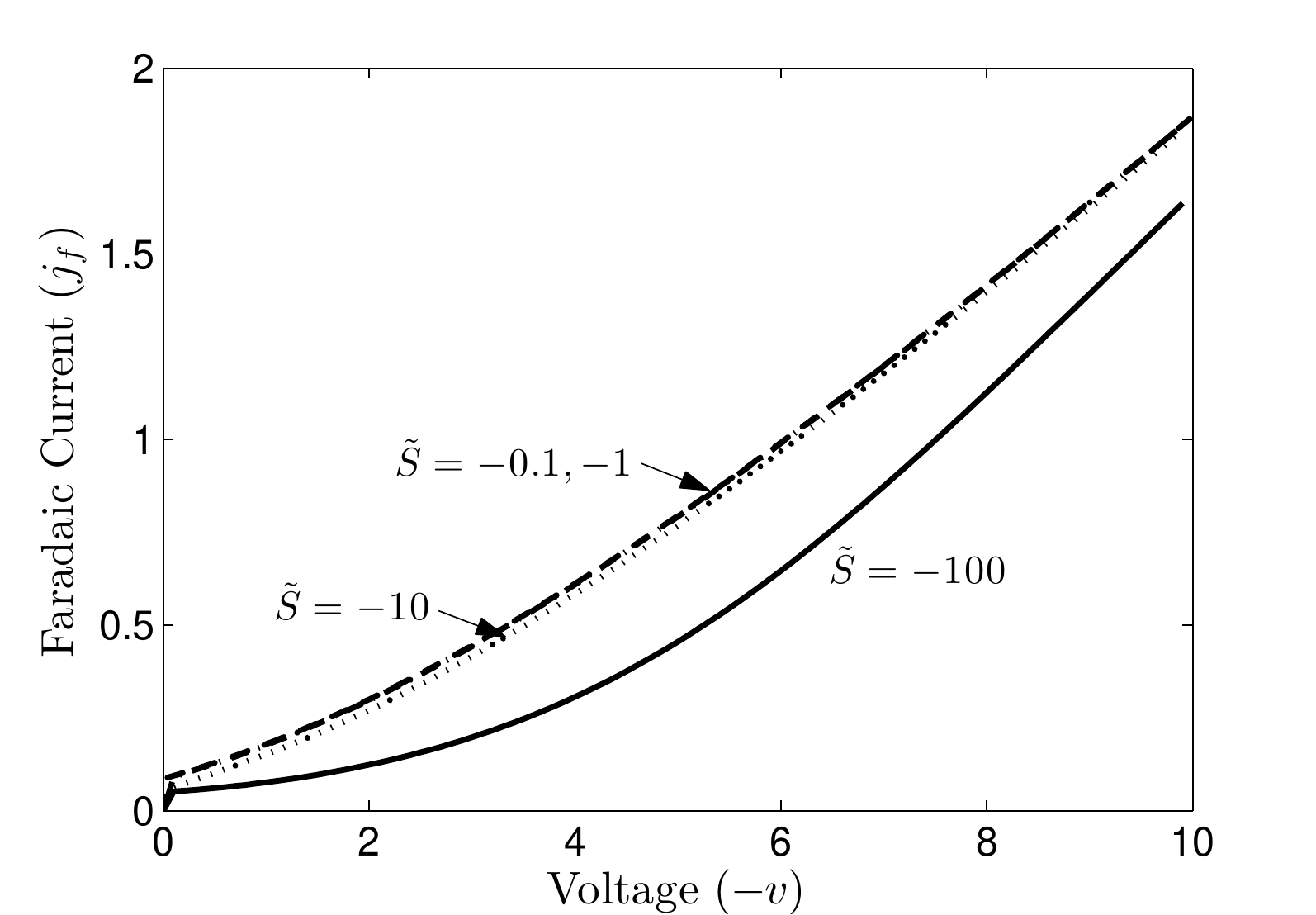}
\caption{Solid electrolyte}\label{solid_RL}
\end{subfigure}
        \caption{Faradaic current vs. voltage for (a) liquid electrolyte with and (b) solid electrolyte two electrodes and with $\tilde S$ varied, with parameters $\epsilon=0.05$, $\delta=1$, $k_{c,a}=50$, $j_{r,a}=100$, $k_{c,c}=0.1$ and $j_{r,c}=0.05$ ($v_0 \approx 1.4$). A current response dominated by reaction limitation is seen when $\tilde S=-100$.}\label{RL}
\end{figure}

To illustrate these points, we first show in Figure \ref{RL} Faradaic current vs voltage for liquid and solid electrolyte with two electrodes and with parameters $\epsilon=0.05$, $\delta=1$, $k_{c,a}=50$, $j_{r,a}=100$, $k_{c,c}=0.1$, and $j_{r,c}=0.05$ ($v_0\approx 1.4$). These parameters represent a situation with slow reactions at the cathode, and we vary the scan rate $\tilde S$ so that reaction limitation will dominate the current. As $\tilde S$ increases, the $j$-$v$ curves for both the liquid and solid electrolyte cases is seen to take on a more pronounced exponential character, thus showing the effect of reaction limitation. When $\tilde S$ is small, the Faradaic current in both cases takes a linear, or predominantly Ohmic character. Note that for plots where the sweep rate $\tilde S$ varies, we only plot the Faradaic part of the current (the second term on the right hand side of equation \eqref{current_conservation_nondim}) since for $\tilde S \gg 1$ there is a significant displacement component.

\begin{figure}[htb!]
        \centering
        \begin{subfigure}{0.5\textwidth}
                \includegraphics[width=\linewidth]{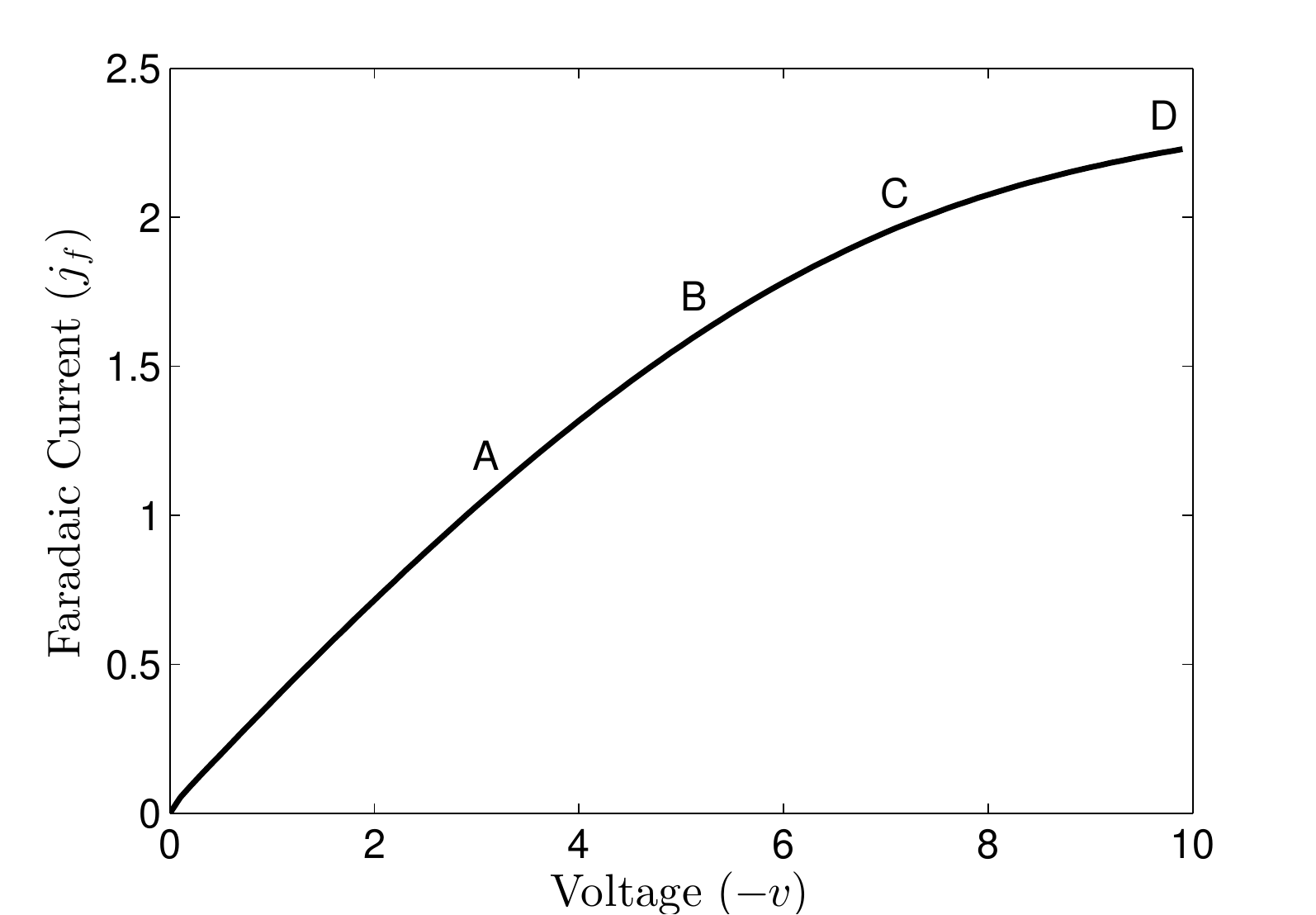}
                \caption{$j$ vs. $v$, Liquid electrolyte}
                \label{DL_liquid_100}
        \end{subfigure}%
        \begin{subfigure}{0.5\textwidth}
                \includegraphics[width=\linewidth]{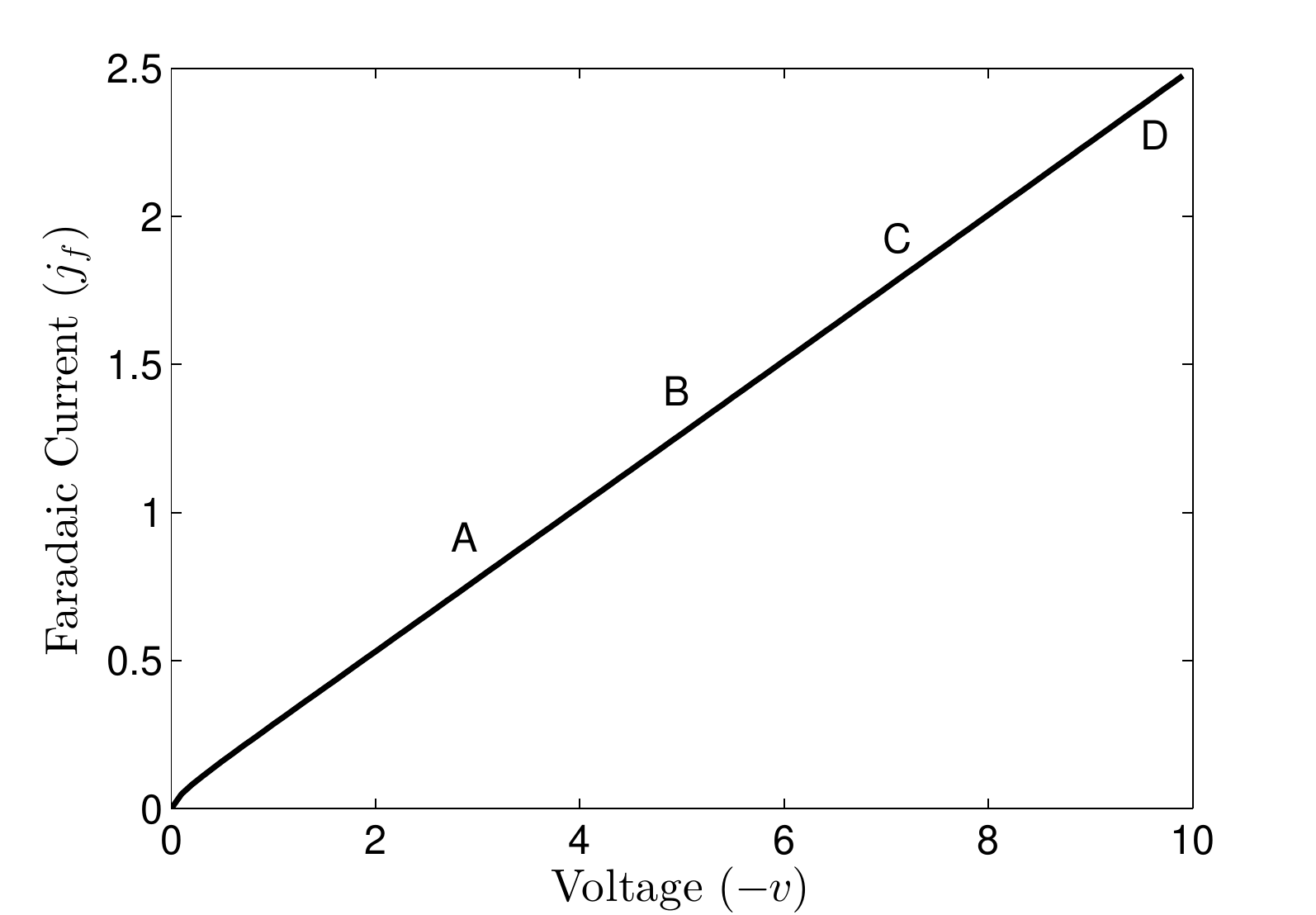}
                \caption{$j$ vs. $v$, Solid electrolyte}
                \label{DL_solid_100}
        \end{subfigure}

        \begin{subfigure}{0.5\textwidth}
                \includegraphics[width=\linewidth]{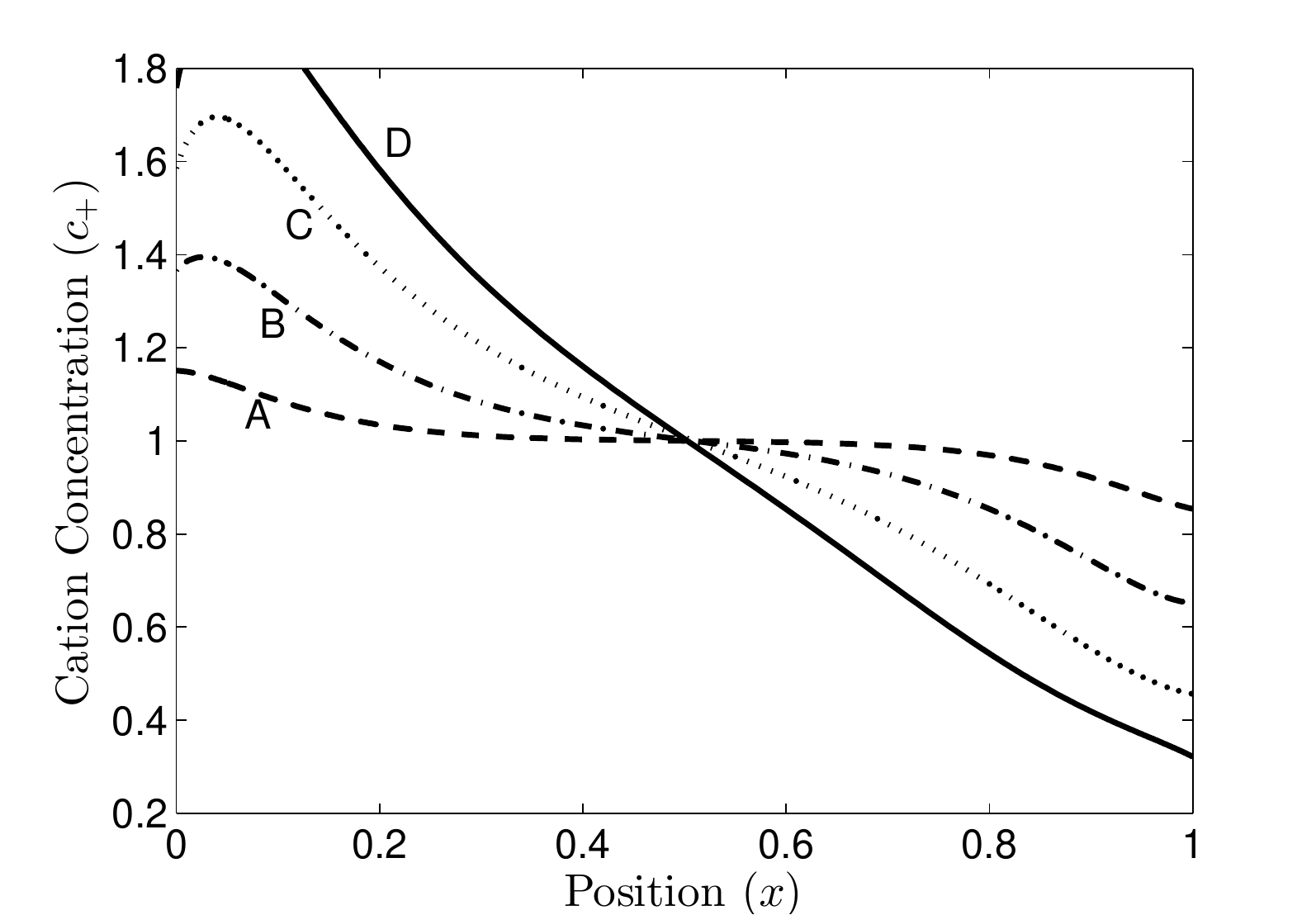}
                \caption{$c_+$, Liquid electrolyte}
                \label{DL_liquid_concentrations}
        \end{subfigure}%
        \begin{subfigure}{0.5\textwidth}
                \includegraphics[width=\linewidth]{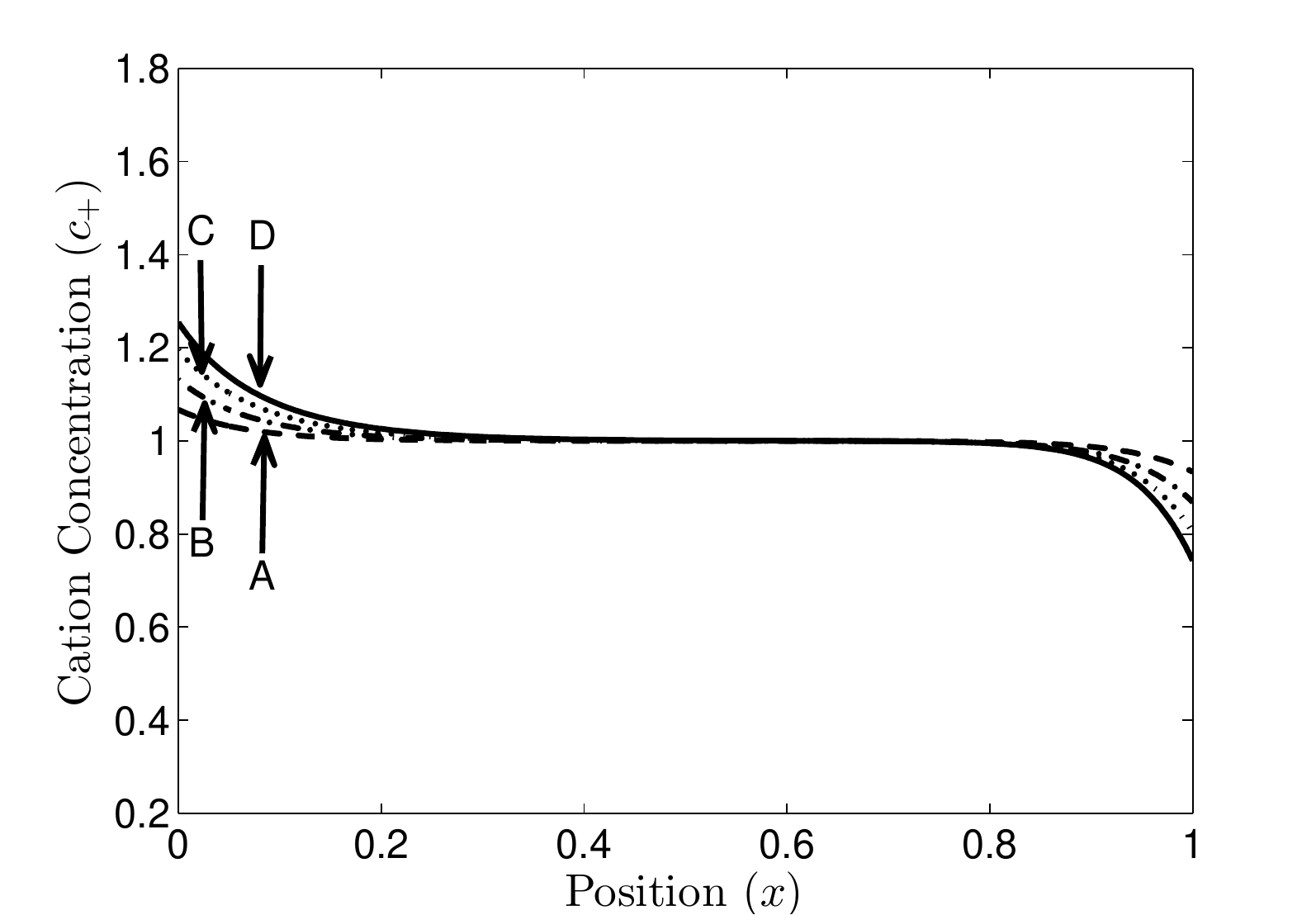}
                \caption{$c_+$, Solid electrolyte}
                \label{DL_solid_concentrations}
        \end{subfigure}
\caption{(a), (b) $j$ vs. $v$ and (c), (d) cation concentrations $c_+$ at $t=0.025$ (dashed line), 0.05 (dash-dotted line), 0.075 (dotted line) and 0.1 (solid line) for $\tilde S=-100$ voltage sweeps on both liquid and solid electrolytes with two electrodes. Other parameters are the same as in Figure \ref{RL}.}
\label{DL_concentrations}
\end{figure}

Next, to demonstrate when and how diffusion limitation plays a role, we show in Figure \ref{DL_concentrations} fast voltage sweeps ($\tilde S=-100$) on both liquid and solid electrolytes with fast reactions ($k=50$), with accompanying concentration profiles. For liquid electrolytes (Figures \ref{DL_liquid_100} and \ref{DL_liquid_concentrations}), the current begins to saturate as the cation are depleted at the cathode. Over the same voltage range, the current in the solid electrolyte (Figures \ref{DL_solid_100} and \ref{DL_solid_concentrations}) remains perfectly linear since the fixed negative ions allow for much less charge separation. 

\subsubsection{Transient Space Charge}\label{TSC_section}
We end our modeling of electrochemical thin films by investigating the development of space charge regions (regions of net charge outside of the double layer where $\rho=c_+-c_-\neq 0$) at large applied voltages. This is a strongly nonlinear effect which occurs in liquid electrolytes as predicted by Bazant, Thornton \& Ajdari \cite{Bazant2004} and solved by Olesen, Bazant \& Bruus using asymptotics and simulations for large sinusoidal voltages \cite{Olesen2010}. In this section\change{, we extend this work by showing} the formation of space charge regions in the context of voltammetry by using triangular voltages. 

\begin{figure}[htb!]
\centering
    \begin{subfigure}[htb!]{0.49\textwidth}
        \centering
        \includegraphics[width=\linewidth]{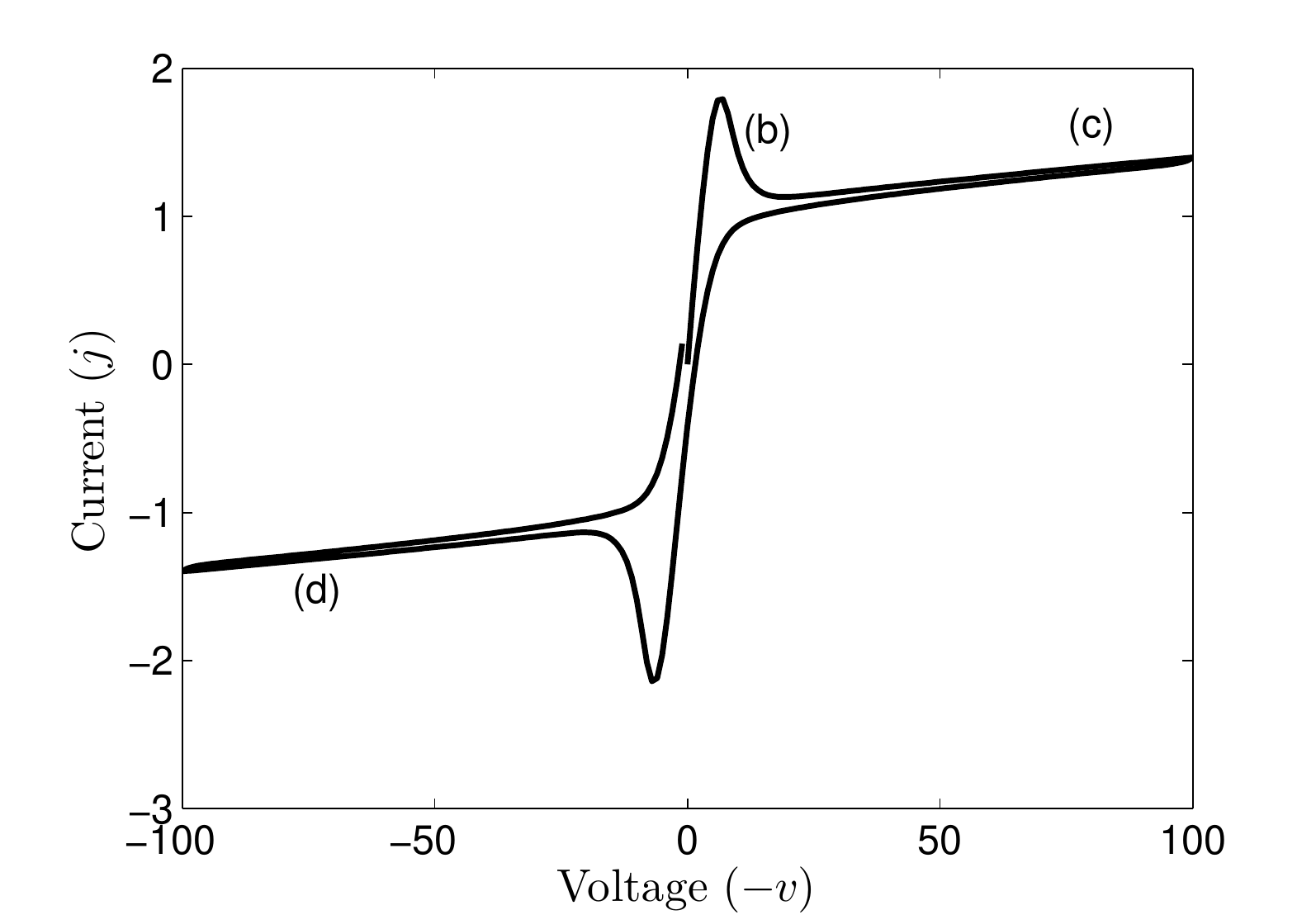}
        \caption{Current \change{vs. Voltage}}\label{TSC_a_voltammogram}
    \end{subfigure}
    \begin{subfigure}[htb!]{0.49\textwidth}
        \centering
        \includegraphics[width=\linewidth]{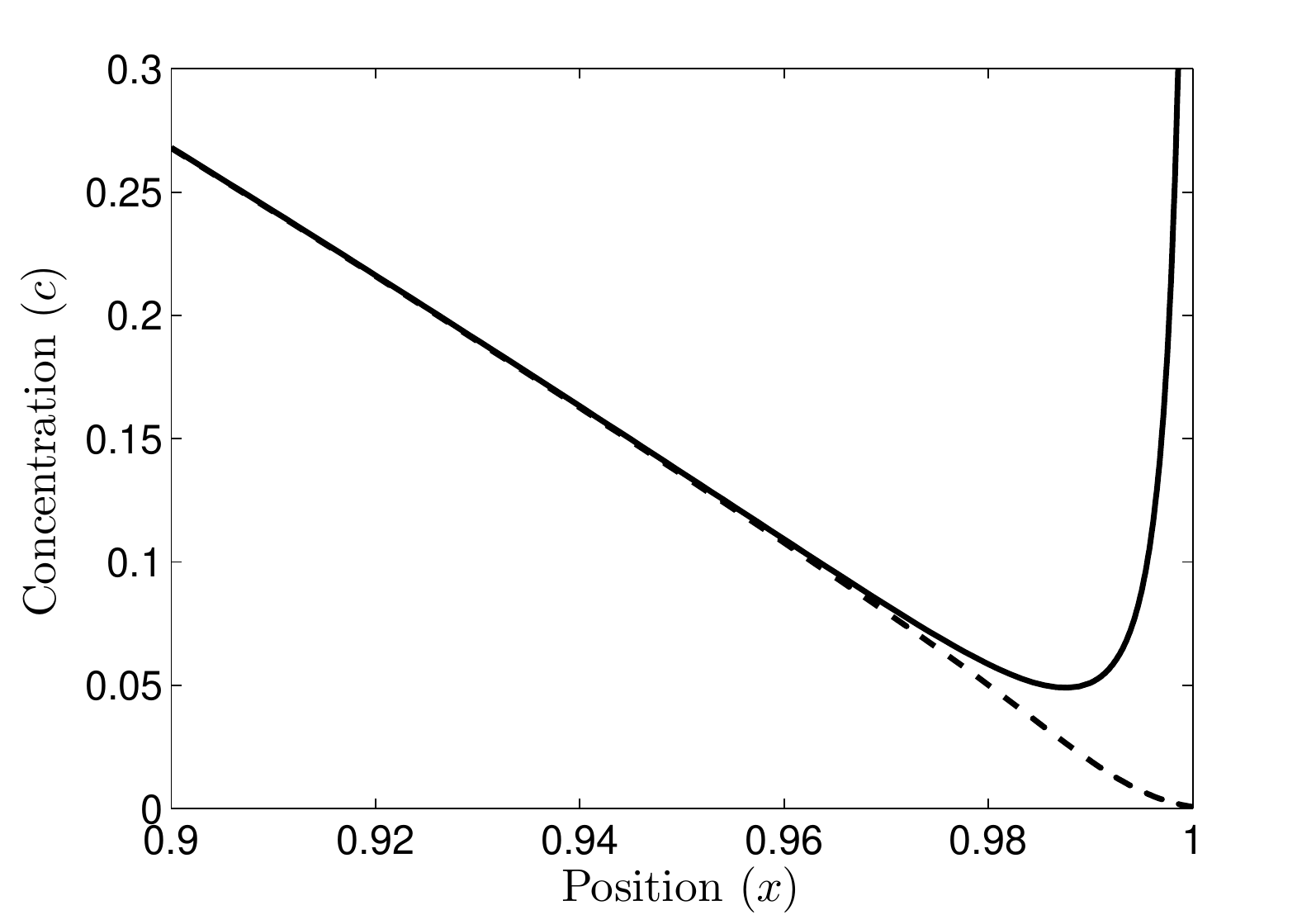}
        \caption{t=0.1}\label{TSC_b}
    \end{subfigure}

    \begin{subfigure}[htb!]{0.49\textwidth}
        \centering
        \includegraphics[width=\linewidth]{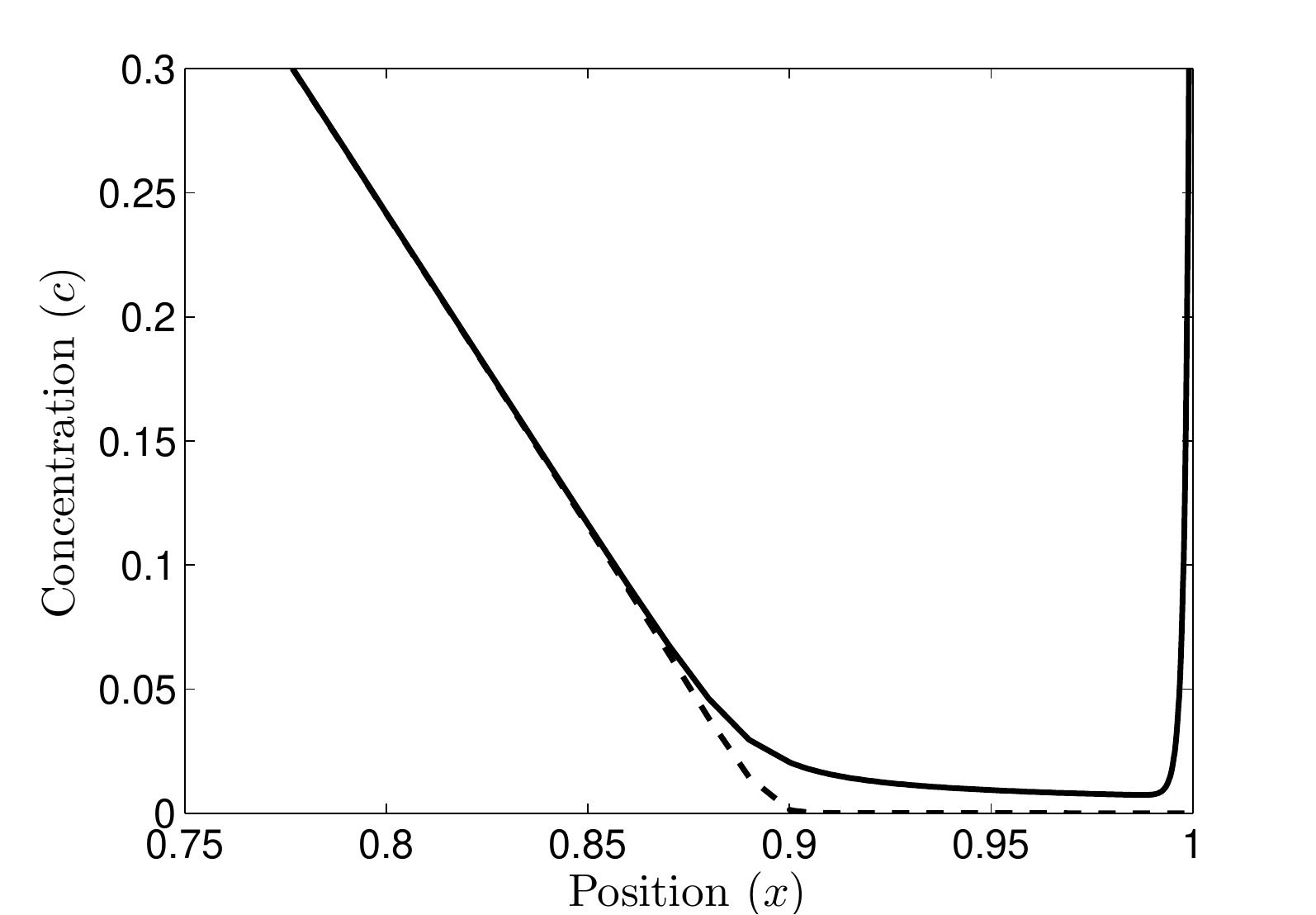}
        \caption{t=0.6}\label{TSC_c}
    \end{subfigure}
    \begin{subfigure}[htb!]{0.49\textwidth}
        \centering
        \includegraphics[width=\linewidth]{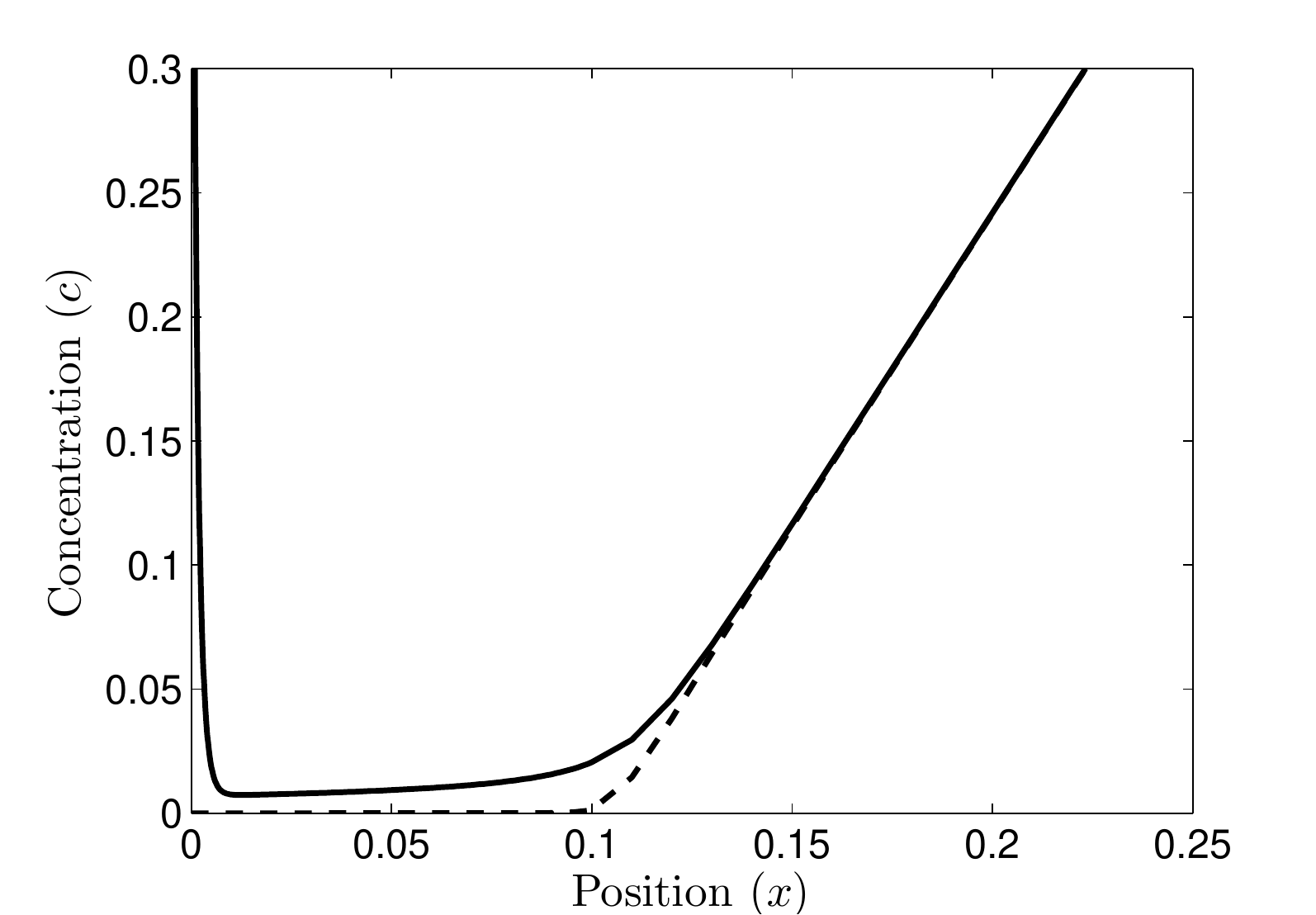}
        \caption{t=2.6}\label{TSC_d}
    \end{subfigure}
    \caption{(a) \change{Voltammogram} and (b-d) resulting cation (solid line) and anion (dashed line) concentrations showing space charge regions developed at the cathode and anode from a triangular applied voltage in a thin EDL liquid electrolyte with two electrodes. Parameters used were $\epsilon=0.001$, $\delta=0.3$, $k=50$, and $|\tilde S|=100$.}\label{TSC}
\end{figure}

Figure \change{\ref{TSC_a_voltammogram}} shows the voltammogram of a two-electrode liquid electrolyte system subjected to a triangular voltage with $\epsilon=0.001$, $\delta=0.3$, $k=50$ and $\tilde S=100$, and \change{Figures \ref{TSC_b}--\ref{TSC_d}} show the development of space charge regions.
%\begin{figure}[htb!]
%\centering
%\includegraphics[width=0.65\linewidth]{TSC_voltammogram.pdf}
%\caption{Voltammogram for a thin EDL liquid electrolyte with two electrodes subjected to a triangular voltage with $\epsilon=0.001$, $%\delta=0.3$, $k=50$ and $|\tilde S|=100$. Concentrations with development of space charge regions are shown in Figure \ref{TSC}.}
%\label{TSC_voltammogram}
%\end{figure}
The current-voltage response during space charge formation is a transient case of a diffusion limited system ($\tilde S, k \gg 1$) being driven above the limiting current \cite{Chu2005, Rubinstein1979, Smyrl1966} and the subsequent breakdown of electroneutrality in the bulk. Figure \ref{TSC} shows the current peaking as diffusion limiting sets in ($t=0.1$) and anion concentration at the cathode reaches zero (Figure \ref{TSC_b}). After this time, a space charge region begins to form outside of the double layer as seen in Figure \ref{TSC_c}, and the current ramps up slowly until the voltage reverses direction. Since the cell is symmetrical, the same thing occurs at the anode during the positive voltage part of the cycle, with current flowing in the other direction (Figure \ref{TSC_d}). The height of the current peak and slope of current during space charge development are dependent on the value of $\tilde S$. Also, though the one-dimensional equations predict it, the formation of large space charge regions may not happen in reality due to hydrodynamic instability caused by electrokinetic effects \cite{BazantSquires2004, Bazant2010, Levitan2005}. 

%\change{\sout{Finally, an effect which is not included in the simulation that could prove important is ionic crowding \cite{Kilic2007b}, as both concentrations reach large values (i.e. $c_-(x=0,t=0.6)\approx 4$) at the electrodes. It may be interesting to extend these simulations to include steric effects in future work.}}

%you could still expand the discussion in a few places. for example the transient space charge section says what has been done before and then points to a single numerical simulation without any discussion, e.g. of when this situation arises, how qualitatively it affects LSV curves, etc. 

\section{Leaky Membranes}\label{leaky}

\subsection{Model Problem}
In this section, we consider the classical description of membranes as having constant, uniform background charge density $\rho_s$, in addition to the mobile ions~\cite{Teorell1935,Meyer1936,Spiegler1971,Abu-Rjal2014}. In this section, we focus on the strongly nonlinear regime of small background charge and large currents in a ``leaky membrane"~\cite{Dydek2013,Yaroshchuk2012}.  This situation can arise as a simple description of micro/nanochannels with charged surfaces, as well as traditional porous media, neglecting electro-osmotic flows. In the case of a microchannel with negative charge on its side walls, surface conduction through the positively charged diffuse layers can sustain over-limiting current (faster than diffusion) \cite{Dydek2011} and deionization shock waves \cite{Mani2011}. This phenomenon has applications to desalination by shock electrodialysis~\cite{Deng2013,Schlumpberger2015}, as well as metal growth by shock electrodeposition~\cite{Han2014,Han2016}, and the following analysis could be used to interpret LSV for such electrochemical systems with bulk fixed charge.

\begin{figure}[htb!]
\centering
\includegraphics[width=0.5\linewidth]{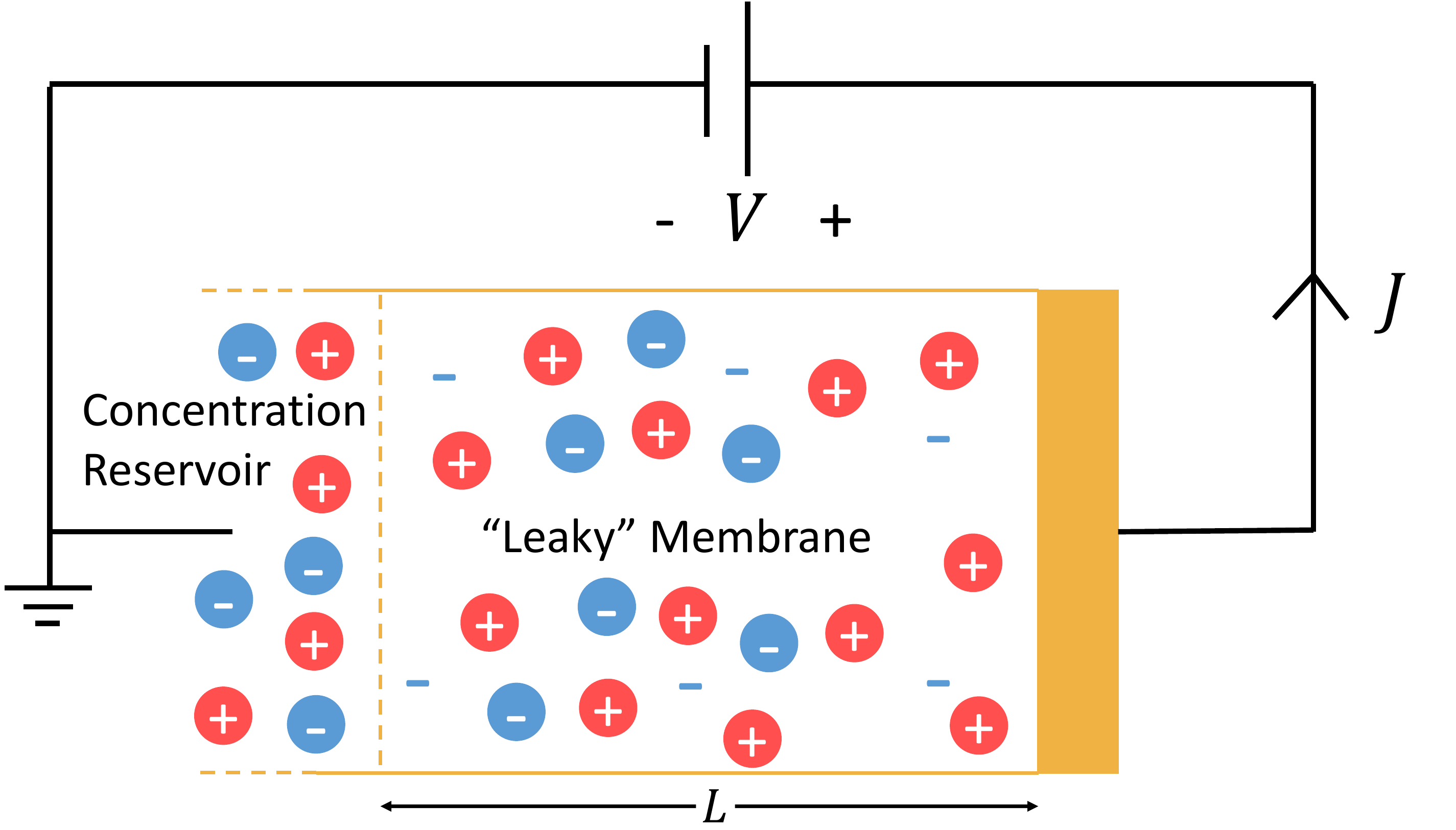}
\caption{Sketch of a ``leaky" membrane. Left hand boundary is an ideal reservoir with constant concentration ($c_+=c_-=1$) and zero potential ($\phi=0$). Right hand boundary is an electrode. Mobile charge is shown with a filled circle, fixed charge (negative in this case) is shown without a circle. Also shown are the directions of the voltage and current density.}\label{voltammetry_batt_diagram_LMM}
\end{figure}

Figure \ref{voltammetry_batt_diagram_LMM} shows a sketch of the model problem. A ``leaky" membrane with a uniform background charge (negative in the figure) and with mobile cations and anions lies between an ideal reservoir on the left hand side ($c_+=c_-=1$, $\phi=0$) and an electrode on the right. We investigate situations with both positive and negative background charge whose concentration is small compared to that of the mobile ions.

The appropriate modification to the Poisson equation for a background charge is equation \eqref{poisson_lmm}, which  is nondimensionalized to
\begin{equation}
\label{poisson_lmm_nondim}
-2\epsilon^2\frac{\partial^2 \phi}{\partial x^2}=z_+c_++z_-c_-+2\tilde \rho_s
\end{equation}
where $\tilde \rho_s=\rho_s/(2C_0F)$. Equations \eqref{poisson_lmm}--\eqref{poisson_lmm_nondim} are equivalent to the ``uniform
potential model" and ``fine capillary model" \cite{Peters2016} and have a long history in membrane science~\cite{Teorell1935, Meyer1936, Spiegler1971}. For example, Tedesco et. al. \cite{Tedesco2016} recently used an electroneutral version of equation \eqref{poisson_lmm_nondim} to model ion exchange membranes for electrodialysis applications.

The time-independent Nernst-Planck equations can be solved along with equation \eqref{poisson_lmm_nondim} in the limit of thin DL's to obtain the steady-state current-voltage relationship~\cite{Dydek2011}, which is given by
\begin{equation}
\label{lmm_iv_steadystate}
j=\frac{1-e^{-|v|/2}-\tilde \rho_s |v|}2
\end{equation}
where the factor of one half is due to a difference in our definition of the scaling current (in equation \eqref{current_conservation_nondim}) from \cite{Dydek2011}. This expression has been successfully fitted to quasi-steady current voltage relations in experiments~\cite{Deng2013,Han2014,Han2016}, which in fact were obtained by LSV at low sweep rates, so it is important to understand the effects of finite sweep rates. In Sections \ref{lmm_neg}--\ref{lmm_pos}, we present simulation results for ramped and cyclic voltammetry on systems with background charge opposite sign (negative background charge) and the same sign (positive background charge) as the reactive ions.

\subsection{Negative Background Charge}\label{lmm_neg}
\begin{figure}[htb!]
\centering
\includegraphics[width=0.65\linewidth]{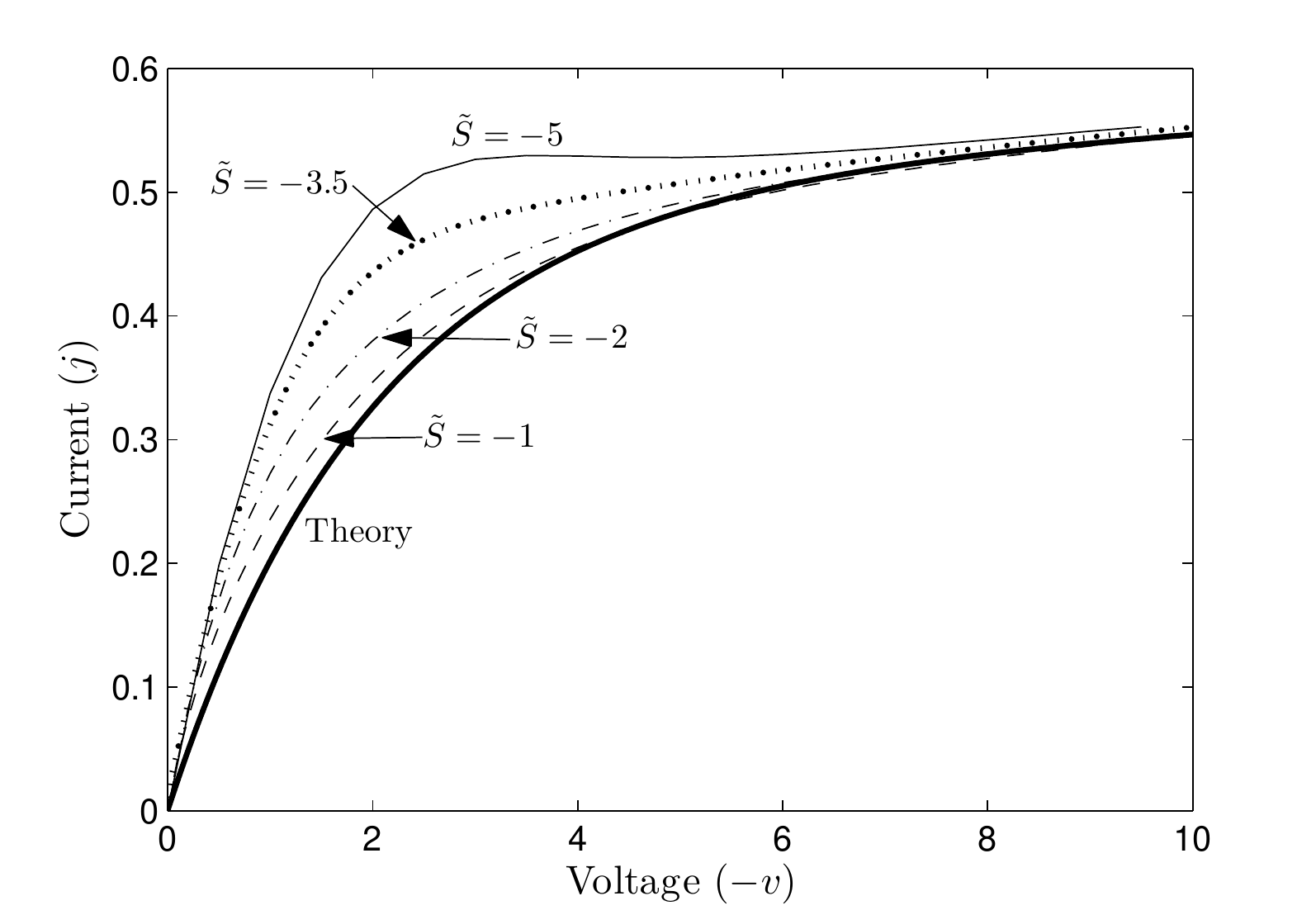}
\caption{Current in response to a voltage ramp in a liquid electrolyte with one electrode and constant background charge. Parameters are $\epsilon=0.005$, $\delta=10$, $\rho_s=-0.01$ and $k=50$. Also shown is the steady state response from equation \eqref{lmm_iv_steadystate}. }
\label{LMM_neg_001}
\end{figure}

First, we consider the case where the sign of the background charge is opposite to that of the reactive cations, which avoid depletion by screening the fixed background charge.  This is the most interesting case for applications of leaky membranes~\cite{Deng2013, Schlumpberger2015, Han2014, Han2016}, since the system can sustain over-limiting current. Figure \ref{LMM_neg_001} shows current in response to voltage ramps with $\epsilon=0.005$, $\delta=10$, $\rho_s=-0.01$ and $k=50$, with various values for $\tilde S$. The limiting behavior for the current for small $\tilde S$ can be predicted by the steady state response from equation \eqref{lmm_iv_steadystate}.  As observed in multiple experiments~\cite{Deng2013,Han2014,Han2016}, a bump of current overshoot occurs prior to steady state for high sweep rates, which we can attribute to diffusion limitation during transient concentration polarization in the leaky membrane. Similar bumps have also been predicted by Moya et al.~\cite{Moya2015} for neutral electrolytes in contact with (non-leaky) ion-exchange membranes with quasi-equilibrium double layers. 

%\begin{figure}[htb!]
%\centering
%\includegraphics[width=0.65\linewidth]{LMM_neg_01.pdf}
%\caption{Current from a voltage ramp applied to a liquid electrolyte with a single electrode and constant background charge. Parameters are $\epsilon=0.005$, $\delta=10$, $\rho_s=-0.1$ and $k=50$. Also shown is the steady state response from equation \eqref{lmm_iv_steadystate}.}
%\label{LMM_neg_01}
%\end{figure}

\begin{figure}[htb!]
\centering
    \begin{subfigure}[htb!]{0.49\textwidth}
        \centering
        \includegraphics[width=\linewidth]{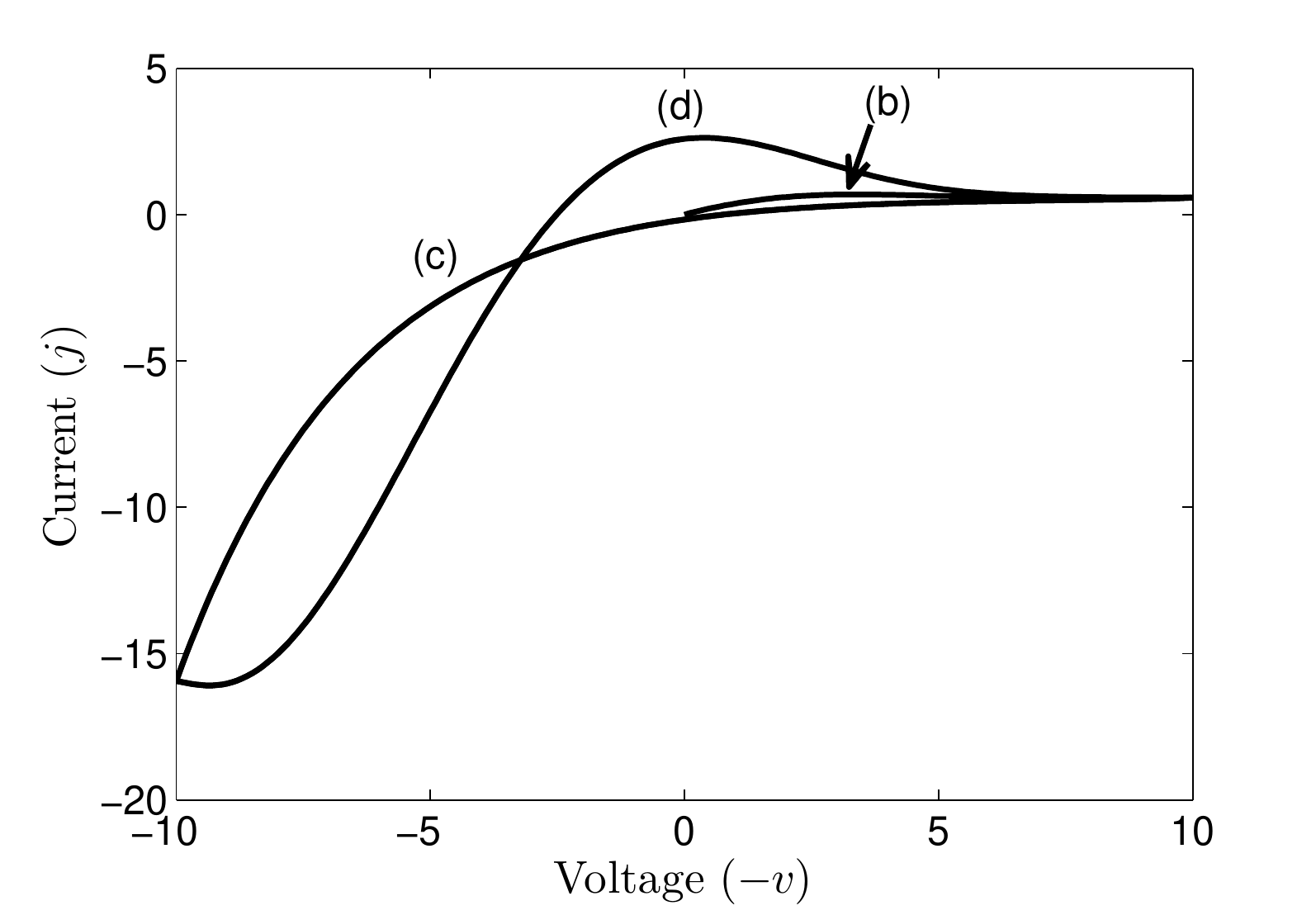}
        \caption{Voltammogram}\label{LMM_voltammogram_neg_001_A}
    \end{subfigure}
    \begin{subfigure}[htb!]{0.49\textwidth}
        \centering
        \includegraphics[width=\linewidth]{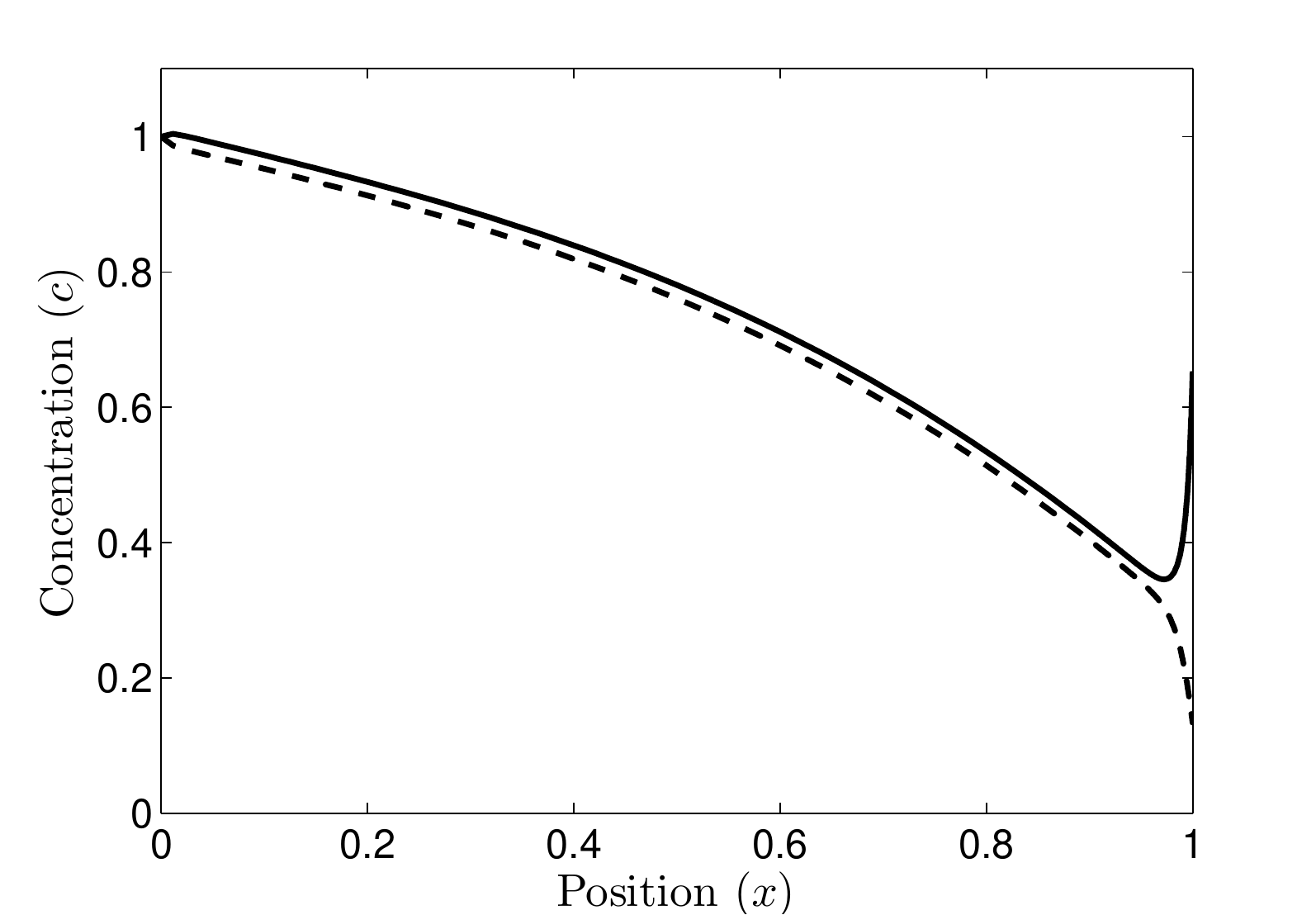}
        \caption{v=-3.3, t=0.33}\label{LMM_voltammogram_neg_001_B}
    \end{subfigure}

    \begin{subfigure}[htb!]{0.49\textwidth}
        \centering
        \includegraphics[width=\linewidth]{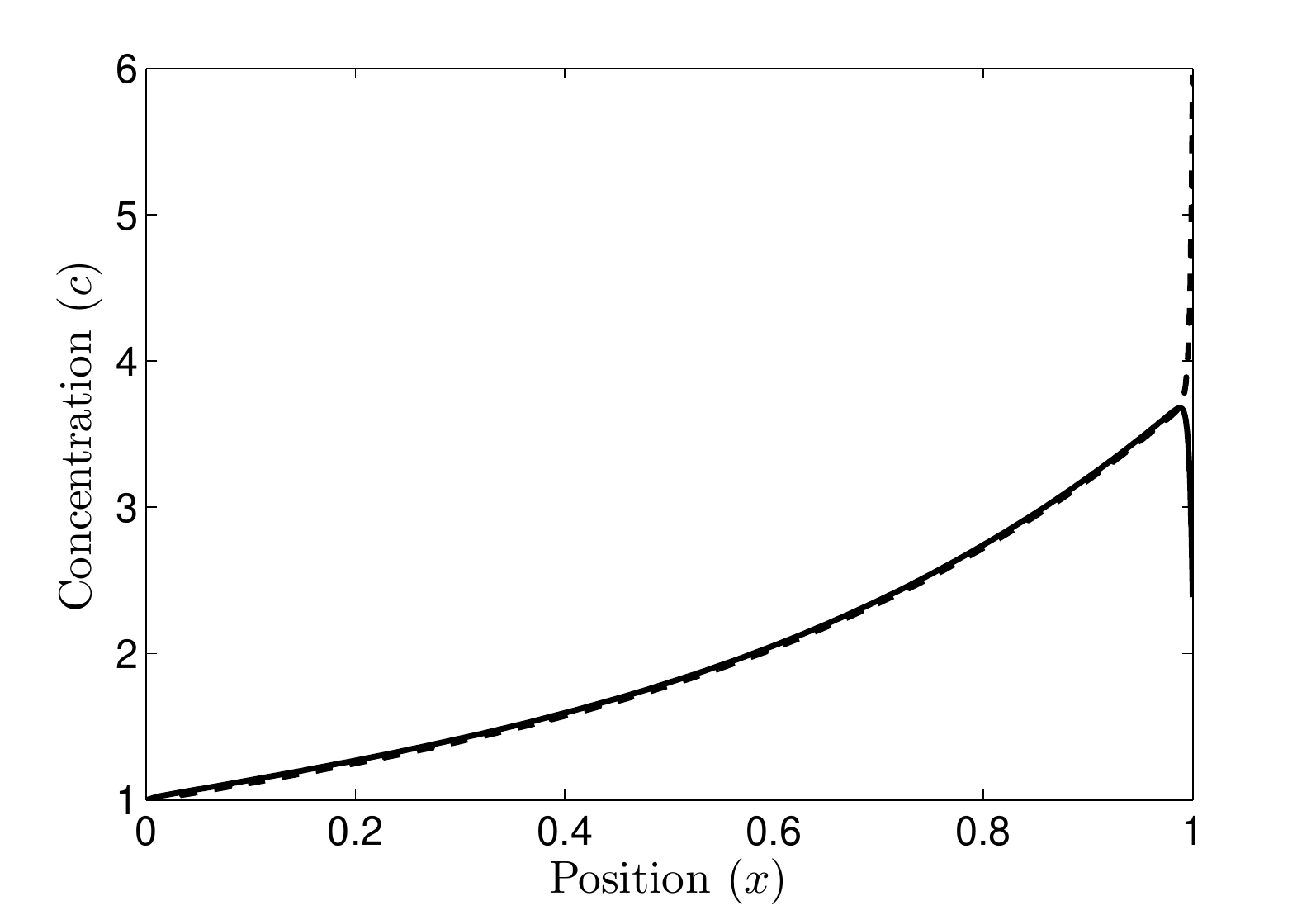}
        \caption{v=5, t=2.5}\label{LMM_voltammogram_neg_001_C}
    \end{subfigure}
    \begin{subfigure}[htb!]{0.49\textwidth}
        \centering
        \includegraphics[width=\linewidth]{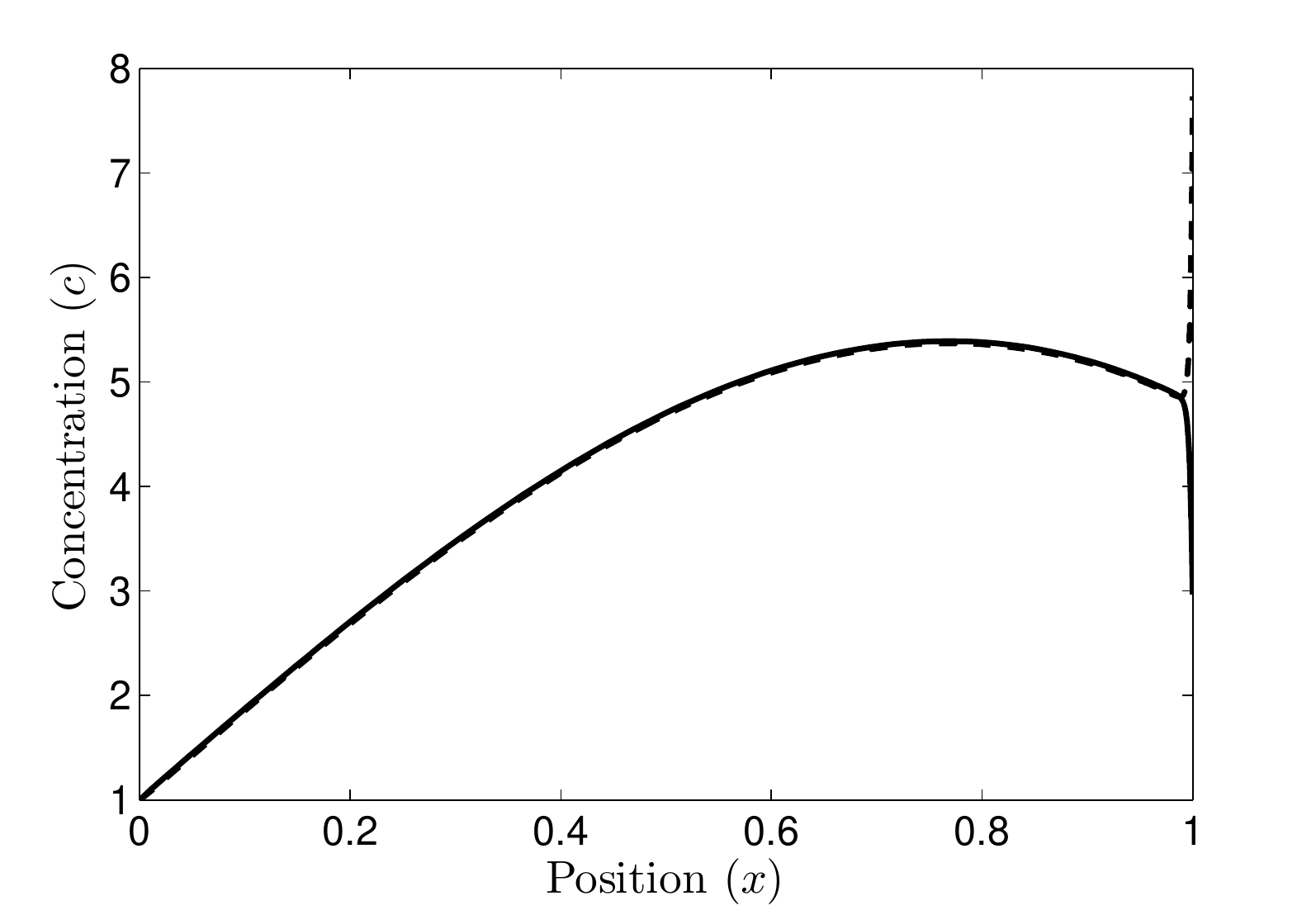}
        \caption{v=0, t=4, second cycle}\label{LMM_voltammogram_neg_001_D}
    \end{subfigure}
    \caption{(a) Voltammogram and (b-d) resulting cation (solid line) and anion (dashed line) concentrations for a thin EDL liquid electrolyte with two electrodes and constant negative background charge. Parameters used were $\epsilon=0.005$, $\delta=1$, $k=50$, $\rho_s=-0.01$ and $|\tilde S|=10$. Two cycles are shown in Figure (a).}\label{LMM_voltammogram_neg_001}
\end{figure}

Next, Figure \ref{LMM_voltammogram_neg_001} shows a cyclic voltammogram with concentration profiles for a background charges of $\rho_s=-0.01$, with $\tilde S=10$. Due to the additional background charge, the concentrations in the bulk are slightly different. For a 1:1 electrolyte, this difference is exactly $-2\rho_s$, as in equation \eqref{poisson_lmm_nondim}. For an electrolyte that is not 1:1, the difference can be obtained using $z_+c_++z_-c_-=-2\rho_s$.

%\begin{figure}[htb!]
%\centering
 %   \begin{subfigure}[htb!]{0.49\textwidth}
   %     \centering
     %   \includegraphics[width=\linewidth]{LMM_voltammogram_neg_01_A.pdf}
      %  \caption{Voltammogram}\label{LMM_voltammogram_neg_01_A}
   % \end{subfigure}
    %\begin{subfigure}[htb!]{0.49\textwidth}
      %  \centering
       % \includegraphics[width=\linewidth]{LMM_voltammogram_neg_01_B.pdf}
      %  \caption{v=-3.3, t=0.33}\label{LMM_voltammogram_neg_01_B}
   % \end{subfigure}
%
  %  \begin{subfigure}[htb!]{0.49\textwidth}
  %      \centering
  %      \includegraphics[width=\linewidth]{LMM_voltammogram_neg_01_C.pdf}
  %      \caption{v=5, t=2.5}\label{LMM_voltammogram_neg_01_C}
%    \end{subfigure}
 %   \begin{subfigure}[htb!]{0.49\textwidth}
%        \centering
  %      \includegraphics[width=\linewidth]{LMM_voltammogram_neg_01_D.pdf}
  %      \caption{v=0, t=4, second cycle}\label{LMM_voltammogram_neg_01_D}
%    \end{subfigure}
%    \caption{(a) Voltammogram and (b-d) resulting cation (solid line) and anion (dashed line) concentrations from a triangular applied voltage applied to a thin EDL liquid electrolyte with two electrodes and constant, negative background charge. Parameters used were $\epsilon=0.005$, $\delta=1$, $k=50$, $\rho_s=-0.1$ and $|\tilde S|=10$. Two cycles are shown in Figure (a).}
%\label{LMM_voltammogram_neg_01}
%\end{figure}

Similar to the cyclic voltammogram in Section \ref{unsupported_thinedl}, the current-voltage relationship in Figure \ref{LMM_voltammogram_neg_001} displays diffusion limited behavior in the negative voltage sweep direction and purely exponential growth (reaction limiting behavior) in the other. This is because there is only one electrode in these simulations with only one of the two species taking part in the reaction. %\change{\sout{Furthermore, as with the space-charge voltammetry simulations in Section \ref{TSC_section}, the large voltage and fast sweep rate causes very large concentrations to develop near the electrodes (for example, $c_-(x=1,t=3)\approx 30$ in Figure \ref{LMM_voltammogram_neg_001}). The addition of steric effects into the model would again make the results more realistic. }}

\subsection{Positive Background Charge}\label{lmm_pos}

\begin{figure}[htb!]
\centering
\includegraphics[width=0.65\linewidth]{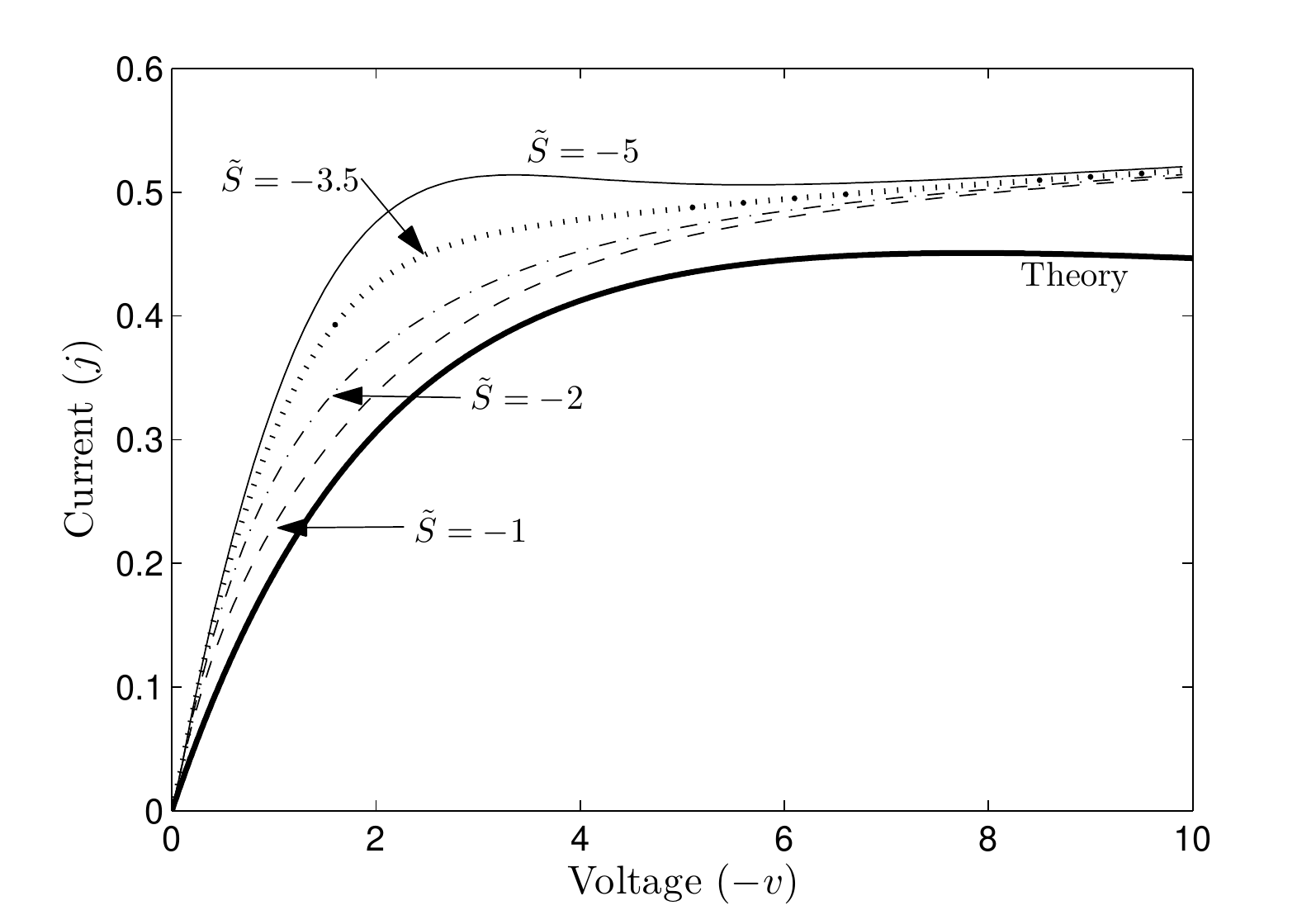}
\caption{Current from a voltage ramp applied to a liquid electrolyte with a single electrode and constant, small, positive background charge. Parameters are $\epsilon=0.005$, $\delta=10$, $\rho_s=0.01$ and $k=50$. Also shown is the steady state response from equation \eqref{lmm_iv_steadystate}, which is shown not to match the simulations.}
\label{LMM_pos_001}
\end{figure}

For positive $\rho_s$, equation \eqref{lmm_iv_steadystate} predicts a decreasing current (or negative steady-state differential resistance) after the exponential portion, a behavior which has been observed in some experiments ~\cite{Han2014,Han2016} and not others \cite{Deng2013}.  Interestingly, when double-layer effects and electrode reaction kinetics are  considered in the model simulations, the region of negative resistance is also not observed, as shown in Figure \ref{LMM_pos_001} for $\rho_s=0.01$. Physically, the interfaces provide overall positive differential resistance, even as the bulk charged electrolyte enters the over-limiting regime with negative local steady-state differential  resistance. 

Lastly, we omit the plot of the two-cycle voltammogram for positive background charge, and just remark that they shows results which are very similar to the voltammogram in Figure \ref{LMM_voltammogram_neg_001_A}.

\section{Blocking Electrodes}\label{blocking}
\subsection{Model Problem}
A blocking, or ideally polarizable electrode, is one where no Faradaic reactions take place. From a modeling perspective, this means setting $k_c$ and $j_r$ in the Butler-Volmer equation to zero, so that current is entirely due to the displacement current term in equation \eqref{current_conservation_nondim}. 

Voltammetry experiments are most often used to probe Faradaic reactions at test electrodes; in this application, nonfaradaic, or charging current is undesireable. With that being said however, linear sweep voltammetry is also a standard approach to measuring differential capacitance. Much of the early work in electrochemistry was centered around matching experimental differential capacitance curves with theory. Gouy \cite{Gouy1910} and Chapman \cite{Chapman1913} independently solved the Poisson-Boltzmann equation to obtain the differential capacitance per unit area for an electrode in a 1:1 liquid electrolyte, which in the present notation can be written as
\begin{equation}
\label{liquid_cap}
\tilde{\mathcal C}_\text{liquid}(\Delta \phi)=\frac{1}{\epsilon}\cosh{\frac{\Delta \phi}{2}}
\end{equation}
where $\Delta \phi$ is the diffuse layer. Later, Kornyshev and Vorotyntsev \cite{Kornyshev1981} performed a similar calculation for a solid electrolyte (an electrolyte where one ion is fixed in position with a homogeneous distribution) to obtain
\begin{equation}
\label{solid_cap}
\tilde{\mathcal C}_\text{solid}(\Delta \phi)=\frac{1}{\epsilon}\frac{1-e^{-\Delta \phi}}{\sqrt{e^{-\Delta \phi}+\Delta \phi-1}} \sgn(\Delta \phi)
\end{equation}
Note that the capacitance for the liquid electrolyte is symmetrical about 0, but the capacitance for the solid electrolyte is not symmetrical due to the fixed charge breaking the symmetry.

In this section, we use ramped voltages to investigate the behaviour of blocking electrodes for liquid and solid electrolytes with thin and thick double layers. Similar work has been done by Bazant et. al. \cite{Bazant2004}, who used asymptotics to study diffuse charge effects in a system with blocking electrodes subjected to a step voltage, Olesen et. al. \cite{Olesen2010} who used both asymptotics and simulations to do the same for sinusoidal voltages\change{, and recently by Feicht et. al. \cite{Feicht2016} who studied dynamics for high-to-low voltage steps. In this section, we present new simulations for ramped voltage boundary conditions.} The results of this section have applications to EDL supercapacitors \cite{Lee2014}, capacitive deionization \cite{Porada2013, Zhao2012} and induced charge electro-osmotic (ICEO) flows \cite{Bazant2009, Bazant2010}. In the simulations in this section, $\delta$ is set to 0.01 so that the capacitance is dominated by the diffuse part of the double layer.

The displacement current in equation \eqref{current_conservation_nondim} is related to the nondimensionalized capacitance through
\begin{equation}
j=-\frac{\epsilon^2}{2}\frac{d\phi_x}{dt}=\frac{\epsilon^2}{2}\frac{dq}{dt}=\frac{\epsilon^2}{2}\frac{dq}{dv}\frac{dv}{dt}
\end{equation}
where $q=-\frac{d \phi}{dx}$ is the surface charge density and $\frac{dv}{dt}=\tilde S$. Since $\mathcal C=\frac{dq}{dv}$, we have that
\begin{equation}
\label{j_and_C}
\frac{j}{\tilde S}=\frac{\epsilon^2}{2}\tilde{\mathcal C}\sim \frac{\epsilon}{2}
\end{equation}
and therefore $\epsilon$ is the natural scale for capacitance when relating to the displacement current. We refer to this as our ``rescaled" capacitance, which we denote using the symbol $\tilde C$.

\begin{figure}[htb!]
  \begin{center}
    \begin{circuitikz}[american voltages]
      \draw (0,0)
      to[short] (0,2)
      to[american voltage source,v=$v(t)$] (6,2)
      to[short,i=$j$] (6,0)
      to[C=$\tilde{C}$,v_<=$\Delta \phi$] (4,0)
      to[R=$\tilde{R}$, v<=$$] (2,0)
      to[C=$\tilde{C}$, v_>=$\Delta \phi$] (0,0);
    \end{circuitikz}
    \caption{Equivalent circuit diagram for system with two blocking electrodes, showing the defined direction of current and polarities of the double layer capacitors. Note that $\tilde C$ is a function of $\Delta \phi$.}
    \label{RC_circuit}
  \end{center}
\end{figure}
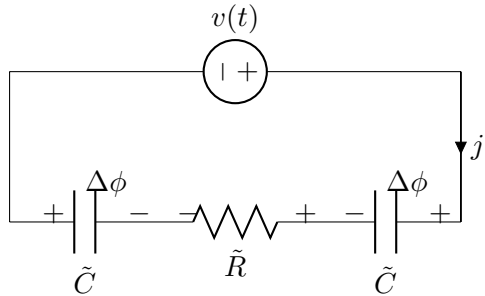

Both solid and liquid electrolyte systems with two blocking electrodes behave like the circuit shown in Figure \ref{RC_circuit}. 
For liquid electrolytes, $\tilde{R}\sim 2$ and $\tilde{C}(0)=\epsilon/2$. For solid electrolytes, $\tilde R \sim 4$ \cite{Bazant2005, Biesheuvel2009} and $\tilde{C}(0)=\frac{\epsilon^2}{2}\lim_{\Delta \phi \rightarrow 0} \tilde{\mathcal C}_\text{solid}(\Delta \phi)=\frac{\sqrt{2}\epsilon}{2}$.

There are two regimes of operation when a ramped voltage is applied to a capacitive system. The first is the small time ($t \ll \epsilon$) behavior, when the double layers are charged with time constant $\tau_\text{RC}=\tilde R \tilde C/2$, where the factor of $1/2$ accounts for the fact that there are two capacitors in series. To predict the behavior during this time, we turn to the ordinary differential equation describing the circuit in Figure \ref{RC_circuit}, which is
\begin{equation}
\label{circuit_equation}
v(t)-2\Delta \phi =\tilde{R}\tilde{C}(0)\frac{d\Delta \phi}{dt}
\end{equation}
where $v(t)=\tilde St$. From equation \eqref{circuit_equation}, the current can be solved in the case of a two electrode liquid electrolyte to be
\begin{equation}
\label{small_times}
\frac{2j_\text{inner}(t)}{\tilde S}=\tilde{C}(0)\left(1-e^{-\frac{v}{\tilde S\tilde{C}(0)}}\right)
\end{equation}
where we have used $\tilde{R}=2$ for a liquid electrolyte. The equivalent expression for solid electrolytes with single and double electrodes will be discussed in Section \ref{cap_results}. The second regime of operation is the large time ($t \gg \epsilon$) behavior. After the RC charging time, the current tracks the capacitance based on equation \eqref{j_and_C}. The relevant equation is
\begin{equation}
\label{j_phi}
j=\frac{\epsilon^2}{2}\frac{dq}{dt}=\tilde C(\Delta \phi) \frac{d\Delta \phi}{dt}
\end{equation}

Since we are only able to control to the potential drop across the cell ($v(t)$) and not the potential drop across the double layer ($\Delta \phi$), we must estimate the value of $\Delta \phi$ by accounting for the potential drop across the bulk. To do this, we can use the equation  $v(t)=2\Delta\phi + \tilde{R}j$. In practice however, $j\ll 1$ for blocking electrodes and so we can instead just use $\Delta \phi \approx v(t)/2=\tilde S t/2$. Equation \eqref{j_phi} can then be rewritten as 
\begin{equation}
\label{j_phi_2}
\frac{2j_\text{outer}}{\tilde S}=\tilde C(v/2)
\end{equation}
From equation \eqref{small_times} we have a solution for small times (the inner solution), and from equation \eqref{j_phi_2} we have a solution for large times (the outer solution). The long time limit for the inner solution must be equal to the small time limit for the outer solution, and so the two solutions can be combined by adding them and subtracting the overlap,
\begin{equation}
\label{matching}
j=j_\text{inner}+j_\text{outer}-j_\text{overlap}
\end{equation}
where $2j_\text{overlap}/\tilde S=\tilde C(0)$, thereby creating a uniformly valid approximation of the capacitance for all values of $v(t)$.

\subsection{Simulation Results} \label{cap_results}
We begin by presenting the uniformly valid approximations (equation \eqref{matching}) for liquid and solid electrolytes. For a liquid electrolyte with two electrodes, the current takes the form
\begin{equation}
\label{matching_liq}
\frac{2j}{\tilde S}=\tilde C\left(1-e^{\frac{v}{\tilde S \tilde C}}\right) + \frac{\epsilon}{2}\cosh{\frac{v}{4}} - \tilde C
\end{equation}
where $\tilde C=\epsilon/2$. The situation for solid electrolytes is more complicated: since the capacitance (equation \eqref{solid_cap}) is not symmetrical about $\Delta \phi=0$, the capacitance for a solid electrolyte system with two blocking electrodes can be represented using two capacitors in series,
\begin{equation}
\label{series_capacitor}
\frac{1}{\tilde C_\text{solid}(\Delta \phi)} = \frac{1}{\tilde{C}_\text{solid}(\Delta\phi)} +  \frac{1}{\tilde{C}_\text{solid}(-\Delta\phi)}
\end{equation}
The uniformly valid approximation for a solid electrolyte with one electrode is
\begin{equation}
\label{matching_solid_SE}
\frac{j}{\tilde S}=\tilde C\left(1-e^{\frac{v}{\tilde S \tilde C}}\right) + \epsilon\frac{1-e^{-v}}{\sqrt{e^{-v}+v-1}}\sgn {(v)} - \tilde C
\end{equation}
where $\tilde C=\sqrt{2}\epsilon$. Using equation \eqref{series_capacitor}, the approximation for two electrodes is
\begin{equation}
\label{matching_solid_DE}
\frac{2j}{\tilde S}=\tilde C\left(1-e^{\frac{v}{2\tilde S \tilde C}}\right) + \frac{\epsilon}{2}\left(\frac{1-e^{-\frac{v}{2}}}{\sqrt{e^{-\frac{v}{2}}+\frac{v}{2}-1}}\sgn {(v)}\bigg | \bigg | \frac{1-e^{\frac{v}{2}}}{\sqrt{e^{\frac{v}{2}}-\frac{v}{2}-1}}\sgn {(-v)}\right) - \tilde C
\end{equation}
where $\tilde C=\frac{\epsilon}{2\sqrt{2}}$ and $||$ indicates the reciprocal of the sum of the reciprocals, i.e. $A||B=1/(1/A+1/B)$. The additional factor of $1/2$ in the exponent in the inner part of equation \eqref{matching_solid_DE} is due to the fact that the electrolyte resistance for solid electrolytes is 4 rather than 2.

\begin{figure}[htb!]
\centering
    \begin{subfigure}[htb!]{0.49\textwidth}
        \centering
        \includegraphics[width=\linewidth]{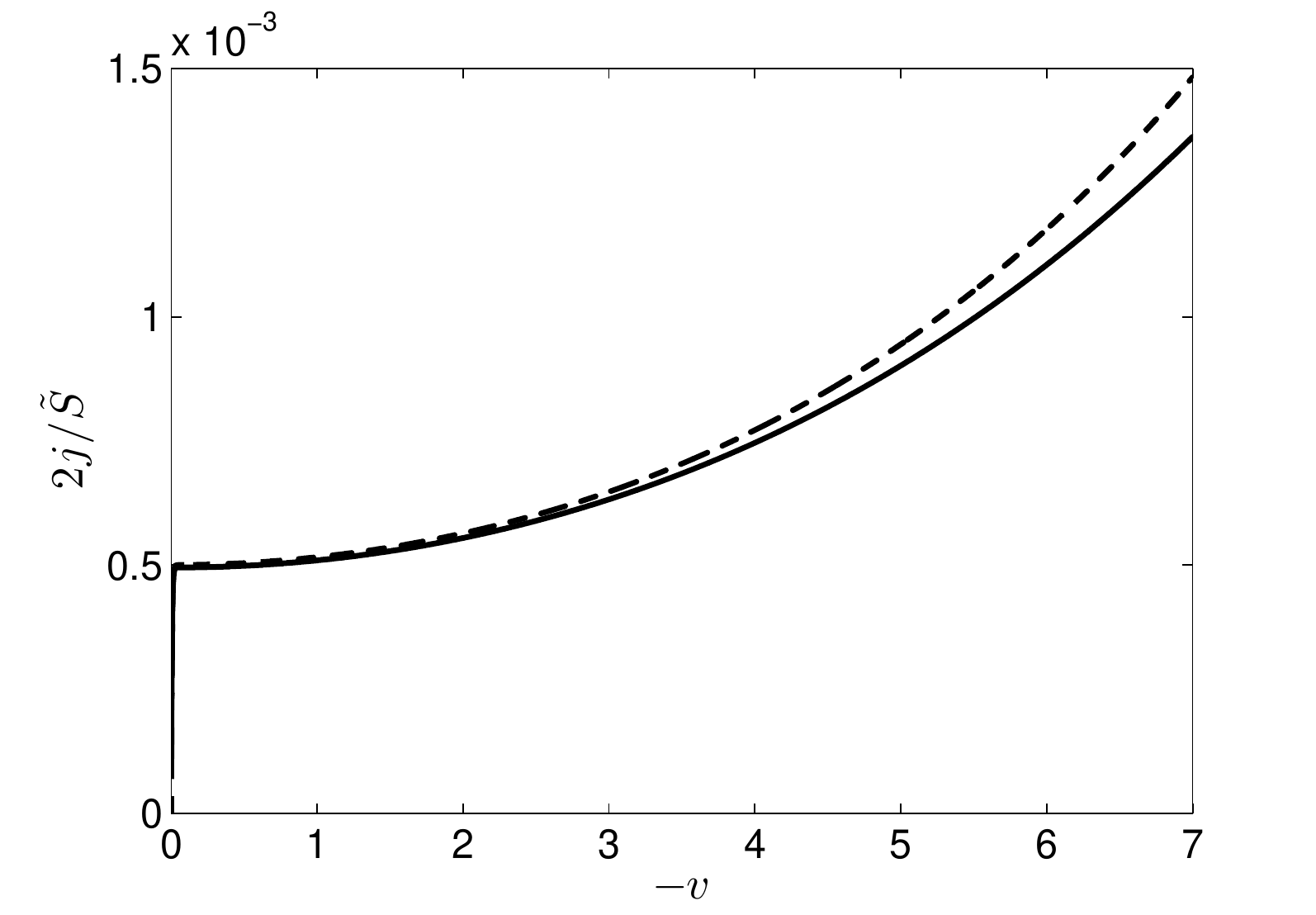}
        \caption{Liquid, Thin EDL}\label{liq_elec_capacitance_10}
    \end{subfigure}
    \begin{subfigure}[htb!]{0.49\textwidth}
        \centering
        \includegraphics[width=\linewidth]{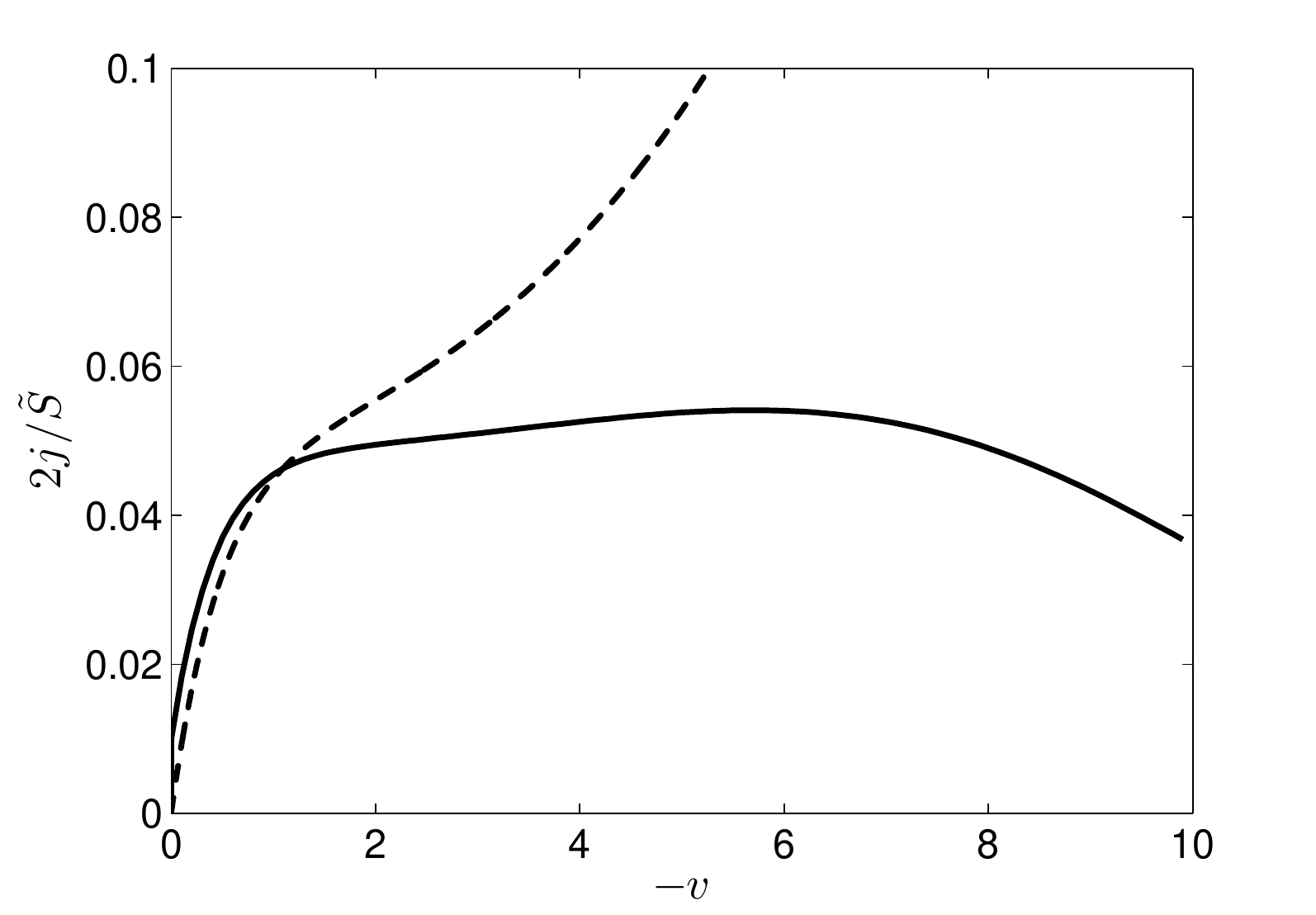}
        \caption{Liquid, Thick EDL}\label{liq_cap_thick_10}
    \end{subfigure}

    \begin{subfigure}[htb!]{0.49\textwidth}
        \centering
        \includegraphics[width=\linewidth]{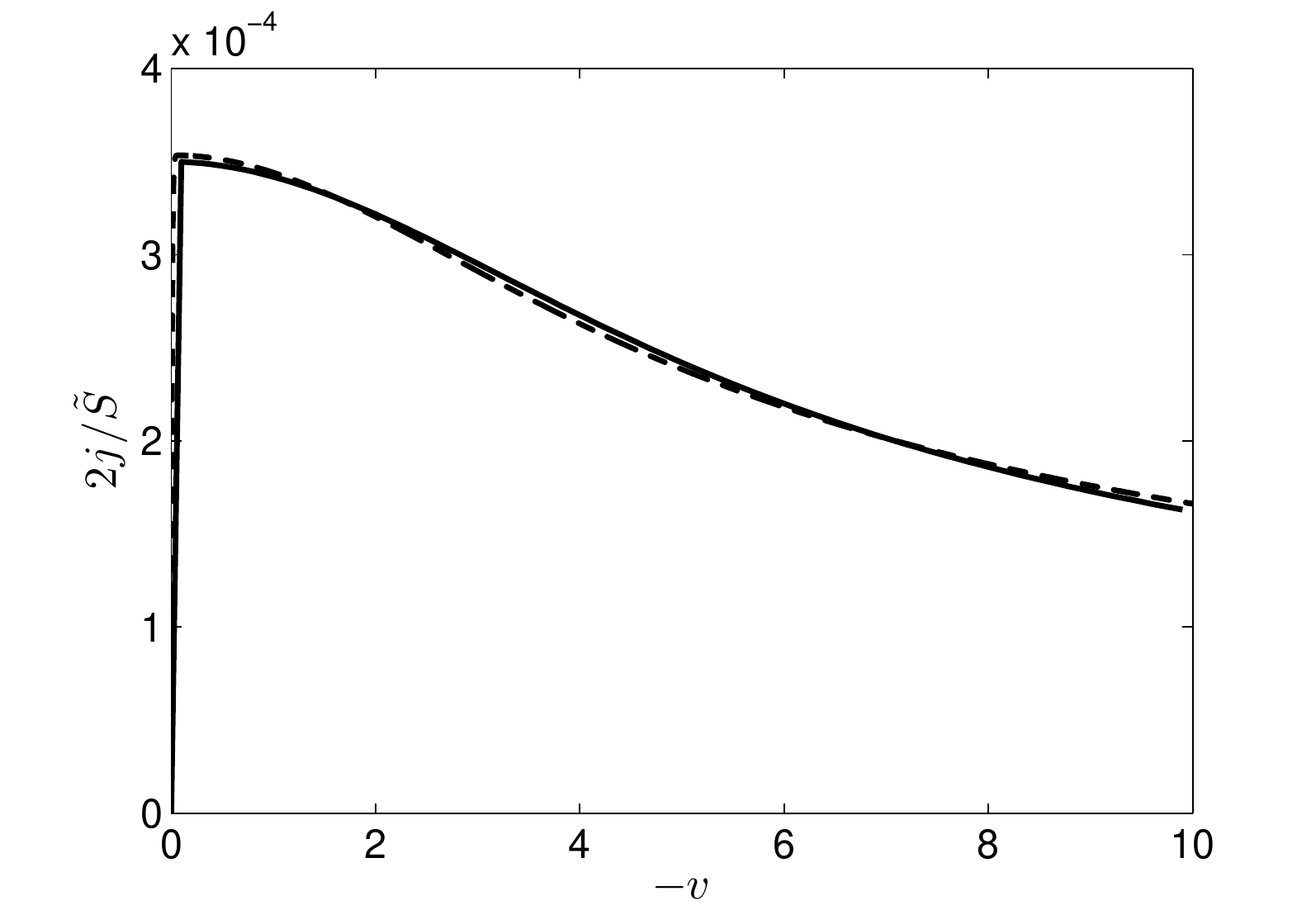}
        \caption{Solid, Thin EDL}\label{sol_elec_capacitance_10}
    \end{subfigure}
    \begin{subfigure}[htb!]{0.49\textwidth}
        \centering
        \includegraphics[width=\linewidth]{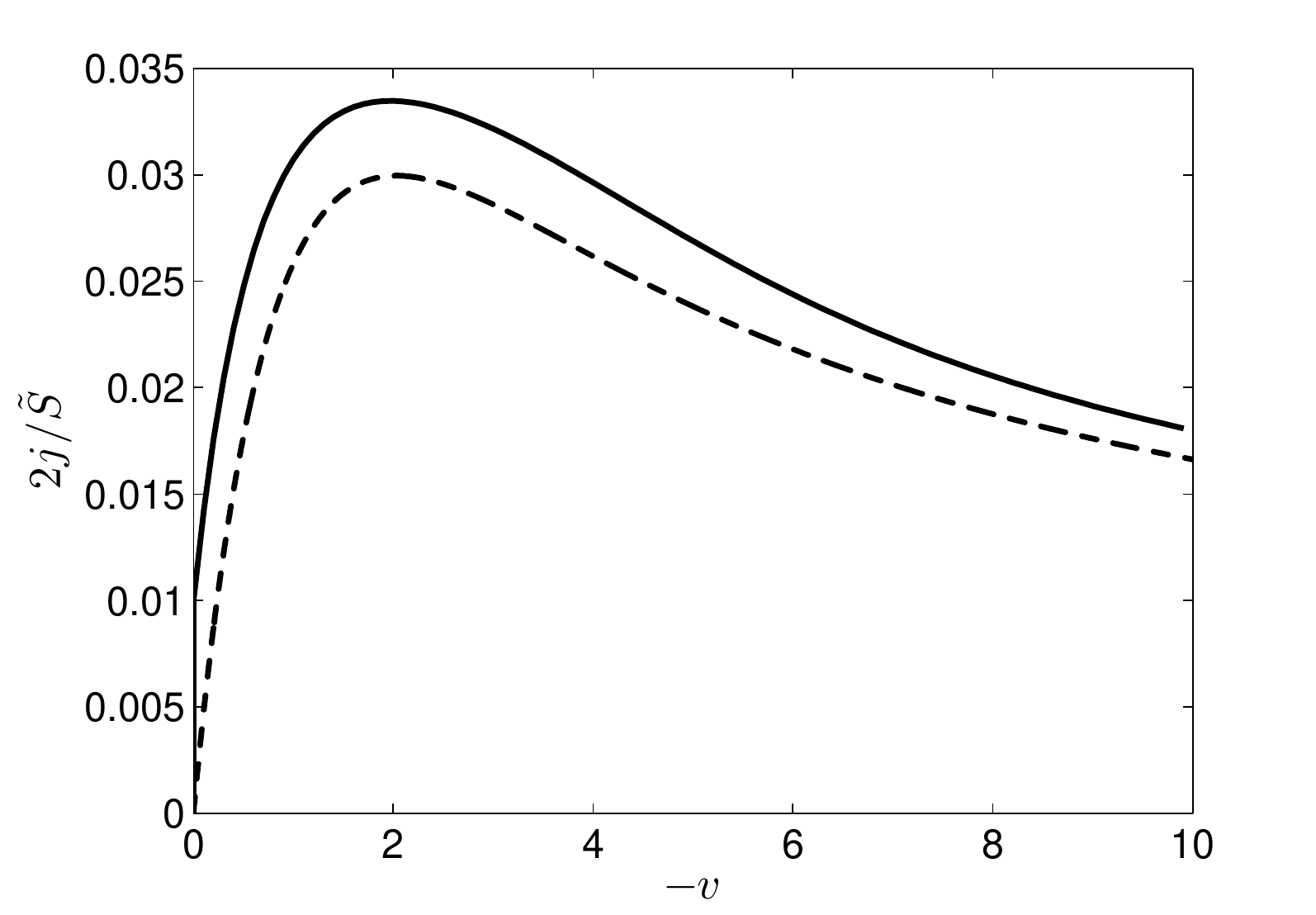}
        \caption{Solid, Thick EDL}\label{sol_cap_thick_DE_10}
    \end{subfigure}
    \caption{Simulated (solid lines) and uniformly valid approximation (dashed lines) $j$ vs $v$ curves for two-electrode liquid electrolyte with (a) thin and (b) thick double layers, as well as two-electrode solid electrolyte with (c) thin and (d) thick double layers. Plots with $\delta=0.01$ and $\tilde S=-10$ are shown. Dashed lines are plotted from equations \eqref{matching_liq} and \eqref{matching_solid_DE}.}\label{twoelectrode_capacitance}
\end{figure}

Figure \ref{twoelectrode_capacitance} shows $j-v$ curves for thin EDL ($\epsilon=0.001$) and thick EDL ($\epsilon=0.1$) liquid and solid electrolyte systems with two electrodes. The plots are compared to the approximations from equations \eqref{matching_liq} and \eqref{matching_solid_DE}, and show generally good agreement, except for Figure \ref{liq_cap_thick_10}, the liquid, thick EDL case.

%I think all I can say is that the approximation we use as the theoretical comparison curve is based on a GC, equilibrium double layer assumption. With thick EDL liquid electrolytes under large voltage forcings, so much charge separation happens that you are too far away from the equilibrium assumption for it to have any validity. You have something like one regime where normal near-equilibrium stuff happens (small v), another regime where the bulk concentration begins to be depleted (medium v), and finally at very large voltages you have complete charge separation. This should be why we see such a complete disagreement with liquid, but not with solid, since the fixed negative charge allows much less charge separation to occur. In fact, it seems that we are only able to see a difference for solid electrolyte when we probe just one electrode (the disagreement is with the larger capacitance positive voltage sweep, the smaller capacitance negative voltage sweep is unaffected. This is because the effect is masked in the two-electrode runs by the fact that for series capacitors, the smaller capacitor dominates).

\begin{figure}[htb!]
\centering
    \begin{subfigure}[htb!]{0.49\textwidth}
        \centering
        \includegraphics[width=\linewidth]{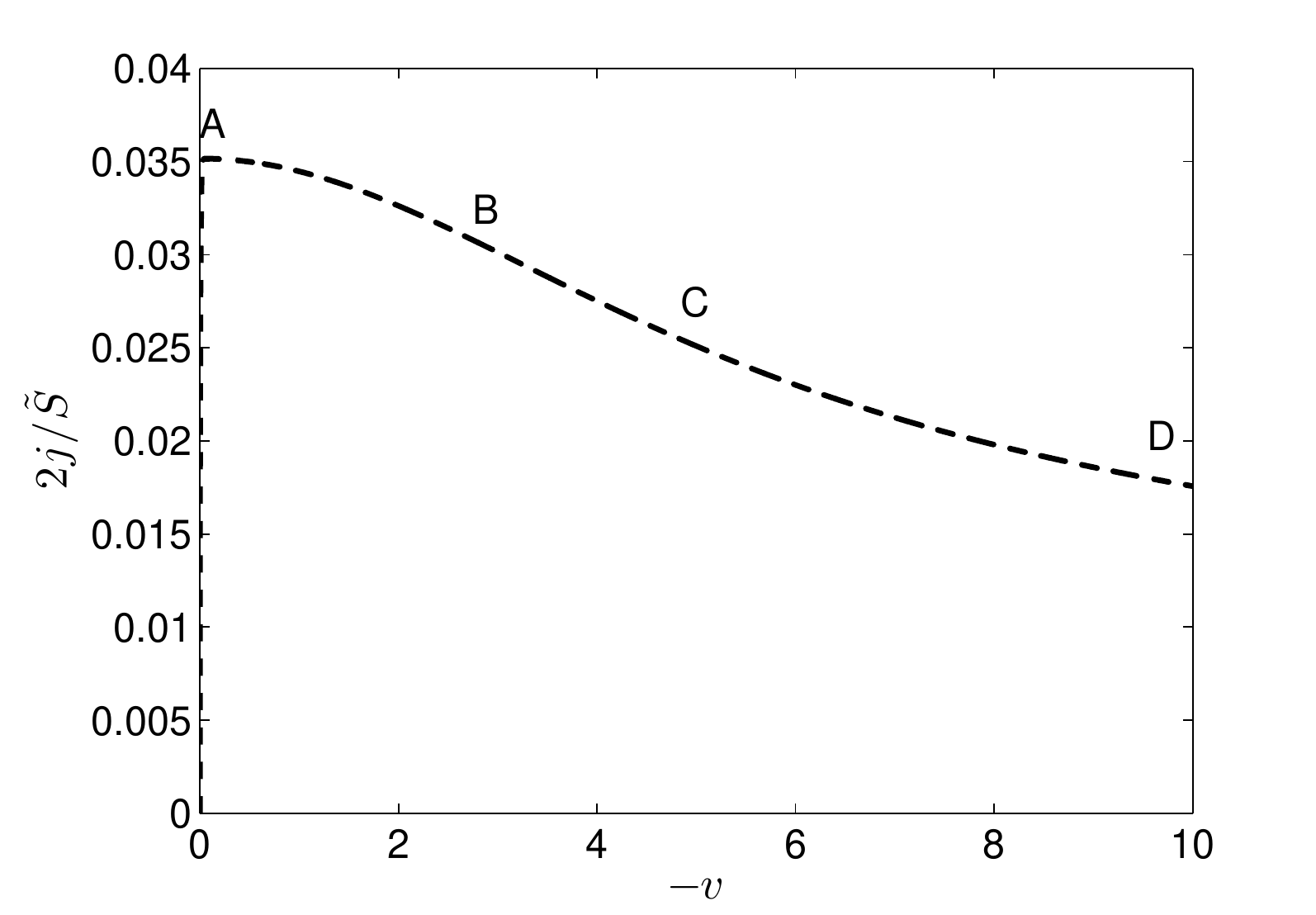}
        \caption{Current vs. voltage, $\tilde S=-0.1$}\label{liq_elec_capacitance_thick3_A}
    \end{subfigure}
    \begin{subfigure}[htb!]{0.49\textwidth}
        \centering
        \includegraphics[width=\linewidth]{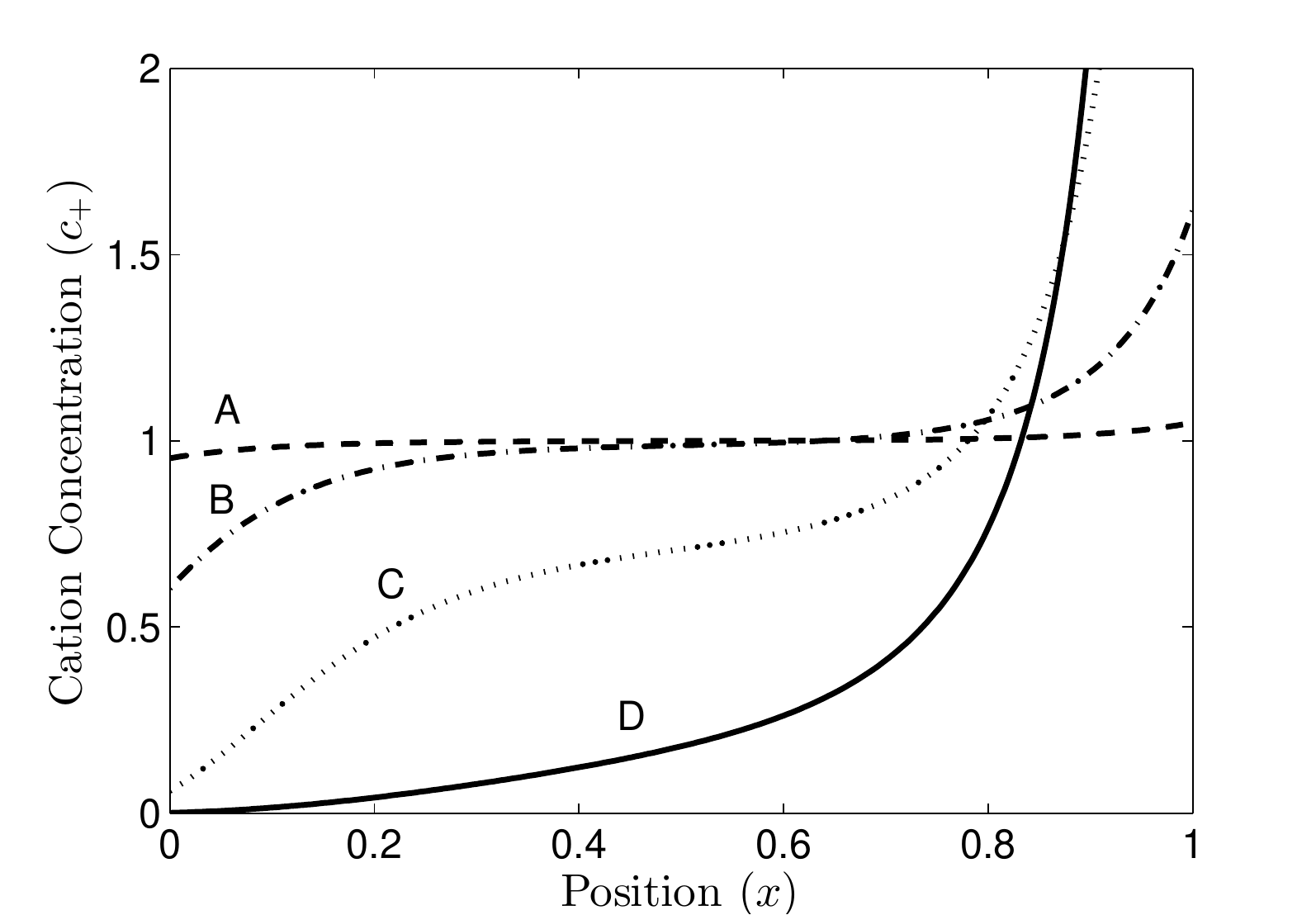}
        \caption{Cation concentrations}\label{liq_elec_capacitance_thick3_B}
    \end{subfigure}
    \caption{(a) $j$ vs. $v$ and (b) cation concentrations for a thick EDL liquid electrolyte with two blocking electrodes and parameters $\epsilon=0.1$, $\delta=0.01$ and scan rate $\tilde S=-0.1$. Labels in (a) correspond to cation concentrations in (b) at various times.} 
\label{liq_elec_capacitance_thick3}
\end{figure}

The reason for this is that the approximation in equation \eqref{matching_liq} is based on an equilibrium (Gouy-Chapman) picture of the EDL. With thick EDL liquid electrolytes under large voltage forcings, so much charge separation occurs that the system is too far from equilibrium for the Gouy-Chapman model to be valid. To investigate further, we plot in Figure \ref{liq_elec_capacitance_thick3} the current resulting from a voltage sweep of $\tilde S=-0.1$ on a liquid electrolyte system with $\epsilon=0.1$ along with cation concentrations at various points during the sweep.
We can separate the behavior in Figure \ref{liq_elec_capacitance_thick3} into three regimes. For small voltages (A), the concentrations show near-equilibrium behavior. As the voltage increases (B, C), the diffuse regions become large and the bulk concentration begins to be depleted, and the two double layers begin to overlap - it may be possible to model behavior at this stage by accounting for the depletion of bulk concentration as in \cite{Bazant2004}. Finally, at large voltages (D), we see complete charge separation, with nearly all of the positive charge located at the cathode (and similarly, with nearly all of the negative charge at the anode). This is result which highlights the need to model of diffuse charge dynamics using the PNP-FBV equations; such a highly nonlinear separation of charge would not be predicted by models which assume electroneutrality or neglect the coupling between diffuse charge dynamics and electrode currents.

\begin{figure}[htb!]
\centering
    \begin{subfigure}[htb!]{0.49\textwidth}
        \centering
        \includegraphics[width=\linewidth]{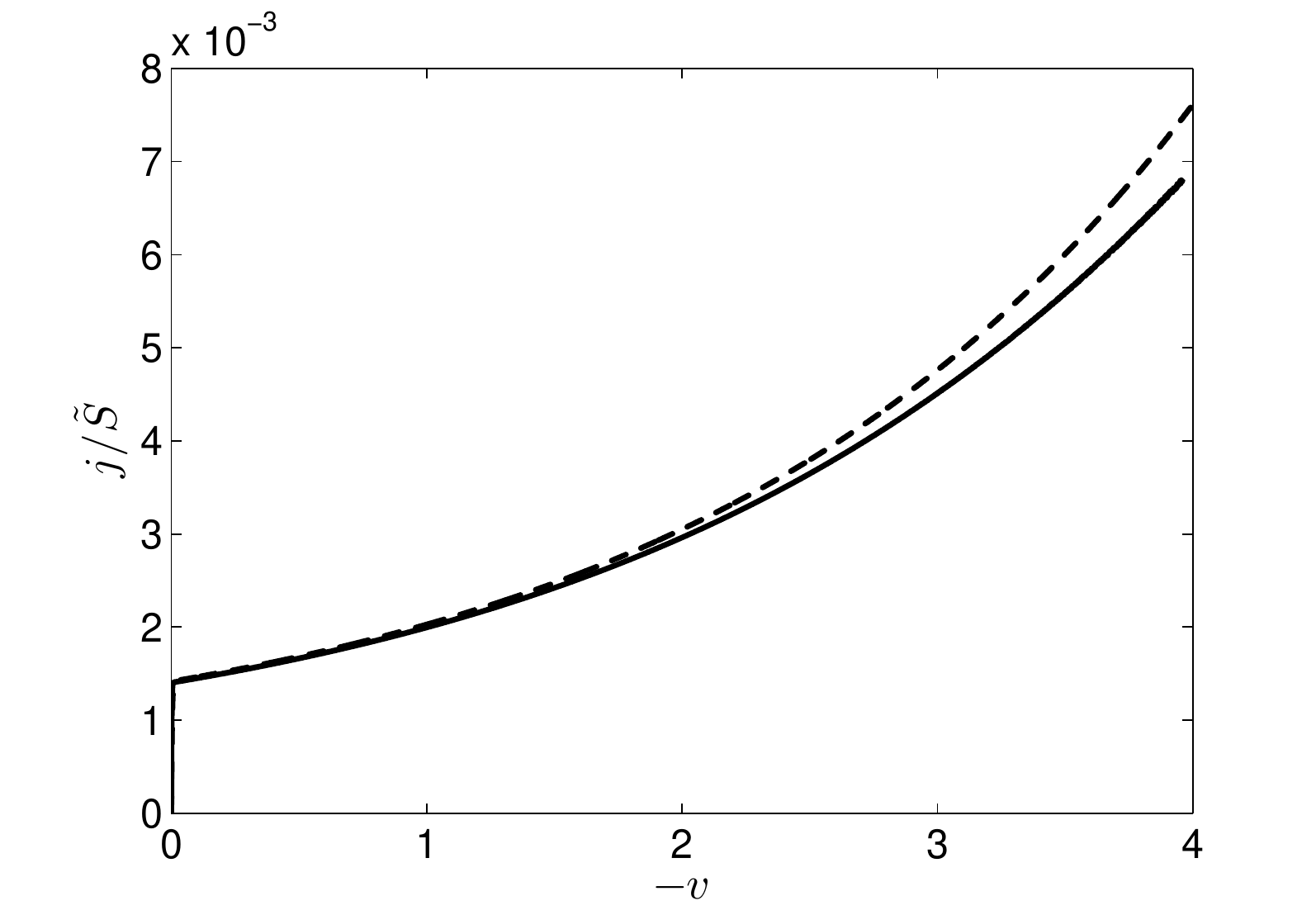}
        \caption{Negative voltage sweep}\label{sol_elec_cap_SE_neg}
    \end{subfigure}
    \begin{subfigure}[htb!]{0.49\textwidth}
        \centering
        \includegraphics[width=\linewidth]{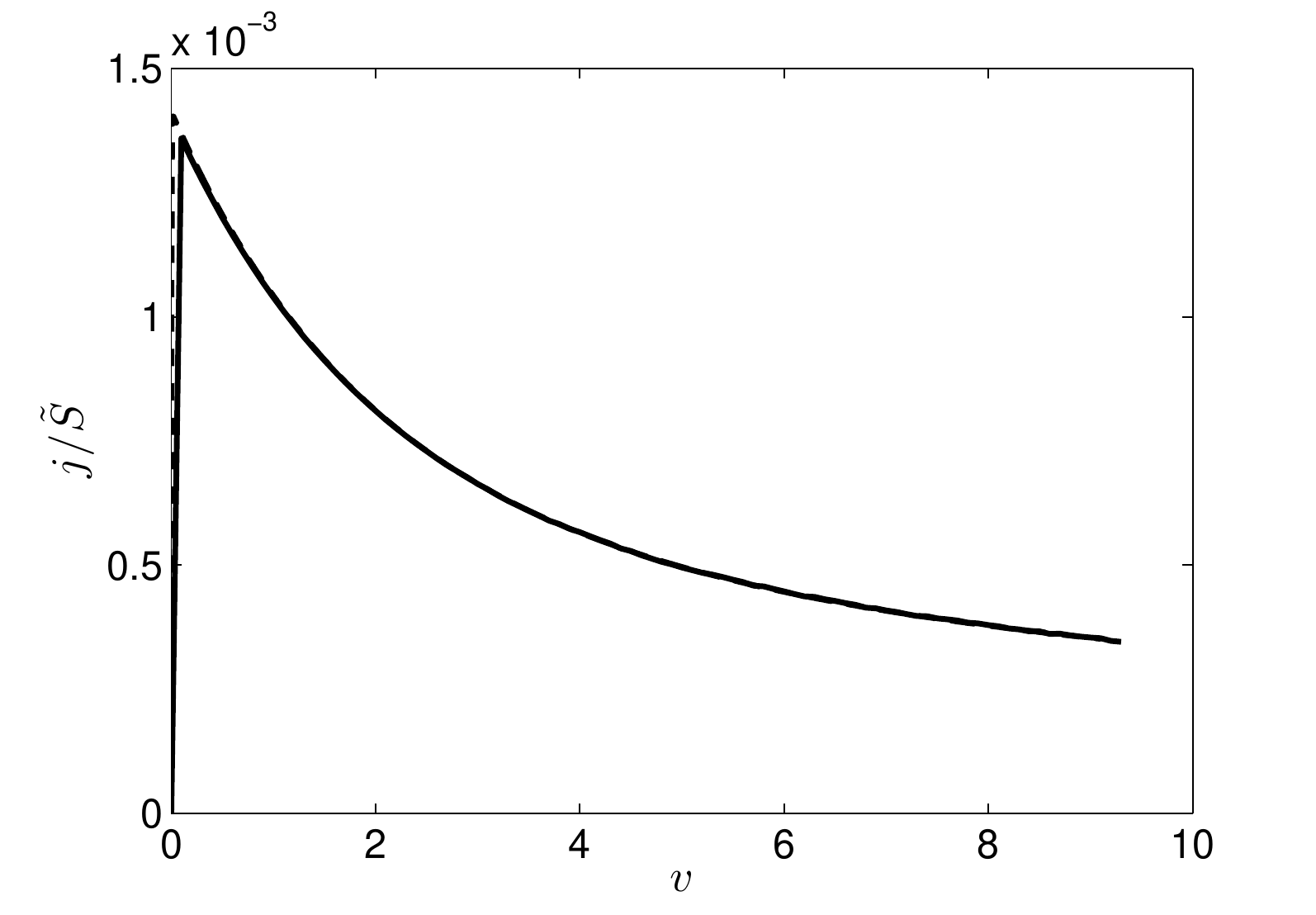}
        \caption{Positive voltage sweep}\label{sol_elec_cap_SE_pos}
    \end{subfigure}

    \begin{subfigure}[htb!]{0.49\textwidth}
        \centering
        \includegraphics[width=\linewidth]{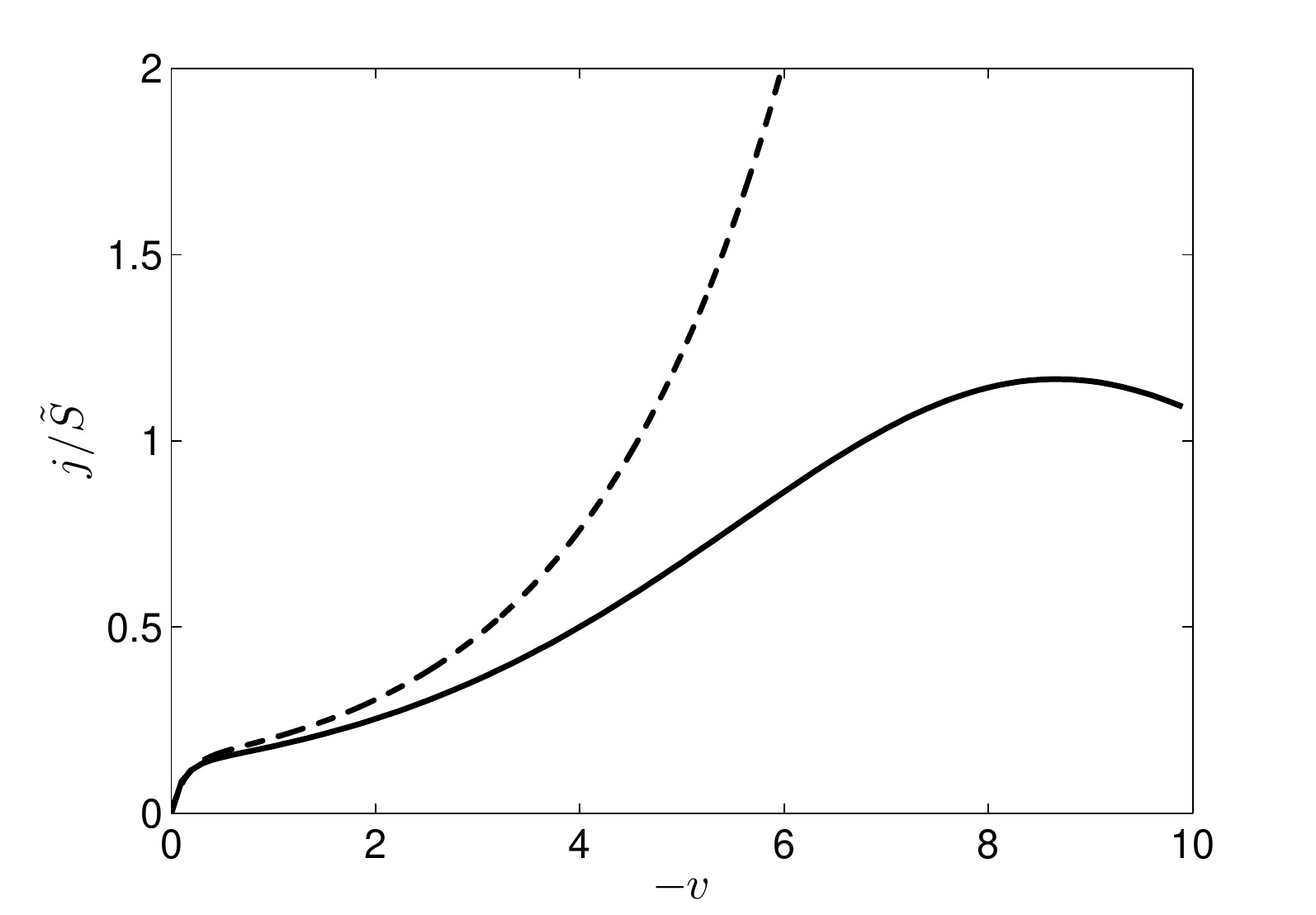}
        \caption{Negative voltage sweep}\label{sol_cap_SE_thick_neg}
    \end{subfigure}
    \begin{subfigure}[htb!]{0.49\textwidth}
        \centering
        \includegraphics[width=\linewidth]{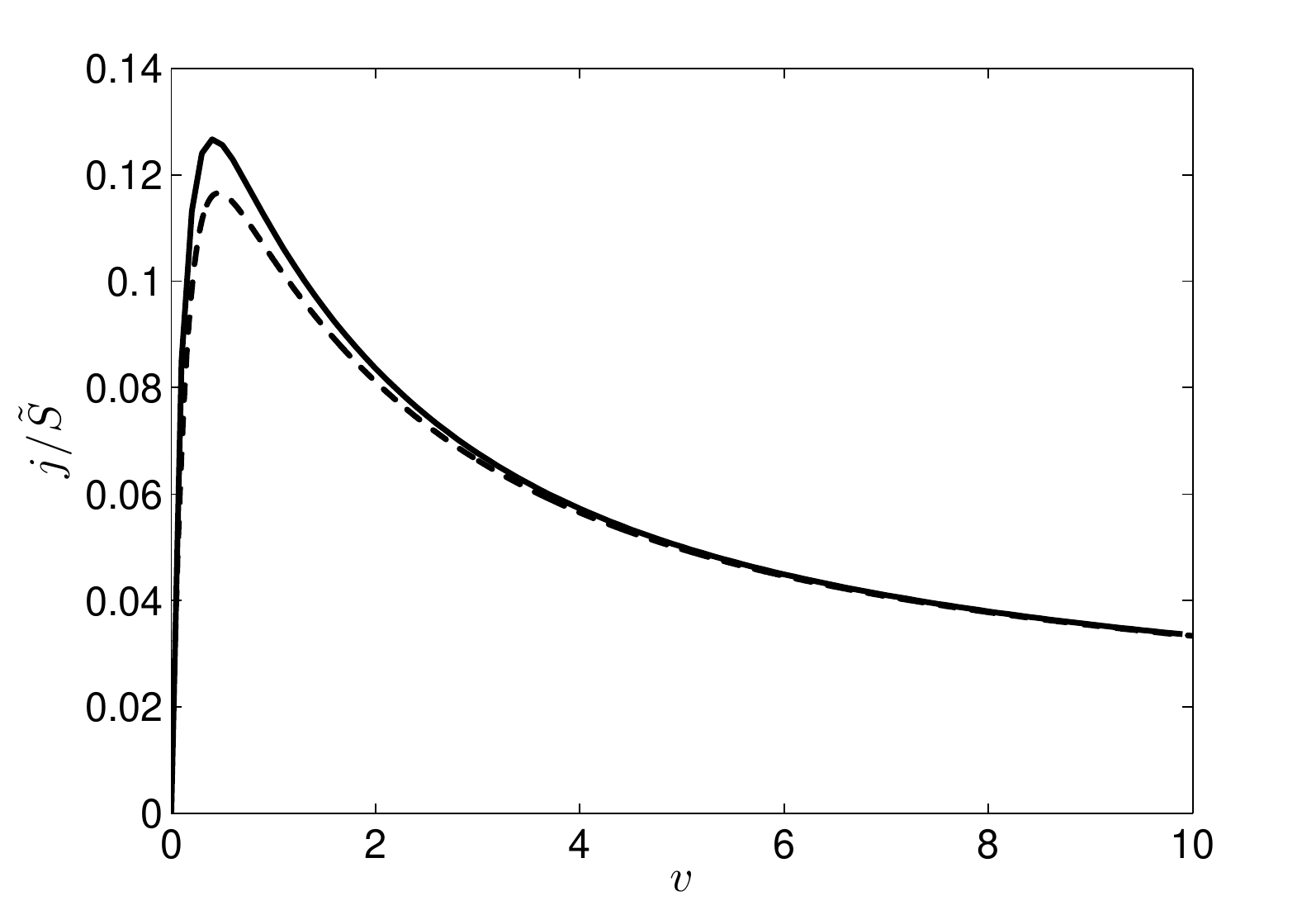}
        \caption{Positive voltage sweep}\label{sol_cap_SE_thick_pos}
    \end{subfigure}
    \caption{Simulated (solid lines) vs. uniformly valid approximation (dashed lines) $j$ vs $v$ curves for (a), (b) thin EDL solid electrolyte and (c), (d) thick EDL solid electrolyte with a single blocking electrode. $\delta=0.01$ and $\tilde S=\pm 1$. Figures (a) and (c) show the negative part of the sweep and Figures (b) and (d) show the positive part. Dashed lines are plotted from equation \eqref{matching_solid_SE}.}\label{sol_cap_SE}
\end{figure}

It is interesting to note that such a departure from equilibrium behavior is not apparent for the thick EDL solid electrolyte shown in Figure \ref{sol_cap_thick_DE_10}. However, if we plot the single electrode, thick EDL curves (shown in Figure \ref{sol_cap_SE_thick_neg} for the negative sweep and Figure \ref{sol_cap_SE_thick_pos} for the positive sweep), we see that there is indeed a departure from the equilibrium capacitance curve for the positive voltage ($\tilde S=1$) sweep, with the simulated capacitance having non-monotonic dependence on voltage. This disagreement is masked by the fact that for two electrodes (which can be thought of as two capacitors in series, as in equation \eqref{matching_solid_DE}), the smaller, positive voltage sweep capacitance, which agrees with the equilibrium approximation, dominates. Figure \ref{sol_cap_SE} also shows the single electrode results for a thin EDL solid electrolyte, which agree with the equilibrium approximation.

We end by remaking that, despite some disagreements in the thick EDL cases, every simulation showed very good agreement with the approximations we used during the RC charging portion of the curve, i.e. the inner solution in equation \eqref{matching}.

%\begin{figure}[htb!]
%\centering
 %   \begin{subfigure}[htb!]{0.49\textwidth}
   %     \centering
     %   \includegraphics[width=\linewidth]{sol_cap_thick_DE_01.pdf}
    %    \caption{$\tilde S=-0.1$}\label{sol_cap_thick_DE_01}
%    \end{subfigure}
  %  \begin{subfigure}[htb!]{0.49\textwidth}
  %      \centering
  %      \includegraphics[width=\linewidth]{sol_cap_thick_DE_1.pdf}
   %     \caption{$\tilde S=-1$}\label{sol_cap_thick_DE_1}
 %   \end{subfigure}
%
 %   \begin{subfigure}[htb!]{0.49\textwidth}
  %      \centering
  %      \includegraphics[width=\linewidth]{sol_cap_thick_DE_10.pdf}
%\caption{$\tilde S=-10$}\label{sol_cap_thick_DE_10}
 %   \end{subfigure}
  %  \begin{subfigure}[htb!]{0.49\textwidth}
   %     \centering
 %       \includegraphics[width=\linewidth]{sol_cap_thick_DE_100.pdf}
%        \caption{$\tilde S=-100$}\label{sol_cap_thick_DE_100}
%    \end{subfigure}
 %   \caption{Simulated (solid line) and uniformly valid approximation (dashed line) $j$ vs. $v$ curves for a thick EDL solid electrolyte with two blocking electrodes and parameters $\epsilon=0.1$, $\delta=0.01$ and various values of $\tilde S$. Dashed lines are plotted from equation \eqref{matching_solid_DE}}\label{sol_cap_thick_DE}
%\end{figure}

\section{Conclusions and Future Work}\label{conclusion}

This paper provides a general theory to enable the use of LSV/CV to characterize electrochemical systems with diffuse charge. Our paper presents \change{and extends} theory and simulations in a variety of situations which extend classical interpretations (unsupported liquid electrolytes and systems with blocking electrodes) and develop an understanding for more complicated situations (thin films, systems where bulk electroneutrality breaks down with space charge formation, and leaky membranes). Following a thorough historical review, we began with single-electrode voltammograms for supported electrolytes, with an analytical expression in the limit of fast reactions, as well as for unsupported electrolytes in the limit of small $\epsilon$. We showed for these systems where inclusion of diffuse charge dynamics plays a larger role, namely when a system with small $\delta$ has a large voltage applied.

Next, we applied ramped voltages to solid and liquid thin films to obtain current-voltage relationships and discussed the effect of various types of polarization, and how their effect on the current differed between liquid and solid electrolytes. For liquid electrolytes, we also observed the formation of space charge regions at large voltages, another prediction which is made possible by using the PNP-FBV equations. For our leaky membrane model simulations, we found that simulations matched a steady-state analytical expression reasonably well when the background charge was opposite in polarity to the reactive ion (negative background charge) but did not match theory with positive background charge. We ended by presenting analytical expressions for the capacitance of liquid and solid electrolyte systems with blocking electrodes, and compared them to simulations with both thin and thick double layers. We found that our analytical approximations worked well for thin double layers, but in some cases disagreed when double layers were thick. In general, we conclude that diffuse charge dynamics becomes important in voltammetry at large applied voltages and/or with thick double layers.

For each type of system we considered, the simulation results we obtained were compared to limiting cases and we showed where simple analytical expressions can be used to predict behavior (such as approximating capacitance curves), and which regimes require more careful analysis and simulation. This is of practical interest for electrochemists and engineers, for whom it can assist in guiding the design of new devices and experiments.

For future work, there many ways to extend the model for additional physics. For example, specific adoption of ions could be added to the boundary conditions, providing an additional mechanism for charge regulation of the surface~\cite{Deng2013, Gentil2006, Behrens2001}, in addition to Faradaic polarization, coupled through the FBV equations. The PNP ion transport equations could be extended to include recombination bulk reactions (ion-ion, ion-defect, etc) \cite{Luntz2015} or various models of ion crowding effects \cite{Bazant2009, Kilic2007b, Kilic2007a, Kornyshev2007} or other non-idealities in concentrated solutions. There is also the possibility of coupled mass transport fluxes (Maxwell Stefan, dusty gas, etc) \cite{Taylor1993}, which become important in concentrated electrolytes ~\cite{Newman2012}\change{, as well as for unequal diffusion coefficients.}  Poisson's equation could also be modified to account for electrostatic correlations \cite{Bazant2011} or dielectric polarization effects~\cite{Hatlo2012}. Furthermore, while we used generalized BV kinetics \cite{Bazant2013ACR} in this work, Marcus kinetics \cite{Bard2001, Zeng2014} may provide a more accurate model of charge transfer reactions. Even keeping the same PNP-FBV framework, it would be interesting to extend the model for 2d or 3d geometries, convection, and moving boundaries, in order to describe conversion batteries, electrodeposition and corrosion.  

\clearpage
\appendix
\section*{Derivation of Modified Randles-Sevcik Equation}\label{app:MRS}
In this appendix, we solve the semi-infinite diffusion equation with a nonhomogeneous boundary condition. The result we obtain is referenced in Section \ref{supported_electrolytes} as an analogue to the original Randles-Sevcik function for a single ion electrodeposition reaction. The method used here  is taken from Chapter 5.5 of O'Neil \cite{ONeil2008}. The equation and boundary conditions are
\begin{equation}
\label{pde_0}
u_t=u_{xx}
\end{equation}
\begin{equation}
\label{ic_0}
u(x,0) = A\,\,\, \text{for}\,\,\, x>0
\end{equation}
\begin{equation}
\label{bc_0}
u(0,t)=f(t) \,\,\, (\text{where} \,\, f(0)=A)
\end{equation}
We begin by considering the same problem with a jump at the boundary at time $t=t_0$:
\begin{equation}
\label{pde}
u_t=u_{xx}
\end{equation}
\begin{equation}
\label{ic}
u(x,0) = A\,\,\, \text{for}\,\,\, x>0
\end{equation}
\begin{equation}
\label{bc}
u(0,t)=\begin{cases}
A & 0<t<t_0 \\
B & t>t_0 \\
\end{cases}
\end{equation}
where $A$, $B$ are positive constants. $u(0,t)$ can be written with the Heaviside step function $H(t)$ as
\begin{equation}
\label{bc_heavi}
u(0,t)=A(1-H(t-t_0)) + BH(t-t_0)
\end{equation}
The PDE \eqref{pde}--\eqref{bc} can be solved via Laplace transform. The Laplace transform of equation \eqref{pde} is
\begin{equation}
\label{pde_lp}
sU(x,s)-A=\partial_{xx} U(x,s)
\end{equation}
where $U(x,s)=\mathscr{L}u(x,t)$. The general solution of equation \eqref{pde_lp}, a constant-coefficient second order ODE for $U$, is
\begin{equation}
\label{gensol}
U(x,s)=a(s)e^{\sqrt{s}x}+b(s)e^{-\sqrt{s}x} + \frac{A}{s}
\end{equation}
We impose the condition $a(s)=0$ because we would like the solution to be bounded as $x\rightarrow \infty$. This leaves us to solve for $b(s)$ using the boundary condition (equation \eqref{bc_heavi}). The Laplace transform of equation \eqref{bc_heavi} is
\begin{equation}
\label{bc_lp}
\mathscr L u(0,t)=\frac{A}{s} - \frac{A}{s}e^{-t_0s} + \frac{B}{s}e^{-t_0s}
\end{equation}
Using equation \eqref{bc_lp}, $b(s)$ can be solved to be
\begin{equation}
b(s)=\frac{B-A}{s}e^{-t_0 s}
\end{equation}
giving us
\begin{equation}
\label{sol_lp}
U(x,s)=\frac{B-A}{s}e^{-t_0 s}e^{-\sqrt{s} x} + \frac{A}{s}
\end{equation}
We can now take the inverse Laplace transform of equation \eqref{sol_lp} to obtain
\begin{equation}
\label{sol_gen}
u(x,t)=A + (B-A)\erfc\left(\frac{x}{2\sqrt{t-t_0}}\right)
\end{equation}
where $\erfc(z)=1-\erf(z)=\frac{2}{\sqrt \pi} \int_z^\infty \exp \left(x^2\right) \, dx$ is the complementary error function. In other words, the solution to the semi-infinite diffusion problem with a jump of $\Delta u(0,t_0)$ relative to $A$ at $x=0$ and time $t=t_0$ is
\begin{equation}
\label{sol_gen2}
u(x,t)=A + \Delta u(0,t_0) \erfc\left(\frac{x}{2\sqrt{t-t_0}}\right)
\end{equation}
Now, using Duhamel's principle, we have for an arbitrary series of steps at $t=t_i$ given by $\Delta u(0,t_i)=f(t_i)$,
\begin{equation}
\label{sol_gen3}
u(x,t)=A + \sum_i\Delta f(t_i) \erfc\left(\frac{x}{2\sqrt{t-t_i}}\right)
\end{equation}
where $\Delta f(t_i)=f(t_{i+1})-f(t_i)$. We can take the continuum limit with $\Delta f(t_i) = \frac{df}{dt}(\tilde t_i) \Delta t_i$, where $t_i<\tilde t_i<t_{i+1}$ to obtain
\begin{equation}
\label{sol_gen4}
u(x,t)=A + \int_0^t \frac{df}{dt}(t') \erfc\left(\frac{x}{2\sqrt{t-t'}}\right) \, dt'
\end{equation}
and finally we can make the substitution $\tau = t-t'$ to write
\begin{align}
\label{sol_gen5}
u(x,t)&=A - \int_t^0 \frac{df}{dt}(t-\tau) \erfc\left(\frac{x}{2\sqrt{\tau}}\right) \, d\tau \nonumber \\
&=A + \int_0^t \frac{df}{dt}(t-\tau) \erfc\left(\frac{x}{2\sqrt{\tau}}\right) \, d\tau
\end{align}
Equation \eqref{sol_gen5} may be numerically integrated. Alternatively, if we are interested in the Nernstian function, $f(t)=e^{-St}$ (for quasi-equilibrium fast reactions, as in equation \eqref{nernst}) and wish to compute the resulting current $j=u_x(0,t)$, we can arrive at an analytical expression for $u_x(0,t)$,
\begin{equation}
\label{analytical0}
u_x(x,t)=Se^{-St}\int_0^t \frac{e^{S\tau}e^{-\frac{x^2}{4\tau}}}{\sqrt{\pi \tau}} \, d\tau
\end{equation}
\begin{equation}
\label{analytical}
u_x(0,t)=\sqrt S e^{-St} \erfi\left(\sqrt{S t}\right)
\end{equation}
where $\erfi (z) = \frac{2}{\sqrt \pi}\int_0^z  \exp \left(x^2\right) \, dx$ is the imaginary error function.

\section*{References}

%\bibliography{LSV_MZB}

\end{document}